\documentclass[%
 reprint,
 amsmath,amssymb,
 aps,
pra,
]{revtex4-2}

\usepackage{graphicx}
\usepackage{dcolumn}
\usepackage{bm}

\usepackage{braket}
\usepackage{physics}
\usepackage{hyperref}
\usepackage{url}
\usepackage{amsmath}
\usepackage{amssymb}
\usepackage{mathtools}
\usepackage{comment}
\usepackage{xcolor} 
\usepackage{array} 
\usepackage{rotating}

\usepackage{mathrsfs}
\usepackage{cancel}
\usepackage{soul}
\usepackage{tikz}
\usetikzlibrary{quantikz2}
\usetikzlibrary{shapes.geometric, arrows, positioning, arrows.meta}
\newcommand{\QLayer}[3]{\gategroup[#1,steps=#2,style={dashed,rounded corners,fill=blue!20, inner xsep=0.2pt},background,
label style={label position=below,anchor=north,yshift=-0.2cm}]{{#3}}}
\usepackage{adjustbox}

\begin{document}

\preprint{APS/123-QED}

\title{Quantum generative modeling for financial time series with temporal correlations}

\author{David Dechant$^{1,2}$}
\author{Eliot Schwander$^{1,2}$}
\altaffiliation{These authors contributed equally to this work}
\author{Lucas van Drooge$^{1,2}$}
\altaffiliation{These authors contributed equally to this work}
\author{Charles~Moussa$^{1,3}$}
\author{Diego Garlaschelli$^{2,4}$}
\author{Vedran Dunjko$^{1,3}$}
\author{Jordi Tura$^{1,2}$}
\affiliation{$^1$ $\langle aQa^L \rangle$ Applied Quantum Algorithms, Universiteit Leiden, the Netherlands}%
\affiliation{$^2$Instituut-Lorentz, Universiteit Leiden, P.O. Box 9506, 2300 RA Leiden, The Netherlands}%
\affiliation{$^3$LIACS, Universiteit Leiden, Niels Bohrweg 1, 2333 CA Leiden, Netherlands}%
\affiliation{$^4$IMT School of Advanced Studies, Lucca, Italy}

\date{\today}

\begin{abstract}
Quantum generative adversarial networks (QGANs) have been investigated as a method for generating synthetic data with the goal of augmenting training data sets for neural networks. 
This is especially relevant for financial time series, since we only ever observe one realization of the process, namely the historical evolution of the market, which is further limited by data availability and the age of the market.
However, for classical generative adversarial networks it has been shown that generated data may (often) not exhibit desired properties (also called \textit{stylized facts}), such as matching a certain distribution or showing specific temporal correlations.
Here, we investigate whether quantum correlations in quantum inspired models of QGANs can help in the generation of financial time series.
We train QGANs, composed of a quantum generator and a classical discriminator, and investigate two approaches for simulating the quantum generator: a full simulation of the quantum circuits, and an approximate simulation using tensor network methods.
We tested how the choice of hyperparameters, such as the circuit depth and bond dimensions, influenced the quality of the generated time series.
The QGAN that we trained generate synthetic financial time series that not only match the target distribution but also exhibit the desired temporal correlations, with the quality of each property depending on the hyperparameters and simulation method.
\end{abstract}

\maketitle

\section{Introduction}\label{sec:introduction}





In recent years, the use of machine learning - in particular neural-network based approaches - has expanded across many domains~\cite{prince2023Understandingdeep}.
A particular approach are the generative adversarial networks (GANs), in which successively a generator and a discriminator are trained~\cite{goodfellow2014GenerativeAdversarial, goodfellow2020Generativeadversarial}.
The generator learns the underlying distribution and samples from it, while the discriminator learns to distinguish real data from generated samples.
Typically, GANs are applied in image generation~\cite{radford2016UnsupervisedRepresentation, karras2018ProgressiveGrowing, karras2019stylebasedgenerator}, but also to other data~\cite{gui2021reviewgenerative}.

The ability of machine learning models to generalize well relies on the availability of large datasets~\cite{prince2023Understandingdeep}.
Data augmentation methods are techniques which increase the training data set in order to alleviate these limitations~\cite{shorten2019surveyImage}. 
Those methods typically involve slightly modifying the training data, but
synthetic data generation by GANs is used as an approach for data augmentation~\cite{dossantostanaka2019DataAugmentation, shorten2019surveyImage}.

This is particularly relevant for finance, a computationally-heavy, yet difficult-to-model field~\cite{dixon2020MachineLearning}. 
Unlike domains where there is an abundance of high-quality data to train on, finance faces a fundamental challenge: the inherent non-repetitive nature of financial events.
For example, the time-series of a specific asset's price can only be observed once. 
A machine learning model that aims to learn properties based on the time-series of a specific asset therefore is heavily limited, as their ability to generalize well relies on large datasets.
By learning the underlying distribution of financial time series as well as its desired temporal properties, one can generate new data that enables the creation of richer training sets~\cite{potluru2024SyntheticData, assefa2020GeneratingSynthetic}.
In particular, temporal correlations such as volatility clustering (periods of large variation are followed by periods of large variation, as do periods of low variation) are important for single time series.

Parallel to the development of machine learning, research in quantum computing and its potential application has increased. 
Motivated by the progress in quantum hardware, quantum algorithms research has been focusing on variational quantum algorithms in the last decade~\cite{preskill2018QuantumComputing}.
These hybrid algorithms consist of succinct calculations on a parameterized quantum circuit (PQC) and a classical optimizer~\cite{cerezo2021Variationalquantum}.
The PQC contains tunable and fixed gates, and the classical optimizer calculates updated parameters based on measurements conducted on the final quantum state of the PQC at each step, until a certain precision or goal is reached.

Previous work proposed replacing the classical generator of a GAN with a parameterized quantum circuit~\cite{lloyd2018QuantumGenerative,dallaire-demers2018Quantumgenerative}. 
Quantum circuits have been proven to enable sampling from distributions which are intractable for classical circuits, and so the set of distributions they access is in general different than what classical models access. 
Consequently, it is expected they may be more effective with some classes of distributions that classical models struggle with~\cite{aaronson2013ComputationalComplexity, wilms2025Quantumreinforcement, abbas2021powerquantum, molteni2025Quantummachine}.
Subsequent studies expanded on this idea and examined if this property can be harnessed in the context of learning financial distributions. 
In particular, in~\cite{coyle2021Quantumclassical}, the idea to use QGANs for synthetic data generation of financial time series has been proposed and tested, using a quantum circuit Born machine, which performed better than a classical restricted Boltzmann machine on learning the distribution of correlated asset pairs with respect to the Wasserstein distance.
However, generating financial time series which do not only follow the same distribution as real-world data, but also show their temporal correlations is challenging~\cite{dogariu2022GenerationRealistic}.

In this work, we develop a quantum GAN for synthetic data generation and examine its ability in generating time series replicating the S\&P~500 index, including its distribution and temporal correlations. 
The discrete time series spans from 20 to 40 points in time, where the expectation value of single-qubit Pauli-$X$ and Pauli-$Z$ operators is interpreted as the log return of the value of the index at each time step. 
The choice of these observables will be motivated in Section~\ref{subsec:pqcsimulation}.
We simulate the quantum circuits both with full-state simulations and with matrix product state (MPS) simulations (also known as the tensor train)~\cite{ostlund1995ThermodynamicLimit, verstraete2008Matrixproduct}. 
While full-state simulations are limited to short time intervals due to their exponential scaling, MPS simulations enable us to model longer time series by exploiting their ability to efficiently replicate linear structures such as financial time series. 
In the MPS simulations, we vary the bond dimension (also referred to as the tensor train rank) to balance computational costs and simulation accuracy.

We show that our model is able to learn the underlying time series close to the real time series distribution.
Further, the samples show temporal correlations, which are qualitatively similar to those observed in real-world data.
Our work shows the potential use of QGANs in learning time series with specific temporal correlations.

This paper is organized as follows.
In Section~\ref{sec:background}, we introduce the concepts of financial time series and its properties, as well as the concept of QGANs. 
Further, we cover related work.
In Section~\ref{sec:implementation}, we present the simulations we used. 
We detail the data pre-processing as well as the quantum generator and the matrix product state simulation.
We show the results, of our simulations in Section~\ref{sec:results}, discuss them in Section~\ref{sec:discussion}, and conclude in Section~\ref{sec:conclusions}.

\begin{figure*}[ht]
\centering
\begin{adjustbox}{width=\linewidth, center}
\begin{tikzpicture}[
  block/.style={draw, thick, minimum height=2cm, minimum width=3cm, align=center},
  arrow/.style={thick, -{Latex[length=3mm]}},
  feedback/.style={thick, dashed, -{Latex[length=3mm]}, font=\small, sloped},
  label/.style={font=\small, align=center},
  node distance=1.5cm and 2cm
]

\node[block, fill=blue!20] (gen) {Generator\\$G(\vec{z})$};

\coordinate (noise_pos) at ([xshift=-1.5cm]gen.west);
\node[align=center] at (noise_pos) (noise) {$\vec{z}$\\Noise};

\node[block, fill=gray!10, right=1.5cm of gen] (fake) {Generated\\time series};
\node[block, fill=gray!10, below=1cm of fake] (real) {Real\\time series};
\node[block, fill=red!20, right=1.5cm of fake] (disc) {Discriminator\\$D(x)$};
\node[block, fill=green!20, right=1.5cm of disc] (output) {Output:\\ Estimation of $W_1(\mathcal{P}_r, \mathcal{P}_g)$};

\draw[arrow] (noise) -- (gen);
\draw[arrow] (gen) -- (fake);
\draw[arrow] (fake) -- (disc);
\draw[arrow] (real) -- (disc);
\draw[arrow] (disc) -- (output);

\draw[feedback, bend right=20] (output.north) to node[above, font=\large] {Generator loss} (gen.north east);
\draw[feedback, bend left=50] (output.south) to node[below, font=\large] {Discriminator loss} (disc.south east);

\end{tikzpicture}%
\end{adjustbox}
\caption{Structure of a generative adversarial network (GAN) used for time series generation. The discriminator takes both the generated and real time series as input and outputs its estimate for the Wasserstein distance $W_1(\mathcal{P}_r, \mathcal{P}_g)$ (see Eqs.~\eqref{eq:Wasserstein_distance} and~\eqref{eq:Estimate_Wasserstein_distance}). Both the generator and discriminator are trained using different loss functions derived from the discriminator's output.}
\label{fig:gan}
\end{figure*}
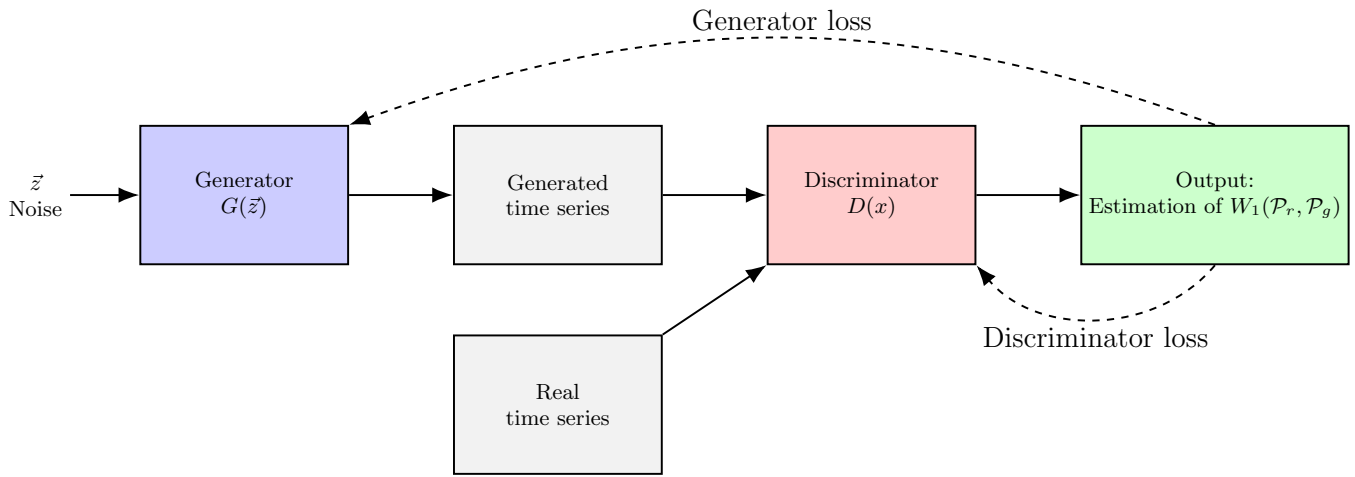

\section{Background}\label{sec:background}
In this section, we will introduce both financial time series and its temporal correlations as well as generative adversarial networks, and their adaptations based on the Wasserstein distance and with a quantum generator. 
Further, we will present related work.
\subsection{Financial time series }\label{subsec:financialtimeseries}
A time series is a set of data points ordered over a given time frame, typically at equally spaced time intervals. 
A financial time series is the set of financial variables such as prices, returns and volatility of assets, indices or other financial instruments.\\
An example of a financial time series is the S\&P~500 index, which includes 500 of most valuable companies that are listed on US stock exchanges~\cite{2025SP500}. 
The price of many instruments, such as the S\&P~500 index, deviates around a mean value that grows in time.
Instead, examine the time series of the log return $r_t$, which is not dependent on this general market growth:
\begin{align}\label{eq:log_return}
    r_t=\log\left(\frac{S_t}{S_{t-1}}\right),
\end{align}
where $S_t$ is the price of the index at time $t$.
Simple models such as the Black-Scholes model~\cite{black1973PricingOptions} assume a normal distribution of these log returns.
However, the distributions observed in the market have more complicated properties.
The returns of assets do not typically follow a normal distribution, their distributions are more narrowly and spiked around the mean and have heavier tails (extreme events are more likely), which is in contrast to the normal distribution of the Black-Scholes model.
Therefore, a main concern of research in finance is about creating models that can mimic time series with higher accuracy, and machine learning approaches have been increasingly explored for this case~\cite{dixon2020MachineLearning}.
Many observed log returns share common properties, also called stylized facts, that stem mostly from behavior of parties that interact with the market~\cite{cont2001Empiricalproperties}. 
These properties can be used in order to assess the quality of models of financial time series.
In this paper, we focus on four stylized facts: non-Gaussianity, the absence of linear autocorrelation, volatility clustering and the leverage effect.
The first of them is in contrast to the Black-Scholes model, as described above.
The latter three stylized facts, which are temporal correlations, can be formulated with the help of the correlation function $corr(X,Y)$, which for the random variables $X$ and $Y$ is defined as:
\begin{align}
    corr(X,Y)=\frac{cov(X,Y)}{\sigma_X\sigma_Y}=\frac{\left\langle\left(X-\mu_{X}\right)\left(Y-\mu_{Y}\right)\right\rangle}{\sigma_X\sigma_Y},
\end{align}
where $cov(X,Y)$ is the covariance, $\sigma_X$ and $\sigma_Y$ the standard deviations of the random variables and $\mu_{X}$ and $\mu_{Y}$ denote their respective means.\\
Current and past values of those time series are typically not linearly autocorrelated.
At time $t$, this means that for all time differences $\tau>0$, the expectation value 
$\mathbb{E}\left[corr(r_t,r_{t+\tau})\right]$
is zero.
Intuitively this comes from the fact that any trend in the return is exploited by traders, which in turn weaken the effect. 
This exploitation of traders is a corollary of the efficient market hypothesis.\\
However, the absolute returns typically exhibit correlation that slowly decays in time. 
This effect is also called volatility clustering, and can be examined by calculating the quantity
$corr(|r_t|,|r_{t+\tau}|)$.
It quantifies the observation that large changes in the price are followed by large changes, and equivalently small changes are followed by small changes.\\
When the price of an asset sinks, the volatility typically rises.
This is called the leverage effect, which can be measured in the quantity $corr(|r_{t+\tau}^2|,r_t)$.\\
The reader can find more details of these properties in~\cite{cont2001Empiricalproperties}.\\
Synthetic data generation of financial time series concerns the generation of artificial time series that observe these stylized facts.
As these properties differ from time series to time series and as they are difficult to compare in general, synthetic time series are typically assessed qualitatively if they are able to capture those properties~\cite{dogariu2022GenerationRealistic}.\\
However, in this work, we provide several quantitative metrics. 
In order to quantify how closely the generated time series reproduce the stylized facts of the S\&P~500 index, we define the following metrics:
\begin{align}
    EMD(\theta)=& \left( \frac{1}{\tau_{max}+1} \sum_{\tau=0}^{\tau_{max}} \left| r_{t+\tau}^{(SP500)} - r^{(\theta)}_{t+\tau} \right| \right) \label{eq:EMD_definition}\\
    E^{ACF}_{id}(\theta)=&\left(\frac{1}{\tau_{max}}\sum_{\tau=1}^{\tau_{max}}corr\left(r^{(\theta)}_t,r^{(\theta)}_{t+\tau}\right)^2\right)^{1/2} \\
    E^{ACF}_{abs}(\theta)=&\bigg(\frac{1}{\tau_{max}}\sum_{\tau=1}^{\tau_{max}}\Big[corr\left(|r^{(SP500)}_t|,|r^{(SP500)}_{t+\tau}|\right)\nonumber\\
    &-corr\left(|r^{(\theta)}_t|,|r^{(\theta)}_{t+\tau}|\right)\Big]^2\bigg)^{1/2} \\
    E_{Lev}(\theta)=&\bigg(\frac{1}{\tau_{max}}\sum_{\tau=1}^{\tau_{max}}\big[corr\left(|r^{(SP500)}_t|^2,r^{(SP500)}_{t+\tau}\right) \nonumber\\
    &-corr\left(|r^{(\theta)}_t|^2,r^{(\theta)}_{t+\tau}\right)\Big]^2\bigg)^{1/2}  \ .\label{eq:lev_definition}
\end{align}
They are quantifying non-Gaussianity, absence of linear autocorrelation, volatility clustering and leverage effect, respectively. The first is based on the earth-movers distance, which is the discretized form of the Wasserstein distance, and the latter quantities are derived from the correlation functions describing these properties, as explained above.
Here, the log returns of the generated time series are written as $r^{(\theta)}_t$, where $\theta$ stand for the parameters and hyperparameters of the simulated QGAN, and the log returns of the S\&P~500 index are written as $r^{(SP500)}_t$.
The lower these metrics are, the closer the stylized facts of the generated time series $r^{(\theta)}_t$ are resembling those of the time series $r^{(SP500)}_t$.

\subsection{Wasserstein QGAN}\label{subsec:gan}
Generative adversarial networks (GANs)~\cite{goodfellow2014GenerativeAdversarial,goodfellow2020Generativeadversarial} are unsupervised machine learning-based methods that are in particular powerful in generating images, but are also successfully applied in the generation of financial time series~\cite{eckerli2021GenerativeAdversarial}. 
They consist of two neural network that compete in a game-like setup.
The generator takes random noise as input and aims to create artificial data that is indistinguishable from real data.
Both real data from the training set and artificial data from the generator is then fed to a discriminator which is being trained to detect the generated data, outputting a probability of the input data being real or fake.
They are trained in an alternating fashion until the generator is able to create data indistinguishable from real data.
GANs face challenges in training instability (one neural network overpowers the other) and mode collapse (The GAN focuses on creating data with limited variety)~\cite{saxena2022GenerativeAdversarial, arjovsky2017WassersteinGenerative}.\\
Those challenges can be mitigated by replacing the discriminator with a critic that learns the Wasserstein distance between the real and generated data distributions, in the so-called Wasserstein GAN~\cite{arjovsky2017WassersteinGenerative}.
The Wasserstein distance between the real and generated distributions $\mathcal{P}_r$ and $\mathcal{P}_g$, respectively, is defined as:
\begin{align}\label{eq:Wasserstein_distance}
    W_1(\mathcal{P}_r, \mathcal{P}_g):=\inf_{\pi\in\Gamma(\mathcal{P}_r, \mathcal{P}_g)}\mathbb{E}_{(x,y)\sim\pi}\left(\|x-y\|\right).
\end{align}
Here, $\Gamma(\mathcal{P}_r, \mathcal{P}_g)$ is the joint probability distribution of $\mathcal{P}_r$ and $\mathcal{P}_g$. 
Calculating this infimum is not feasible in practice, but the Kantorovich-Rubinstein duality delivers a quantity that can be used in a machine learning context~\cite{villani2009Wassersteindistances,arjovsky2017WassersteinGenerative}: 
\begin{align}\label{eq:Estimate_Wasserstein_distance}
    W_1(\mathcal{P}_r, \mathcal{P}_g)=\sup_{\|f\|_L\leq 1}\left(\mathbb{E}_{x\sim\mathcal{P}_r}\left(f(x)\right)-\mathbb{E}_{\tilde{x}\sim\mathcal{P}_g}\left(f(\tilde{x})\right)\right),
\end{align}
where $\sup_{\|f\|_L\leq 1}$ is the supremum over all $1$-Lipschitz functions.
The role of the critic $D$ is to maximise the loss function \begin{align}
\label{eq:loss-discriminator}
L_{D}(D,\mathcal{P}_g)=&\mathbb{E}_{x\sim\mathcal{P}_r}\left(D(x)\right)-\mathbb{E}_{\tilde{x}\sim\mathcal{P}_g}\left(D(\tilde{x})\right)\\
&+\lambda \mathbb{E}_{\hat{x}\sim\mathcal{P}_{\hat{x}}}\left(\left(\|\nabla_{\hat{x}} D(\hat{x})\|_2-1\right)^2\right)\ ,
\end{align}
where $D(x)$ is trained to approximate $f(x)$ in Eq.~\eqref{eq:Estimate_Wasserstein_distance}. 
The latter term is a gradient penalty regularization~\cite{gulrajani2017Improvedtraining} that enforces the $1$-Lipschitz condition by a scaling parameter $\lambda$ and where $\hat{x}=\epsilon x+(1-\epsilon)\tilde{x}$ with the random parameter $\epsilon\sim U[0,1]$. This loss function will train the critic to approximate the Wasserstein distance between the probability distributions of real and generated data. 
In contrast, the role of the generator is to maximize 
\begin{align}
L_{G}
\label{eq:generator-loss}(D,\mathcal{P}_g)=\mathbb{E}_{\tilde{x}\sim\mathcal{P}_g}\left(D(\tilde{x})\right).
\end{align}
See Figure~\ref{fig:gan} for a sketch of a Wasserstein GAN. \\
Quantum generative adversarial networks (QGANs) are GANs in which the classical generator and/or the classical discriminator are replaced by a quantum circuit~\cite{lloyd2018QuantumGenerative,dallaire-demers2018Quantumgenerative}.\\ 
They are motivated by the fact that quantum circuits can learn distributions efficiently that are hard to model by classical means~\cite{aaronson2013ComputationalComplexity, wilms2025Quantumreinforcement, abbas2021powerquantum, molteni2025Quantummachine}.
The proofs of advantage showcase that learning can not be achieved when the distribution is hard (generated by a quantum process). 
Market data is manifestly not so. 
However, in other models it was shown that one can have learning separations even if the generation of the data is classically tractable~\cite{sweke2021QuantumClassical}.
It remains an open question if such separations can also hold for estimation value sampler as we use here. 
However, even without separations it may be the case that quantum models simply have more convenient inductive biases than classical models and understanding those is valuable, and the interests of this work go into this direction.

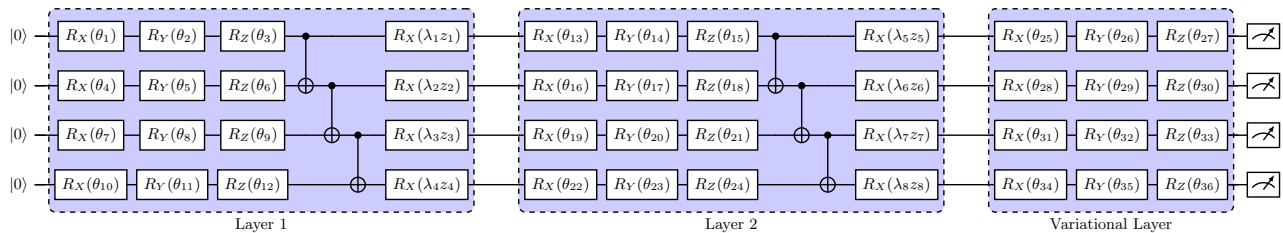
\begin{figure*}
\centering
\begin{adjustbox}{width=0.95\textwidth}
\begin{quantikz}[scale=0.6, every node/.style={transform shape},column sep=0.2cm, row sep={1cm,between origins}]
        \lstick{$\ket{0}$} 
        & \qw    & 
       \gate{R_X{\left(\theta_{1}\right)}} \QLayer{4}{8}{Layer 1} & \gate{R_Y{\left(\theta_{2}\right)}} & \gate{R_Z{\left(\theta_{3}\right)}}  &\ctrl{1} &&& & \gate{R_X{\left(\lambda_{1} z_{1}\right)}}
        &  \phantomgate{large}     & 
       \gate{R_X{\left(\theta_{13}\right)}} \QLayer{4}{8}{Layer 2} & \gate{R_Y{\left(\theta_{14}\right)}}& \gate{R_Z{\left(\theta_{15}\right)}}      &\ctrl{1} &&& & \gate{R_X{\left(\lambda_{5} z_{5}\right)}} 
        &  \phantomgate{large}  & 
        \gate{R_X{\left(\theta_{25}\right)}} \QLayer{4}{3}{Variational Layer} & \gate{R_Y{\left(\theta_{26}\right)}}& \gate{R_Z{\left(\theta_{27}\right)}}    
        & \qw    & \meter{} \\
        \lstick{$\ket{0}$} 
        & \qw    &   
       \gate{R_X{\left(\theta_{4}\right)}} & \gate{R_Y{\left(\theta_{5}\right)}}& \gate{R_Z{\left(\theta_{6}\right)}}   &  \targ{} & \ctrl{1}   & && \gate{R_X{\left(\lambda_{2} z_{2}\right)}}
        &  \phantomgate{large}     &
       \gate{R_X{\left(\theta_{16}\right)}} & \gate{R_Y{\left(\theta_{17}\right)}}& \gate{R_Z{\left(\theta_{18}\right)}}   &  \targ{} & \ctrl{1}   & && \gate{R_X{\left(\lambda_{6} z_{6}\right)}}
        &  \phantomgate{large}  &  
       \gate{R_X{\left(\theta_{28}\right)}} &\gate{R_Y{\left(\theta_{29}\right)}}& \gate{R_Z{\left(\theta_{30}\right)}}
        & \qw    & \meter{}\\
          \lstick{$\ket{0}$} 
        & \qw    &   
         \gate{R_X{\left(\theta_{7}\right)}} & \gate{R_Y{\left(\theta_{8}\right)}}& \gate{R_Z{\left(\theta_{9}\right)}}  & & \targ{} & \ctrl{1}   & & \gate{R_X{\left(\lambda_{3} z_{3}\right)}}
        &  \phantomgate{large}     &    
         \gate{R_X{\left(\theta_{19}\right)}} & \gate{R_Y{\left(\theta_{20}\right)}}& \gate{R_Z{\left(\theta_{21}\right)}}  && \targ{} & \ctrl{1}   & & \gate{R_X{\left(\lambda_{7} z_{7}\right)}}
        &     \phantomgate{large} &    
       \gate{R_X{\left(\theta_{31}\right)}}&\gate{R_Y{\left(\theta_{32}\right)}}& \gate{R_Z{\left(\theta_{33}\right)}}
        & \qw    & \meter{}\\
        \lstick{$\ket{0}$} 
        & \qw    &  
         \gate{R_X{\left(\theta_{10}\right)}} & \gate{R_Y{\left(\theta_{11}\right)}}& \gate{R_Z{\left(\theta_{12}\right)}}  &&  & \targ{} && \gate{R_X{\left(\lambda_{4} z_{4}\right)}}
        &  \phantomgate{large}     &       
         \gate{R_X{\left(\theta_{22}\right)}} & \gate{R_Y{\left(\theta_{23}\right)}}& \gate{R_Z{\left(\theta_{24}\right)}}  && & \targ{} && \gate{R_X{\left(\lambda_{8} z_{8}\right)}}
        &  \phantomgate{large}   &       
       \gate{R_X{\left(\theta_{34}\right)}} & \gate{R_Y{\left(\theta_{35}\right)}}& \gate{R_Z{\left(\theta_{36}\right)}}
        & \qw    &\meter{}
    \end{quantikz}
\end{adjustbox}
\caption{Example of a parameterized quantum circuit with $4$ qubits and $2$ layers used as the generator in the QGAN. Each layer consists of single-qubit Pauli rotations, CNOT gates and data uploading gates, which upload particular realizations of the random noise $\vec{z}$. Each element of $\vec{z}$ is uniformly sampled in $[0,2\pi]$. Before measurement, there is a final variational layer consisting of single-qubit Pauli rotations. The measurements are conducted single-qubit Pauli-X and Pauli-Z basis. The single-qubit Pauli rotations are tunable with parameters $\theta_i$ and the data-uploading gates are tunable with parameters $\lambda_i$.
This Ansatz is a hardware-efficient Ansatz, which is commonly chosen in the field~\cite{cerezo2021Variationalquantum, kandala2017Hardwareefficientvariational}.}
\label{fig:pqc}
\end{figure*}

A parameterized quantum circuit (PQC) consists of tunable and fixed gates, and measurements at the end. 
The parameters of the tunable gates are updated based on the loss function, which consists of the measurement results and the output of the discriminator.\\
The PQC can be used in different ways, for example as a quantum circuit Born machine~\cite{benedetti2018generativemodeling, liu2018DifferentiableLearning, coyle2021Quantumclassical} or as an expectation value sampler~\cite{romero2021VariationalQuantum, anand2021NoiseRobustness, barthe2024Parameterizedquantum, shen2024ShadowFrugalExpectationValueSampling}.
The latter approach is the one which we use for our QGANs.
In the former case, a quantum circuit is used to learn an underlying distribution and every single sample forms a bitstring corresponding to the learned distribution. 
The probability of each sample depends on the amplitudes of the final quantum state. 
However, the precision of the generated values is limited by the discrete nature of this approach.
In contrast, the expectation value sampler identifies expectation values of quantum circuit measurement outcomes with samples of a distribution.
The underlying randomness comes from classical noise uploaded to the quantum circuit.\\
A sketch of the PQC architecture used in this paper is given in Figure~\ref{fig:pqc}.

\subsection{Related work}\label{subsec:relatedwork}
In recent years, classical algorithms for generating synthetic financial data have been proposed and explored, in GAN settings~\cite{dogariu2022GenerationRealistic, eckerli2021GenerativeAdversarial, takahashi2019Modelingfinancial, wiese2020QuantGANs, zhang2019StockMarket}, and with other approaches~\cite{dogariu2022GenerationRealistic, takahashi2024Generationsynthetic}. 
A common challenge is the generation of time series that exhibit all stylized facts sufficiently well~\cite{dogariu2022GenerationRealistic}.\\
QGANs were introduced in~\cite{lloyd2018QuantumGenerative} and~\cite{dallaire-demers2018Quantumgenerative}, substituting generator and discriminator with quantum circuits.
Other approaches for QGANs, such as combining a classical discriminator with a quantum generator, and their applications have been explored as well~\cite{tian2022RecentAdvances, ngo2023SurveyRecent}. In~\cite{zoufal2019QuantumGenerative} the generation of certain probability distributions and in~\cite{mourya2025ContextualQuantum}, the generation of correlated stocks has been examined. \\
Our work was motivated by~\cite{coyle2021Quantumclassical}, which compared the performance of generating synthetic financial data of correlated asset pairs by the two models of restricted Boltzmann machines and quantum circuit Born machines, observing an advantage of the quantum circuit Born machine for comparable model sizes. 
Here we go beyond learning time-aggregated distributions of financial time series as in~\cite{coyle2021Quantumclassical}, by additionally examining the temporal correlations of generated time series.
In contrast to the models used in~\cite{coyle2021Quantumclassical}, our method is based on the expectation value sampler, which was introduced in~\cite{romero2021VariationalQuantum}, proven to be universal in~\cite{barthe2024Parameterizedquantum} and further generalized in~\cite{shen2024ShadowFrugalExpectationValueSampling}.
In Appendix~\ref{appendix:fx_comparison}, we adapted our approach for learning the distribution of correlated pairs of foreign exchanges and compare our results with the results of~\cite{coyle2021Quantumclassical}.\\
QGANs have also been used for other applications, such as image generation~\cite{huang2021ExperimentalQuantum, zhou2023Hybridquantum, silver2023MosaiQQuantum, tsang2023HybridQuantum} and other discrete distributions~\cite{situ2020Quantumgenerative}, in generative chemistry~\cite{kao2023ExploringAdvantages}, fraud detection~\cite{herr2021Anomalydetection}, option pricing~\cite{fuchs2023HybridQuantum}, and high-energy physics~\cite{baglio2024Dataaugmentation, dimeglio2024QuantumComputing}.
Further, other quantum machine learning strategies have been used in learning financial time series~\cite{paquet2022QuantumLeapHybrid}.\\

The Wasserstein QGAN, proposed in~\cite{kiani2022Learningquantum, chakrabarti2019QuantumWasserstein}, shows improvement in training stability and efficiency compared to QGANs based on other metrics.
The full-state simulation of the Wasserstein-QGAN with gradient penalty as an application to generate financial time series has been explored in~\cite{schwander2022Quantumgenerative} and compared to classical GANs.

\section{Implementation}\label{sec:implementation}
For this paper, we trained a Wasserstein QGAN for generating time series based on training data originating in the time series of the S\&P~500 index. 
We use a hybrid approach, with a classical neural network as a discriminator and a parameterized quantum circuit (PQC) as a generator. \\
We use a convolutional neural network as a discriminator, motivated by~\cite{takahashi2019Modelingfinancial}, where it was used as the discriminator of a GAN that generates financial time series. 
All its activation functions are rectified linear units, except the last single neuron in the critic,
which uses a linear activation function. 
The architecture of the discriminator is detailed in Appendix~\ref{appendix:cnn_achitecture}.\\
We simulated the QGAN based on a full simulation of the PQC by using the Tensorflow software library~\cite{abadi2016TensorFlowsystem}, and the QGAN based on the MPS approximation of the PQC with the JAX~\cite{jax2018github} and Quimb~\cite{gray2018quimbpython} software libraries.
The gradients are calculated via automatic differentiation.
As an optimizer for the training of the QGAN, we chose the Adam optimizer with a learning rate of $10^{-3}$.\\
In the following, we outline the data pre-and post-processing, the setup of the quantum generator and the MPS simulation applied in this paper.

\subsection{Data pre-and post-processing}
\label{subsec:pre-and-postprocessing}
The outputs of expectation value samplers are the expectation values of Pauli strings which lie in the range $[-1,1]$. 
As this is a key difference to the raw time-series data in the form of log returns, which have unbounded support (see Eq.~\eqref{eq:log_return}), we perform data pre-and post-processing.
For that, we follow the same approach as taken in
Refs.~\cite{wiese2020QuantGANs, hogenboom2025CasperHogenboomWGAN_financial_timeseries}. 
We first describe the approach for the pre-processing, transforming training data to lie in the interval $[-1,1]$. 
This process consists of four steps: (i) data normalization, (ii) the inverse Lambert-W transform, (iii) data normalization and (iv) a rolling window.\\
(i) We normalize our training data to have a mean of $0$ and a variance of $1$:
\begin{align}
    r_{t,(i)}:=\frac{r_t-\mu_r}{\sigma_r},
\end{align}
where $r_t$ is the original log returns, and $\mu_r$ and $\sigma_r$ estimate the
mean and standard deviation of all log returns.\\
(ii) As learning a heavy-tailed distribution can be challenging due to a limited number of samples in the tails, we implement the inverse Lambert-W transform on the normalized log returns. 
This transformation will bring the heavy-tailed distributed data closer to a Gaussian distribution. 
Given Lambert’s $W$ function, which is the inverse of $z = u \exp(u)$ with $z : \mathbb{R}\to \mathbb{R}$, we can define the following transform on the normalized heavy-tailed data set $V$:
\begin{align}
    W_{\delta}(\nu):=\text{sgn}(\nu)\left(\frac{W(\delta\nu^2)}{\delta}\right)^{1/2}
\end{align}
with $\nu\in V$, $\delta\geq 0$ a tunable parameter, $\text{sgn}(\nu)$ the sign of $\nu$ and $W$ the Lambert’s $W$ function. 
The inverse of this function is given by
$v = W_{\delta}(v) \exp\left(\frac{\delta}{2} W_{\delta}(\nu)^2\right)$~\cite{goerg2015LambertWay}.
Throughout this paper, we pick $\delta=0.5$.\\
(iii) As the inverse transform can sometimes be poorly behaved in specific data points, we implement clipping in order to avoid outliers.\\
(iv) After transforming the S\&P 500 time series, we divide it into smaller batches. 
We achieve this by applying a rolling window of window length $m$ and stride $s$ to the time series, which divides it into multiple subsequences.
Each subsequence becomes a new sample in our dataset.
Note that this stride is often shorter than the length of the window, which results in partially overlapping and thus correlated samples. 
Although this correlation between training samples is not ideal, the more extensive set of training samples can be beneficial for model performance.
Throughout all simulations shown in this paper, we used a stride of $s=5$ and a window length of $m=20$ and $m=40$.\\
We apply (i)-(iv) as a data pre-processing of the time series used for training. 
The generated samples consisting of the expectation values of Pauli operators (see Section~\ref{subsec:pqcsimulation}) are then post-processed by transforming steps (iv)-(i) backwards to form samples of log returns before we calculate model performance metrics.
\subsection{Quantum generator and full-state simulation}\label{subsec:pqcsimulation}
As a generator of the QGAN, we chose a PQC with an architecture that is sketched in Figure~\ref{fig:pqc}, based on the hardware efficient Ansatz~\cite{cerezo2021Variationalquantum, kandala2017Hardwareefficientvariational}. 
The qubits are initialized in the $\ket{0}$ state. 
Each layer consists of single-qubit Pauli rotations, CNOT gates connecting nearest neighbors and noise encoding gates.
The latter encode each a uniformly distributed noise sample with single-qubit rotations with trainable parameters. 
Such circuit architectures suffer from barren plateaus when scaled up in the number of qubits and layers~\cite{larocca2025Barrenplateaus}. 
Therefore, we do not consider them to be scalable in their current form. 
Instead, our investigation should be understood as establishing lower bounds on what can be achieved with quantum circuits: if these perform well at small scales, it motivates efforts to refine them for improved trainability at larger scales. 
Conversely, if they fail to perform even at small scales, this indicates that the application may be less promising than one might have hoped.\\
In Section~\ref{sec:results}, we present results from training circuits with $10$ and $20$ qubits and between $1$ and $18$ layers. 
This choice of the number of qubits and layers makes the QGANs classically simulatable.
After the $n$-th layer, we apply single-qubit Pauli rotations and we measure each qubit in two bases: the Pauli-$Z$ basis and the Pauli-$X$ basis.
We chose these measurements in order to enable the simulation of longer time series using fewer qubits.
The exact consequences in terms of expressivity and potential greater sampling costs was discussed in~\cite{barthe2024Parameterizedquantum, shi2024EffectsObservable}. 
The expectation values $\{\langle X\rangle_1,\langle Z\rangle_1,\langle X\rangle_2,\langle Z\rangle_2,... \}$, where $\langle X\rangle_i,\langle Z\rangle_i$ are the expectation values of the measurement in the Pauli-$X$ and Pauli-$Z$ basis on the $i$-th qubit, respectively, are post-processed as described in Section~\ref{subsec:pre-and-postprocessing}.
The obtained set $\{p(\langle X\rangle_1),p(\langle Z\rangle_1),p(\langle X\rangle_2),p(\langle Z\rangle_2),... \}$, where $p(\cdot)$ stands as the post-processing map, is identified with the generated time series $r_{t,\text{gen}}$ in the following way:
\begin{align}
    r_{t,\text{gen}}&=\{r_{1},r_2,r_3,r_4,...\}\nonumber\\
    &=\{p(\langle X\rangle_1),p(\langle Z\rangle_1),p(\langle X\rangle_2),p(\langle Z\rangle_2),... \}\ .
\end{align}
A circuit of $n$ qubits thus generates a time series of length $2n$.\\
The first part of our simulations is based on the full-state simulation of PQC. 
Let $\ket{\psi_{\mathrm{full}}}$ be the quantum state that describes the state at the end of the parameterized quantum circuit (PQC) used in our quantum generative adversarial network (QGAN). 
In the computational basis, it takes the form
\begin{align}
\ket{\psi_{\mathrm{full}}}\approx\sum_{i_1,i_2,...,i_n}c_{i_1,i_2,...,i_n}\ket{{i_1,i_2,...,i_n}}\ ,
\end{align}
where $i_j\in\{0,1\}$.
The number of coefficients $c_{i_1,i_2,...,i_n}$ of this state, and thus the memory requirement, scales exponentially with the number of qubits $n$.
Moreover, the time cost of the classical full-state simulation scales linearly with the number of layers.
This makes the full-state simulation quickly infeasible.\\

\subsection{Matrix product state simulation}\label{subsec:mpssimulation}
\begin{figure}[h]
\centering
\scalebox{0.75}{
\begin{tikzpicture}[scale=1, every node/.style={font=\small},
    tensor/.style={draw, rectangle, minimum size=1cm, fill=blue!20}, 
    merged/.style={draw, rectangle, minimum width=4.4cm, minimum height=1cm, fill=blue!20},
    every node/.style={font=\small},
    thickline/.style={-Latex, thick},
    virtual/.style={midway, above}
]

\node[merged] (psi) at (-3.5,0) {$|\psi\rangle$};

\foreach \i/\x in {1/-1.8, 2/-0.6, 3/0.6, 4/1.8} {
    \draw[-] (psi.south) ++(\x,0) -- ++(0,-0.8);
    \node at ($(psi.south)+(\x,-1.0)$) {$i_{\i}$};
}

\node at (-0.9,0) {$\approx$};
\node[tensor] (T1) at (0,0) {$A^{[1]}$};
\node[tensor] (T2) at (1.6,0) {$A^{[2]}$};
\node[tensor] (T3) at (3.2,0) {$A^{[3]}$};
\node[tensor] (T4) at (4.8,0) {$A^{[4]}$};

\draw[-] (T1.east) -- (T2.west) node[virtual] {$\chi$};
\draw[-] (T2.east) -- (T3.west) node[virtual] {$\chi$};
\draw[-] (T3.east) -- (T4.west) node[virtual] {$\chi$};

\draw[-] (T1.south) -- ++(0,-0.8) node[below] {$i_1$};
\draw[-] (T2.south) -- ++(0,-0.8) node[below] {$i_2$};
\draw[-] (T3.south) -- ++(0,-0.8) node[below] {$i_3$};
\draw[-] (T4.south) -- ++(0,-0.8) node[below] {$i_4$};

\end{tikzpicture}
}
\caption{A matrix product state (MPS) consists of a chain of local tensors $A^{[j]}$ connected by virtual bonds of dimension $ \chi $, each with a physical leg representing a qubit index. A virtual bond represents a sum over a particular index of the tensors. The left term represents the state $\psi$, which lives in a $4$-qubit space, and the right term shows its MPS approximation.}
\label{fig:mps-detail}
\end{figure}
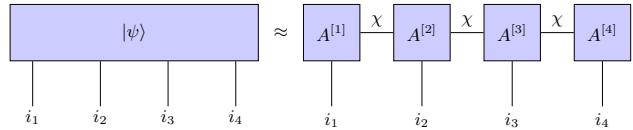

For being able to simulate PQC with a higher number of layers and qubits, we use matrix product states (MPS) as efficient approximation methods under some circumstances~\cite{ostlund1995ThermodynamicLimit,verstraete2008Matrixproduct, cirac2021Matrixproduct}, which in the context of machine learning is also called the tensor-train decomposition~\cite{oseledets2011TensorTrainDecomposition}.
They provide a compact representation of quantum states with limited entanglement and have been extensively used in physics~\cite{cirac2021Matrixproduct}.
This makes them well suited for simulating quantum states that are prepared by the PQC used in our QGAN.\\
An MPS represents an $n$-qubit quantum state $\ket{\psi_{\mathrm{full}}}$ as a product of local tensors~\cite{cirac2021Matrixproduct}:
\begin{align}
\ket{\psi_{\mathrm{full}}} = \sum_{i_1, \dots, i_n} A^{[1]}_{i_1} A^{[2]}_{i_2} \cdots A^{[n]}_{i_n} \ket{i_1 i_2 \cdots i_n},
\end{align}
where each $A^{[k]}_{i_k}$ is a $\chi_{k-1}\times \chi_k$-dimensional tensor.
We call $\chi_k$ the bond dimensions of the MPS that controls the amount of entanglement the MPS can represent.
The contraction of these tensors yields the amplitude corresponding to each computational basis state.
For simplicity, we choose $A^{[1]}_{i_1}$ as $1\times \chi$-dimensional, $A^{[n]}_{i_n}$ $\chi\times 1$-dimensional, and each remaining tensor $A^{[k]}_{i_k}$ with the dimensions $\chi \times \chi$.
Such an MPS is described by $(2n-1)\chi^2+2\chi$ coefficients, which for constant $\chi$ scales linearly in the number of qubits, making it more efficient than the full-state simulation with $2^n$ coefficients.
Figure~\ref{fig:mps-detail} provides a sketch of an MPS representation.\\

We simulate the quantum circuit using MPS in the following way: We start with the trivial tensor corresponding to the initial state of the circuit $\ket{0}^n$.
Then, we apply each layer sequentially to the MPS, on which single-qubit Pauli rotations are trivially applied. 
After each CNOT gate, we recalculate the tensors by singular value decompositions (SVD)~\cite{berezutskii2025Tensornetworks}. 
We partition the system into left and right parts and apply SVD, truncate the number of singular values to the bond dimension $\chi_k$, and the left unitary multiplied with the truncated singular value matrix forms the tensor $A^{[k]}_{i_k}$.
After applying every layer in this way, we get an MPS approximation of the output state $\ket{\psi_{\text{full}}}$ of the PQC.

\begin{figure}[ht]
    \centering
    \includegraphics[width=1.05\linewidth]{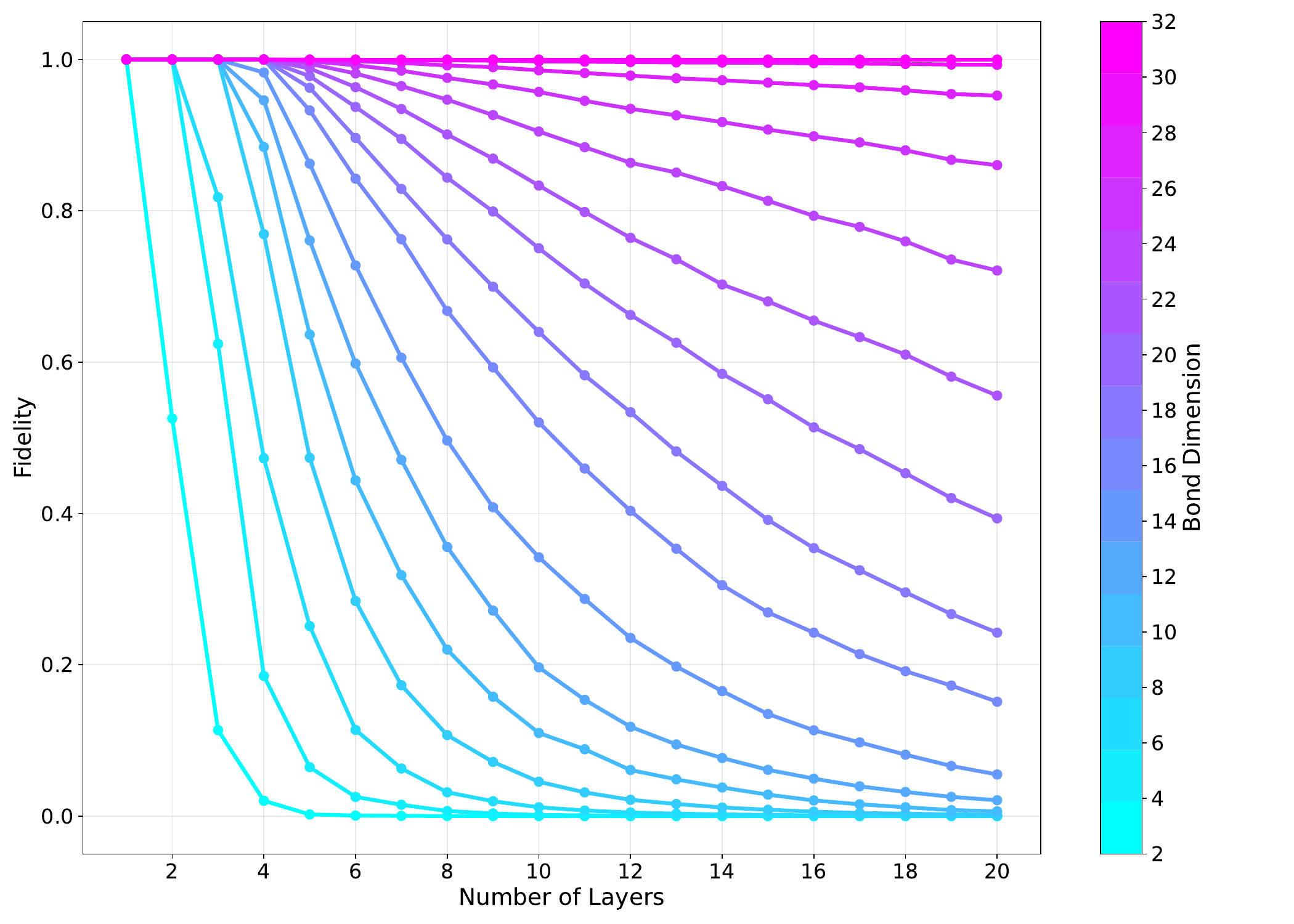}
    \caption{Fidelity between the exact quantum state prepared by a PQC consisting of $10$ qubits, in the architecture as sketched in Figure~\ref{fig:pqc} and the MPS approximation as a function of the depth of the PQC, for different bond dimensions \(\chi\).}
    \label{fig:fidelity_vs_bond}
\end{figure}

To assess the quality of the MPS approximation for systems that can efficiently be simulated with the full-state, we compute the fidelity between the full quantum state $\ket{\psi_{\mathrm{full}}}$, obtained from an exact state vector simulation, and the MPS-approximated state $\ket{\psi_{\mathrm{MPS}}}$:
\begin{align}
F = |\bra{\psi_{\mathrm{full}}}\ket{\psi_{\mathrm{MPS}}}|^2.
\end{align}
We evaluate this fidelity for various values of the maximum bond dimension $\chi$. 
As shown in Figure~\ref{fig:fidelity_vs_bond}, the fidelity increases with $\chi$, indicating improved approximation accuracy.
For a higher number of layers, the PQC increases the entanglement across the qubits, which decreases the fidelity for fixed bond dimension.
For sufficiently large $\chi$ ($\chi=32$ for $10$ qubits), the MPS becomes numerically indistinguishable from the full-state~\cite{verstraete2004DensityMatrix}.\\
In general, for a higher number of layers and qubits, it is not possible anymore to calculate this fidelity as it is not feasible to simulate the full-state.
Instead, one can calculate the fidelity between MPS simulations of different bond dimensions, and choose the lowest bond dimension at which this fidelity does not increase anymore.

This tunability of the bond dimension provides a practical trade-off between simulation efficiency and accuracy. 
In the context of training the QGAN, where the PQC must be evaluated many times, moderate bond dimensions already yield sufficiently high fidelity while significantly reducing computational cost.

\section{Results}\label{sec:results}
In this section, we describe the results of our simulations. 
We simulated the parameterized quantum circuit (PQC) with the architecture shown in Figure~\ref{fig:pqc} in two ways, as described in Section~\ref{sec:implementation}. 
First, we performed full-state simulations for systems with up to $10$ qubits and up to $6$ layers. Second, we used matrix product state (MPS) simulations for systems with $10$ and $20$ qubits and between $1$ and $18$ layers.
We performed $5$ runs for each of these simulations, and describe their results in separate subsections.
The results are discussed in Section~\ref{sec:discussion}.

\subsection{Full-state simulation}\label{subsec:wasserstein loss}
\begin{figure*}[htbp!]
    \centering
    \begin{tikzpicture}
        \node[inner sep=0pt] (img) at (0,0)
            {\includegraphics[width=0.8\linewidth]{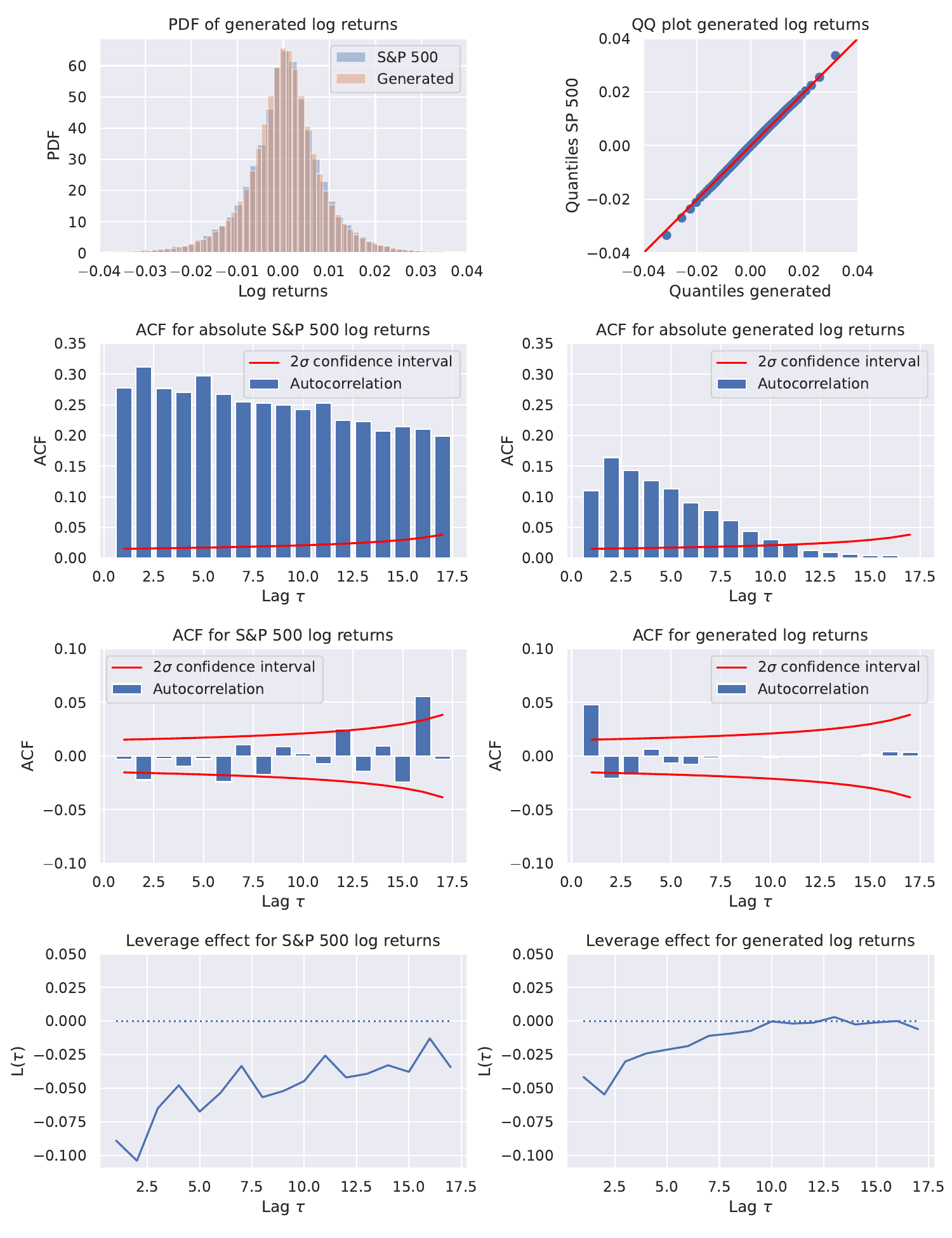}};

        \node at (-5.8,8.9) {\textbf{(a)}};
        \node at (1.2,8.9) {\textbf{(b)}};

        \node at (-5.8,4.5) {\textbf{(c)}};
        \node at (1.2,4.5) {\textbf{(d)}};

        \node at (-5.8,-0.1) {\textbf{(e)}};
        \node at (1.2,-0.1) {\textbf{(f)}};

        \node at (-5.8,-4.7) {\textbf{(g)}};
        \node at (1.2,-4.7) {\textbf{(h)}};

    \end{tikzpicture}
    \caption{Metrics of the stylized facts for a synthetic time series of window size $20$ generated by a QGAN, compared to the metrics of the S\&P~500 index. 
    The generator of the QGAN is a PQC based on the architecture shown in Figure~\ref{fig:pqc} consisting of $10$ qubits and $8$ layers, simulated with the full-state approach.
    In \textbf{(a)}, we plot the probability density functions and in \textbf{(b)} the quantile-quantile plot of both the S\&P~500 index and the generated time series.
    In \textbf{(c)}-\textbf{(h)}, we plot the metrics absolute autocorrelation, linear autocorrelation and the leverage effect, as an indication of the stylized facts as described in Section~\ref{subsec:financialtimeseries}. 
    The Subfigures \textbf{(c)},  \textbf{(e)} and \textbf{(g)} show the metrics of the S\&P~500 index and the Subfigures \textbf{(d)}, \textbf{(f)} and \textbf{(h)} the metrics of the generated time series, respectively. Confidence intervals are calculated as in~\cite{chatfield1975time}.}
    \label{fig:full-state-simulation}
\end{figure*}
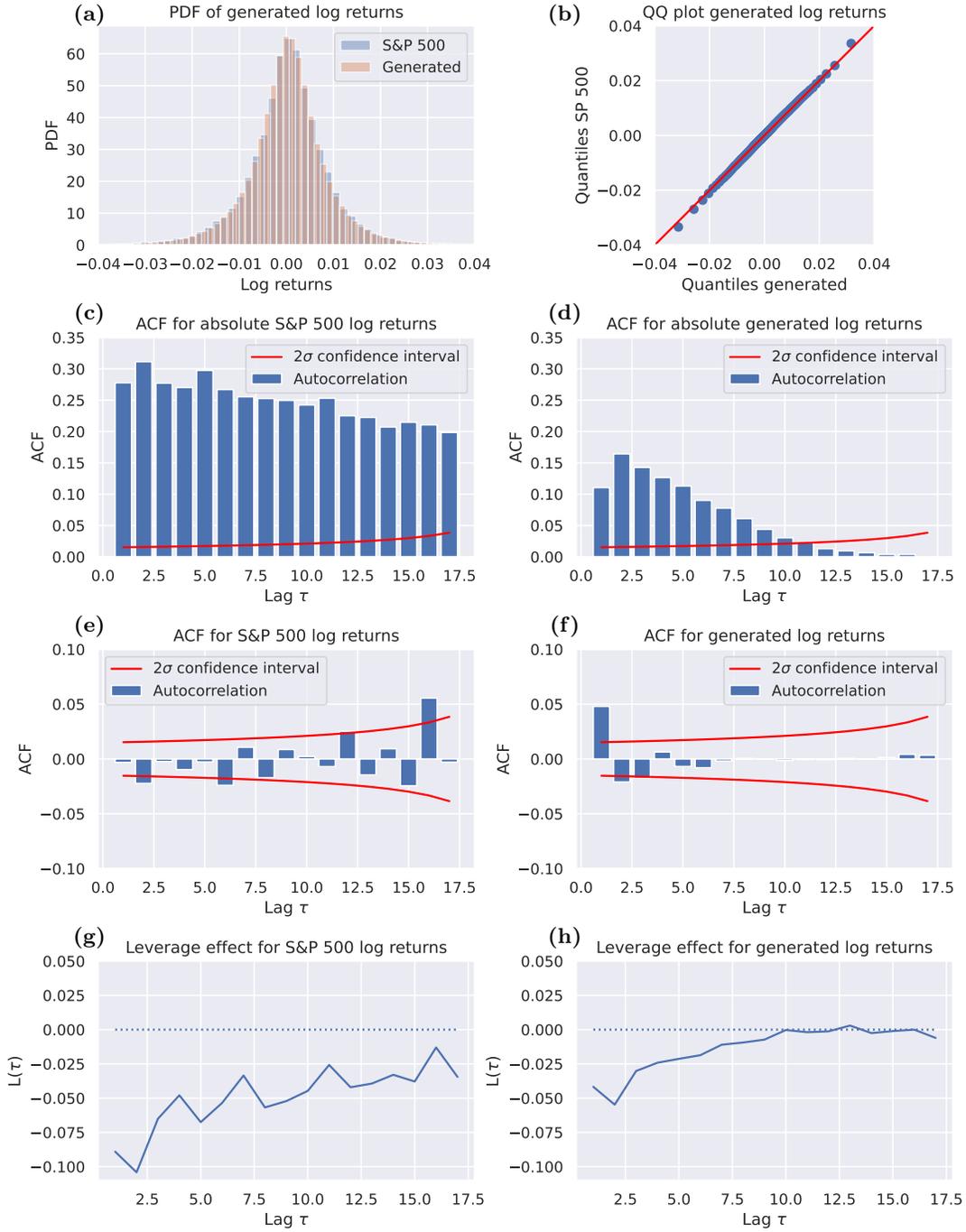

\begin{figure}[hbtp!]
    \centering
    \includegraphics[width=1\linewidth]{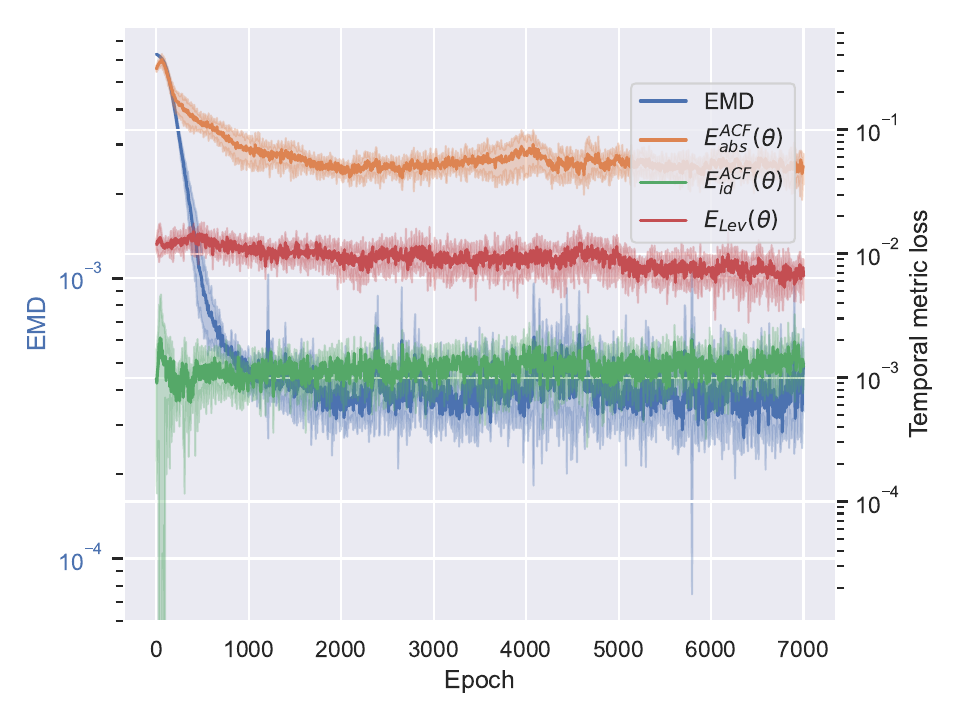}
    \caption{Wasserstein loss as defined in Equation \eqref{eq:loss-discriminator} (here called the EMD) and metrics corresponding to the temporal correlations as described in Section~\ref{subsec:financialtimeseries} in the training of the QGAN in the full-state simulation with $10$ qubits and $8$ layers (see Figure~\ref{fig:full-state-simulation} for the final metrics), depending on the number of epochs. We show mean and standard deviation of $5$ training runs.}
    \label{fig:full-state-metrics-per-epoch}
\end{figure}

For the full-state simulation, we chose a PQC consisting of $10$ qubits and $8$ layers.
Therefore, for the generation of the training set from the historical S\&P~500 time series, we set the window size to $20$. 
We trained the QGAN for $8000$ epochs, and plotted the metrics of a generated time series in Figure~\ref{fig:full-state-simulation}.
The generated time series closely resembles the distribution of the S\&P~500 index, as shown in Subfigures \textbf{(a)} and \textbf{(b)}.
Similar to the S\&P~500 index, the generated time series shows a weaker, but decaying absolute autocorrelation (Subfigures \textbf{(c)} and \textbf{(d)}) and does not show linear autocorrelation (Subfigures \textbf{(e)} and \textbf{(f)}).
The leverage effect, which is negative and increasing in the S\&P~500 index (Subfigure \textbf{(g)}, is also reproduced in a weaker way in the generated time series (Subfigure \textbf{(h)}).
As can be seen in Figure~\ref{fig:full-state-metrics-per-epoch}, both the loss function and the temporal metrics decrease with the number of epochs, indicating stable training of the QGAN.\\
We show the metrics of the best out of $5$ runs that correspond to the time series as shown in Figure~\ref{fig:full-state-simulation} in column \textbf{(a)} of Table~\ref{tab: table of metrics of differnt sims}.

\begin{table}[htbp!]
    \centering
    \begin{tabular}{|c|c|c|c|c|}
        \hline
        Metrics & \textbf{(a)} & \textbf{(b)} & \textbf{(c)} & \textbf{(d)} \\ \hline
        EMD & $2.4\cdot 10^{-4}$ & $5\cdot 10^{-4}$ & $3.1\cdot 10^{-4}$ & $4.2\cdot 10^{-3}$ \\ \hline
        $E^{ACF}_{id}(\theta)$ & $7.8\cdot 10^{-4}$ & $3.6\cdot 10^{-4}$ & $3.9\cdot 10^{-4}$ & $1.1\cdot 10^{-3}$ \\ \hline
        $E^{ACF}_{abs}(\theta)$ & $0.15$ & $0.17$ & $0.29$ & $0.99$ \\ \hline
        $E_{Lev}(\theta)$ & $4.9\cdot 10^{-3}$ & $7.1\cdot 10^{-3}$ & $2.8\cdot 10^{-2}$ & $4.4\cdot 10^{-2}$ \\ \hline
    \end{tabular}
        \caption{Comparison of different metrics defined in Equations \eqref{eq:EMD_definition}-\eqref{eq:lev_definition} for the best out of $5$ runs of \textbf{(a)} the full-state simulation with $10$ qubits and $8$ layers as in Figures~\ref{fig:full-state-simulation} and~\ref{fig:full-state-metrics-per-epoch}, \textbf{(b)} the full-state simulation with $10$ qubits and $4$ layers with a different circuit architecture as described in Appendix~\ref{appendix:full-state-simulation-connected}, \textbf{(c)} the MPS simulation with $10$ qubits, $18$ layers and a bond dimension $32$ as in Figures~\ref{fig:MPS-simulation-10Q-L18-B32} and~\ref{fig:MPS-metrics-per-epoch-10Q-18L-B32}, \textbf{(d)} the MPS simulation with $20$ qubits, $6$ layers and a bond dimension $70$ as in Figure~\ref{fig:MPS-simulation-20Q-L6-B70}.}
        \label{tab: table of metrics of differnt sims}
\end{table}

The Wasserstein QGAN does not explicitly account for temporal effects, so any such structure in the generated time series must result from other aspects of the model.
To investigate the influence of the PQC architecture on these temporal effects, we trained a QGAN with a different PQC and present the results in Appendix~\ref{appendix:full-state-simulation-connected}. 
We indeed see that the absolute autocorrelation of the generated time series increases at larger time lags, in contrast to the time series generated in Figure~\ref{fig:full-state-simulation}.\\
Additionally, to compare with the results of a GAN based on a quantum circuit Born machine~\cite{coyle2021Quantumclassical}, we trained the QGAN on generating currency pairs; the results are shown in Appendix~\ref{appendix:fx_comparison}.\\
We analyze these results in Section~\ref{sec:discussion}.\\
As explained in Section~\ref{sec:implementation}, full-state simulation of PQCs quickly becomes infeasible as the number of layers and qubits increases.
In the following, we describe MPS-based simulations, which make it feasible to simulate PQCs with larger numbers of layers and qubits.
\subsection{MPS simulation}\label{subsec:mps simulation}
For the MPS simulation, we first chose a PQC of $10$ qubits, with a varying number of layers ($1$, $5$, $10$ and $18$) and different bond dimensions ($1$, $8$, $16$, $24$ and $32$) of the MPS.
The MPS simulation of PQCs with $10$ qubits generates time series with the same window size as the full-state simulation, making the results directly comparable. For higher numbers of layers and bond dimensions below $\chi=32$, the MPS simulation is also faster than the full-state simulation.\\
For higher bond dimensions and number of layers, the training time in the MPS simulation increases, and the stylized facts of the generated time series vary considerably for each choice of the number of layers and bond dimension.
See in Appendix~\ref{appendix: table of mps sims} for a comparison of the Wasserstein distance and metrics for the temporal effects for simulations of different numbers of layers and bond dimensions.
In Figure~\ref{fig:MPS-simulation-10Q-L18-B32}, we show the metrics of a generated time series from a well-performing QGAN that is trained for $7032$ epochs, whose PQC consists of $18$ layers and is simulated as an MPS with bond dimension $32$. 
The metrics of this generated time series are shown in column \textbf{(c)} of Table~\ref{tab: table of metrics of differnt sims}.
We chose to show the results for this particular model, as they match the stylized facts of the time series of the S\&P 500 index qualitatively well,
and as it proves that it is possible to train a QGAN for which the PQCs in the MPS simulation has more layers than what would be feasible with the full-state simulation.

The quantile-quantile plot shows that the generated time series matches the distribution of the S\&P~500 index closely.
In contrast to the time series generated with the full-state simulation shown in Figure~\ref{fig:full-state-simulation}, the absolute autocorrelation (Subfigure \textbf{(d)}) that indicates volatility clustering is positive and decreasing for all time lags. 
This behavior is closer to the absolute autocorrelation of the S\&P~500 index (Subfigure \textbf{(c)}), however the values are lower.
In contrast, the leverage effect is weaker than in the time series generated by the full-state simulation.
The quantitative metrics decrease more slowly during training compared to the full-state simulation, as can be seen in Figure~\ref{fig:MPS-metrics-per-epoch-10Q-18L-B32}.

\begin{figure}[htbp!]
    \centering
    \includegraphics[width=1\linewidth]{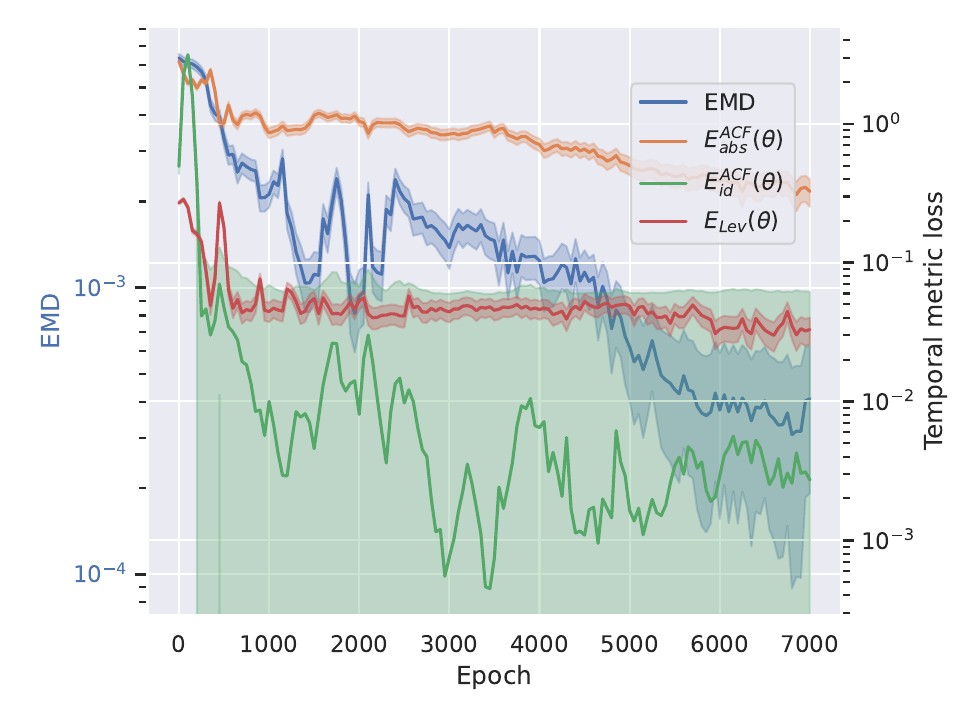}
    \caption{Wasserstein loss as defined in Equation \eqref{eq:loss-discriminator} (here called the EMD) and metrics corresponding to the temporal correlations as described in Section~\ref{subsec:financialtimeseries} in the training of the QGAN in the MPS simulation with $10$ qubits, $18$ layers and a bond dimension of $32$ (see Figure~\ref{fig:MPS-simulation-10Q-L18-B32} for the final metrics), depending on the number of epochs. We show mean and standard deviation of $5$ training runs.}
    \label{fig:MPS-metrics-per-epoch-10Q-18L-B32}
\end{figure}
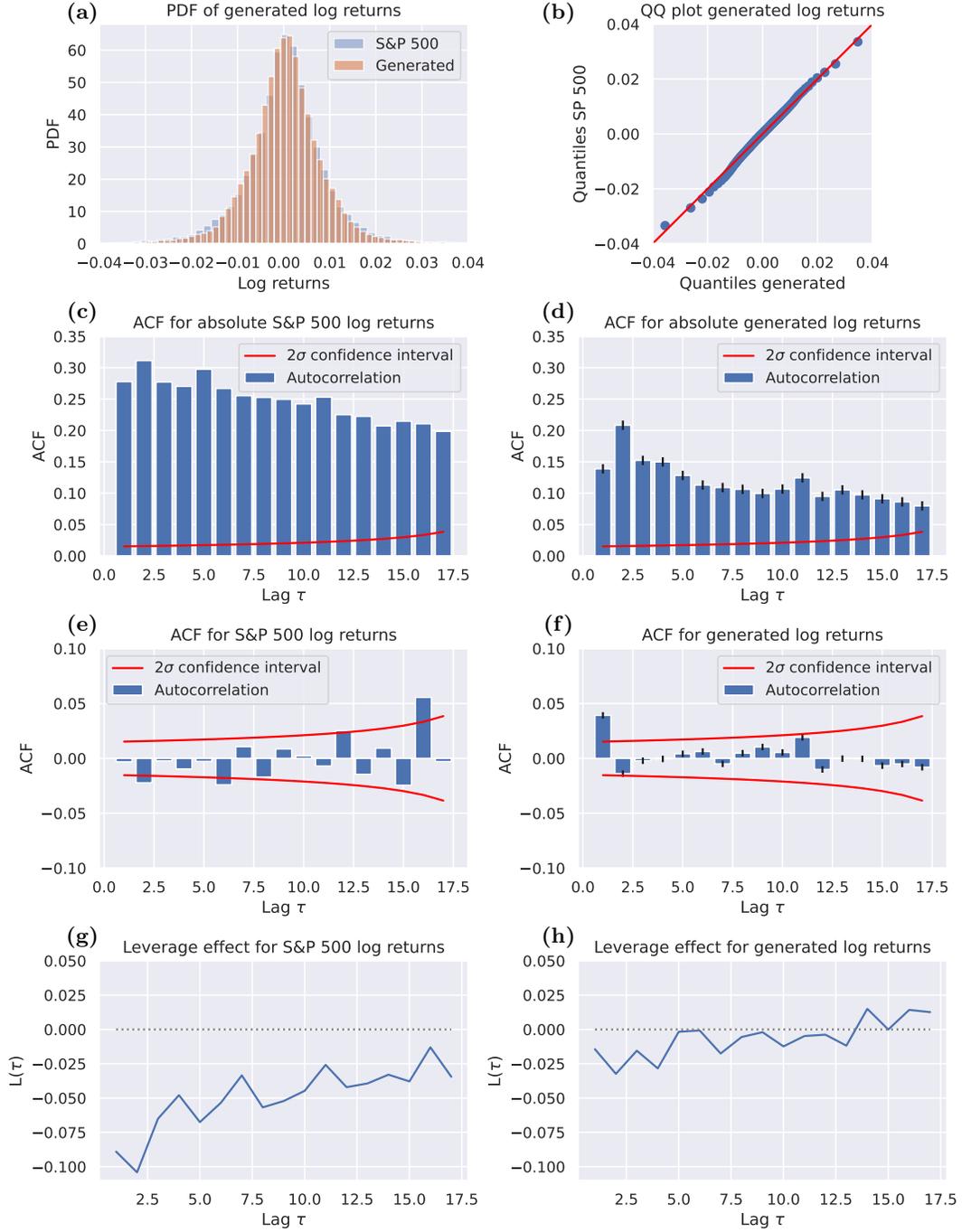
\begin{figure*}[htbp!]
    \centering
    \begin{tikzpicture}
        \node[inner sep=0pt] (img) at (0,0)
            {\includegraphics[width=0.8\linewidth]{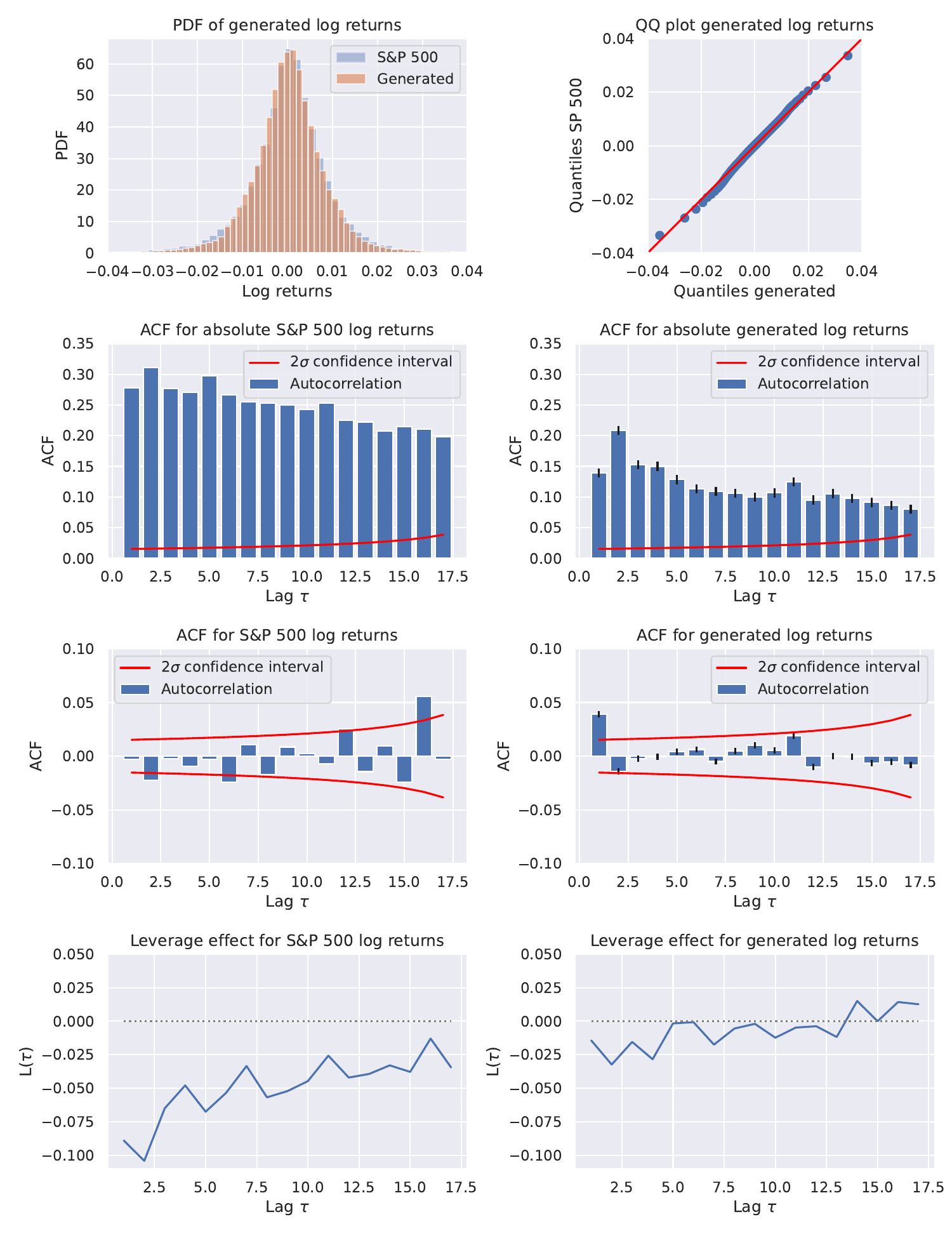}};

        \node at (-5.8,8.9) {\textbf{(a)}};
        \node at (1.2,8.9) {\textbf{(b)}};

        \node at (-5.8,4.5) {\textbf{(c)}};
        \node at (1.2,4.5) {\textbf{(d)}};

        \node at (-5.8,-0.1) {\textbf{(e)}};
        \node at (1.2,-0.1) {\textbf{(f)}};

        \node at (-5.8,-4.7) {\textbf{(g)}};
        \node at (1.2,-4.7) {\textbf{(h)}};

    \end{tikzpicture}
    \caption{Metrics of the stylized facts for a synthetic time series of window size $20$ generated by a QGAN, compared to the metrics of the S\&P~500 index. 
    The generator of the QGAN is a PQC consisting of $10$ qubits and $18$ layers, simulated with the MPS approach with bond dimension $32$. 
    In \textbf{(a)}, we plot the probability density functions and in \textbf{(b)} the quantile-quantile plot of both the S\&P~500 index and the generated time series.
    In \textbf{(c)}-\textbf{(h)}, we plot the metrics absolute autocorrelation, linear autocorrelation and the leverage effect, as an indication of the stylized facts as described in Section~\ref{subsec:financialtimeseries}. 
    The Subfigures \textbf{(c)},  \textbf{(e)} and \textbf{(g)} show the metrics of the S\&P~500 index and the Subfigures \textbf{(d)}, \textbf{(f)} and \textbf{(h)} the metrics of the generated time series, respectively. Confidence intervals are calculated as in~\cite{chatfield1975time}.}
    \label{fig:MPS-simulation-10Q-L18-B32}
\end{figure*}

\begin{figure*}[htbp!]
    \centering
    \begin{tikzpicture}
        \node[inner sep=0pt] (img) at (0,0)
            {\includegraphics[width=0.8\linewidth]{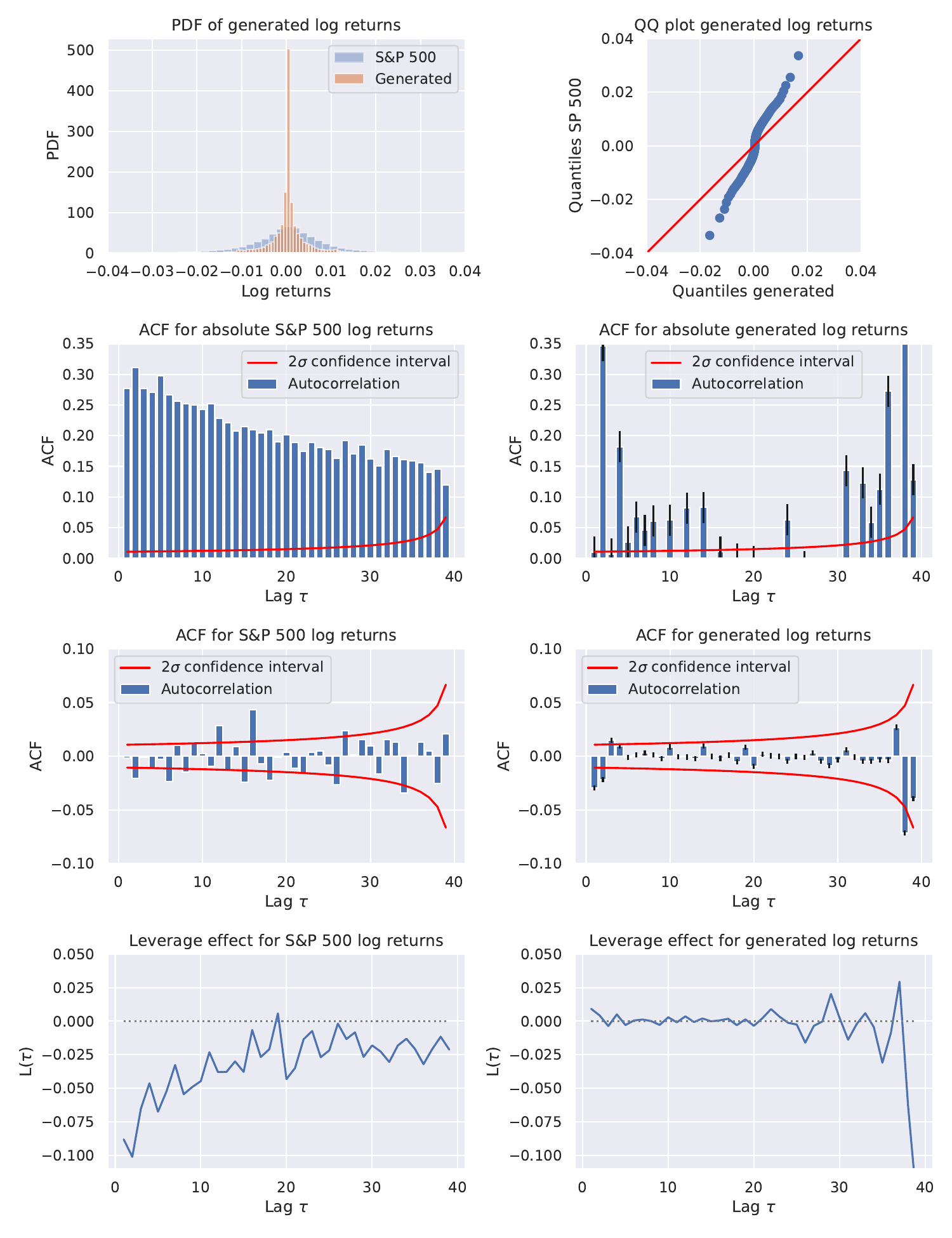}};

        \node at (-5.8,8.9) {\textbf{(a)}};
        \node at (1.2,8.9) {\textbf{(b)}};

        \node at (-5.8,4.5) {\textbf{(c)}};
        \node at (1.2,4.5) {\textbf{(d)}};

        \node at (-5.8,-0.1) {\textbf{(e)}};
        \node at (1.2,-0.1) {\textbf{(f)}};

        \node at (-5.8,-4.7) {\textbf{(g)}};
        \node at (1.2,-4.7) {\textbf{(h)}};

    \end{tikzpicture}
    \caption{Metrics of the stylized facts for a synthetic time series of window size $40$ generated by a QGAN, compared to the metrics of the S\&P~500 index. 
    The generator of the QGAN is a PQC consisting of $20$ qubits and $7$ layers, simulated with then MPS approach with bond dimension $70$. 
    In \textbf{(a)}, we plot the probability density functions and in \textbf{(b)} the quantile-quantile plot of both the S\&P~500 index and the generated time series.
    In \textbf{(c)}-\textbf{(h)}, we plot the metrics absolute autocorrelation, linear autocorrelation and the leverage effect, as an indication of the stylized facts as described in Section~\ref{subsec:financialtimeseries}. 
    The Subfigures \textbf{(c)},  \textbf{(e)} and \textbf{(g)} show the metrics of the S\&P~500 index and the Subfigures \textbf{(d)}, \textbf{(f)} and \textbf{(h)} the metrics of the generated time series, respectively. Confidence intervals are calculated as in~\cite{chatfield1975time}.}
    \label{fig:MPS-simulation-20Q-L6-B70}
\end{figure*}

Across the QGANs trained with different numbers of layers and bond dimensions in the MPS simulation, we generally observe that the generated time series reproduces the distribution, absence of linear autocorrelation, and volatility clustering, while the leverage effect is less pronounced.\\
In order to show that MPS can also be used for simulating QGANs that can generate time series with a larger window, we trained a QGAN with the MPS simulation of a PQC that consists of $20$ qubits. 
Such a simulation would be infeasible with full-state simulation.
We show the results of this simulation in Figure~\ref{fig:MPS-simulation-20Q-L6-B70} and in column \textbf{(d)} of Table~\ref{tab: table of metrics of differnt sims}.
Since increasing the number of qubits and the bond dimension raises the time required to train each epoch, the QGAN is trained for only $650$ epochs.\\
In the following section, we will analyze and compare the results of the different simulations shown here.

\section{Analysis of the results}\label{sec:discussion} 

In all simulations, the probability distributions of the generated time series closely resembles the distribution of the S\&P~500 index.
The temporal correlations show significant differences between the simulations. 
While the absence of linear autocorrelations is visible in all simulated time series, their absolute autocorrelation (indicating volatility clustering) and the leverage are of different quality. 
The full-state simulation (see Figure~\ref{fig:full-state-simulation}) shows both effects, even though the absolute autocorrelation is weaker and only visible for shorter time lags and the leverage effect is weaker than in the S\&P~500 index.\\
The MPS simulation with $10$ qubits, $18$ layers, and bond dimension $32$ (see Figure~\ref{fig:MPS-simulation-10Q-L18-B32}) shows absolute autocorrelation over longer lags, but both this and the leverage effect remain weaker than in the S\&P~500 index.\\
In the full-state simulation, adding a CNOT gate between the first and last qubit in each layer increases qubit correlation (see Appendix~\ref{appendix:full-state-simulation-connected}).
This might be a reason for the observation that the absolute autocorrelation of the generated time series increases at larger time lags.
We further benchmarked the QGAN with a full-state simulation against a quantum circuit Born machine in modeling the time-aggregated distribution of foreign exchange pairs yielding a better approximation of those distributions (see Appendix~\ref{appendix:fx_comparison}).\\
Using the MPS simulation, we also trained a QGAN with $20$ qubits, $5$ layers, and a bond dimension of $70$ (see Figure~\ref{fig:MPS-simulation-20Q-L6-B70}). 
This demonstrates that MPS can handle QGANs of greater complexity than those feasible with full-state simulation.
However, the training of QGANs with PQCs of a higher number of layers and qubits and MPS of higher bond dimensions increases the number of epochs needed in the training.
Further, each training epoch takes a longer time for these more complex models. 
For an equal computational cost, the generated time series therefore does not resemble the distributions and temporal effects of the target time series as closely as in the simulations with $10$ qubits.
But, by using a PQC with $20$ qubits, it is possible to simulate time series with a larger window size of $40$.\\
We remark that the loss landscapes differ significantly between full-state and MPS simulations due to their different approximation and simulation structure.
The difference in the quality of the generated time series can be partially attributed to the different features of the loss landscape.\\
Compared to the classical GAN experiments in~\cite{schwander2022Quantumgenerative}, which use multi-layer dense neural networks as generators and either a multi-layer dense or convolutional neural network as discriminator (with the same specifications as described in Appendix~\ref{appendix:cnn_achitecture}), both of our quantum simulation methods yield qualitatively improved results, particularly with respect to the Wasserstein distance and volatility clustering, as observed in the plots of the stylized facts. 
Note that the window size used in the classical experiments differs from ours, which may influence the comparison.

\section{Conclusions}\label{sec:conclusions}
We constructed a Wasserstein quantum generative adversarial network (QGAN) with a classical convolutional network as a discriminator and an expectation value sampler based on a parameterized quantum circuit (PQC) as a generator, in order to assess whether these quantum architectures have suitable inductive biases for generating synthetic financial time series known to be problematic for classical models.
This approach leverages the PQC architecture to intrinsically capture temporal correlations in the time series, while the training itself is performed solely on the Wasserstein distance computed over the aggregated time series distribution.
We simulated a PQC with $10$ qubits and $8$ layers with a full-state simulation and a PQC with $10$ and $20$ qubits and with up to $18$ layers as an approximation by a matrix product state (MPS) simulations with bond dimensions of up to $70$.
The latter approach allowed us to simulate PQCs with a higher number of layers and qubits, which makes it possible to train the generation of longer time series.\\
We compare the generated time series qualitatively with the S\&P~500 index by their distributions and their temporal correlations, also called the stylized facts.
These stylized facts are typically assessed qualitatively rather than quantitatively~\cite{dogariu2022GenerationRealistic}.\\
Our trained QGANs generate time series that match the desired distributions and exhibiting some of the temporal correlations seen in financial time series, such as in the S\&P~500 index.
Simulating the PQC with full-state simulations and MPS simulations yield different results, with circuit depth and the MPS bond dimension further influencing the performance. 
Both simulation methods motivate the study of quantum hardware in their ability to generate financial time series with stylized facts.
Our work has motivated studies in which the effect of such generated data on the training of neural networks has been explored~\cite{orlandi2024EnhancingFinancial} and~\cite{komninos2023Quantumcomputing}.\\
The application of this QGAN as subroutines for applications such as option pricing~\cite{rebentrost2018Quantumcomputational} and risk analysis~~\cite{woerner2019Quantumrisk} can be explored as well.
Further, a possible extension of our method is to train the model to replicate correlated stocks of the S\&P 500 index, motivated by research in community detection~\cite{macmahon2015CommunityDetection}. 
This could possibly be achieved by either learning the underlying distributions (in a similar way as done in Appendix~\ref{appendix:fx_comparison}), or by learning the individual time series similar to the ones in Section~\ref{sec:results}. 
As the number of qubits restricts the number of time steps and the number of stocks that can be generated, one could examine if quantum generators consisting of circuits on qudits can be successful, as that enables more independent measurements on each qudit.
Specifically for qudits, not only superconducting qubits form a suitable experimental platform, but also trapped ions, neutral atoms and integrated photonics are excellent candidates for manipulating higher-dimensional quantum information~\cite{wang2020QuditsHighDimensional, ringbauer2022universalqudit}. \\
An improvement of the training of the QGAN could be achieved in several ways.
Firstly, the effects of shot noise~\cite{shen2024ShadowFrugalExpectationValueSampling} in the training of the quantum generator could be explored. 
Secondly, different design choices, like choosing a different classical or quantum discriminator in the QGAN, diffusion model~\cite{takahashi2024Generationsynthetic}, or quantum long-short time memory models~\cite{chen2022QuantumLong} might lead to different results.
Thirdly, as the QGAN is trained with Wasserstein loss functions (see Equations \eqref{eq:loss-discriminator} and \eqref{eq:generator-loss}) that are taking the distribution of the time series into account, but not the temporal effects, an adaption of the training to consider them as well might lead to a better recovery of those temporal effects.
Lastly, one could try different definition of quantum Wasserstein distance~\cite{beatty2025OrderQuantum} that give theoretical improvements over the qualitative accuracy. 

\section{Acknowledgments \label{sec:acknowledgments}}
The authors thank Adrian Perez Salinas, Stefano Polla, Patrick Emonts, Liubov Markovich and Felix Frohnert for useful discussions during the project.
D.D., J.T. and V.D. acknowledge the support received from the Dutch National Growth Fund (NGF), as part of the Quantum Delta NL program.
V.D. acknowledges the support of the Dutch Research Council (NWO/ OCW),
as part of the Quantum Software Consortium program (project number 024.003.037).
J.T. acknowledges the support received from the European Union’s Horizon Europe research and innovation program through the ERC StG FINE-TEA-SQUAD (Grant No. 101040729).
V.D. acknowledges the support received from the European Union’s Horizon Europe program through the ERC CoG BeMAIQuantum (Grant No. 101124342).
This publication is part of the ‘Quantum Inspire — the Dutch Quantum Computer in the Cloud’ project (NWA.1292.19.194) of the NWA research program ‘Research on Routes by Consortia (ORC)’, which is funded by the Netherlands Organization for Scientific Research (NWO).
This work is supported by the European Union under the scheme HORIZON-INFRA-2021-DEV-02-01 – Preparatory phase of new ESFRI research infrastructure projects, Grant Agreement n.101079043, “SoBigData RI PPP: SoBigData RI Preparatory Phase Project”.
This work is also supported by the project ``Reconstruction, Resilience and Recovery of Socio-Economic Networks'' RECON-NET EP\_FAIR\_005 - PE0000013 ``FAIR'' - PNRR M4C2 Investment 1.3, financed by the European Union – NextGenerationEU.
The views and opinions expressed here are solely those of the authors and do not necessarily reflect those of the funding institutions. Neither of the funding institution can be held responsible for them. \\
This work was performed using the Xmaris and ALICE compute resources provided by Leiden University.
\section{Code Availability}
The code supporting this work is available at the repository: \url{https://github.com/LucasAugustusvd/Quantum-Finance}

\bibliography{bib}

\begin{thebibliography}{83}%
\makeatletter
\providecommand \@ifxundefined [1]{%
 \@ifx{#1\undefined}
}%
\providecommand \@ifnum [1]{%
 \ifnum #1\expandafter \@firstoftwo
 \else \expandafter \@secondoftwo
 \fi
}%
\providecommand \@ifx [1]{%
 \ifx #1\expandafter \@firstoftwo
 \else \expandafter \@secondoftwo
 \fi
}%
\providecommand \natexlab [1]{#1}%
\providecommand \enquote  [1]{``#1''}%
\providecommand \bibnamefont  [1]{#1}%
\providecommand \bibfnamefont [1]{#1}%
\providecommand \citenamefont [1]{#1}%
\providecommand \href@noop [0]{\@secondoftwo}%
\providecommand \href [0]{\begingroup \@sanitize@url \@href}%
\providecommand \@href[1]{\@@startlink{#1}\@@href}%
\providecommand \@@href[1]{\endgroup#1\@@endlink}%
\providecommand \@sanitize@url [0]{\catcode `\\12\catcode `\$12\catcode `\&12\catcode `\#12\catcode `\^12\catcode `\_12\catcode `\%12\relax}%
\providecommand \@@startlink[1]{}%
\providecommand \@@endlink[0]{}%
\providecommand \url  [0]{\begingroup\@sanitize@url \@url }%
\providecommand \@url [1]{\endgroup\@href {#1}{\urlprefix }}%
\providecommand \urlprefix  [0]{URL }%
\providecommand \Eprint [0]{\href }%
\providecommand \doibase [0]{https://doi.org/}%
\providecommand \selectlanguage [0]{\@gobble}%
\providecommand \bibinfo  [0]{\@secondoftwo}%
\providecommand \bibfield  [0]{\@secondoftwo}%
\providecommand \translation [1]{[#1]}%
\providecommand \BibitemOpen [0]{}%
\providecommand \bibitemStop [0]{}%
\providecommand \bibitemNoStop [0]{.\EOS\space}%
\providecommand \EOS [0]{\spacefactor3000\relax}%
\providecommand \BibitemShut  [1]{\csname bibitem#1\endcsname}%
\let\auto@bib@innerbib\@empty
\bibitem [{\citenamefont {Prince}(2023)}]{prince2023Understandingdeep}%
  \BibitemOpen
  \bibfield  {author} {\bibinfo {author} {\bibfnamefont {S.~J.}\ \bibnamefont {Prince}},\ }\href@noop {} {\emph {\bibinfo {title} {Understanding Deep Learning}}}\ (\bibinfo  {publisher} {MIT press},\ \bibinfo {year} {2023})\BibitemShut {NoStop}%
\bibitem [{\citenamefont {Goodfellow}\ \emph {et~al.}(2014)\citenamefont {Goodfellow}, \citenamefont {{Pouget-Abadie}}, \citenamefont {Mirza}, \citenamefont {Xu}, \citenamefont {{Warde-Farley}}, \citenamefont {Ozair}, \citenamefont {Courville},\ and\ \citenamefont {Bengio}}]{goodfellow2014GenerativeAdversarial}%
  \BibitemOpen
  \bibfield  {author} {\bibinfo {author} {\bibfnamefont {I.~J.}\ \bibnamefont {Goodfellow}}, \bibinfo {author} {\bibfnamefont {J.}~\bibnamefont {{Pouget-Abadie}}}, \bibinfo {author} {\bibfnamefont {M.}~\bibnamefont {Mirza}}, \bibinfo {author} {\bibfnamefont {B.}~\bibnamefont {Xu}}, \bibinfo {author} {\bibfnamefont {D.}~\bibnamefont {{Warde-Farley}}}, \bibinfo {author} {\bibfnamefont {S.}~\bibnamefont {Ozair}}, \bibinfo {author} {\bibfnamefont {A.}~\bibnamefont {Courville}},\ and\ \bibinfo {author} {\bibfnamefont {Y.}~\bibnamefont {Bengio}},\ }\bibfield  {title} {\bibinfo {title} {Generative {{Adversarial Nets}}},\ }in\ \href@noop {} {\emph {\bibinfo {booktitle} {Advances in {{Neural Information Processing Systems}}}}},\ Vol.~\bibinfo {volume} {27}\ (\bibinfo  {publisher} {Curran Associates, Inc.},\ \bibinfo {year} {2014})\BibitemShut {NoStop}%
\bibitem [{\citenamefont {Goodfellow}\ \emph {et~al.}(2020)\citenamefont {Goodfellow}, \citenamefont {{Pouget-Abadie}}, \citenamefont {Mirza}, \citenamefont {Xu}, \citenamefont {{Warde-Farley}}, \citenamefont {Ozair}, \citenamefont {Courville},\ and\ \citenamefont {Bengio}}]{goodfellow2020Generativeadversarial}%
  \BibitemOpen
  \bibfield  {author} {\bibinfo {author} {\bibfnamefont {I.}~\bibnamefont {Goodfellow}}, \bibinfo {author} {\bibfnamefont {J.}~\bibnamefont {{Pouget-Abadie}}}, \bibinfo {author} {\bibfnamefont {M.}~\bibnamefont {Mirza}}, \bibinfo {author} {\bibfnamefont {B.}~\bibnamefont {Xu}}, \bibinfo {author} {\bibfnamefont {D.}~\bibnamefont {{Warde-Farley}}}, \bibinfo {author} {\bibfnamefont {S.}~\bibnamefont {Ozair}}, \bibinfo {author} {\bibfnamefont {A.}~\bibnamefont {Courville}},\ and\ \bibinfo {author} {\bibfnamefont {Y.}~\bibnamefont {Bengio}},\ }\bibfield  {title} {\bibinfo {title} {Generative adversarial networks},\ }\href {https://doi.org/10.1145/3422622} {\bibfield  {journal} {\bibinfo  {journal} {Communications of the ACM}\ }\textbf {\bibinfo {volume} {63}},\ \bibinfo {pages} {139} (\bibinfo {year} {2020})}\BibitemShut {NoStop}%
\bibitem [{\citenamefont {Radford}\ \emph {et~al.}(2016)\citenamefont {Radford}, \citenamefont {Metz},\ and\ \citenamefont {Chintala}}]{radford2016UnsupervisedRepresentation}%
  \BibitemOpen
  \bibfield  {author} {\bibinfo {author} {\bibfnamefont {A.}~\bibnamefont {Radford}}, \bibinfo {author} {\bibfnamefont {L.}~\bibnamefont {Metz}},\ and\ \bibinfo {author} {\bibfnamefont {S.}~\bibnamefont {Chintala}},\ }\href {https://doi.org/10.48550/arXiv.1511.06434} {\bibinfo {title} {Unsupervised {{Representation Learning}} with {{Deep Convolutional Generative Adversarial Networks}}}} (\bibinfo {year} {2016}),\ \Eprint {https://arxiv.org/abs/1511.06434} {arXiv:1511.06434 [cs]} \BibitemShut {NoStop}%
\bibitem [{\citenamefont {Karras}\ \emph {et~al.}(2018)\citenamefont {Karras}, \citenamefont {Aila}, \citenamefont {Laine},\ and\ \citenamefont {Lehtinen}}]{karras2018ProgressiveGrowing}%
  \BibitemOpen
  \bibfield  {author} {\bibinfo {author} {\bibfnamefont {T.}~\bibnamefont {Karras}}, \bibinfo {author} {\bibfnamefont {T.}~\bibnamefont {Aila}}, \bibinfo {author} {\bibfnamefont {S.}~\bibnamefont {Laine}},\ and\ \bibinfo {author} {\bibfnamefont {J.}~\bibnamefont {Lehtinen}},\ }\href {https://doi.org/10.48550/arXiv.1710.10196} {\bibinfo {title} {Progressive {{Growing}} of {{GANs}} for {{Improved Quality}}, {{Stability}}, and {{Variation}}}} (\bibinfo {year} {2018}),\ \Eprint {https://arxiv.org/abs/1710.10196} {arXiv:1710.10196 [cs]} \BibitemShut {NoStop}%
\bibitem [{\citenamefont {Karras}\ \emph {et~al.}(2019)\citenamefont {Karras}, \citenamefont {Laine},\ and\ \citenamefont {Aila}}]{karras2019stylebasedgenerator}%
  \BibitemOpen
  \bibfield  {author} {\bibinfo {author} {\bibfnamefont {T.}~\bibnamefont {Karras}}, \bibinfo {author} {\bibfnamefont {S.}~\bibnamefont {Laine}},\ and\ \bibinfo {author} {\bibfnamefont {T.}~\bibnamefont {Aila}},\ }\bibfield  {title} {\bibinfo {title} {A style-based generator architecture for generative adversarial networks},\ }in\ \href@noop {} {\emph {\bibinfo {booktitle} {Proceedings of the {{IEEE}}/{{CVF}} Conference on Computer Vision and Pattern Recognition}}}\ (\bibinfo {year} {2019})\ pp.\ \bibinfo {pages} {4401--4410}\BibitemShut {NoStop}%
\bibitem [{\citenamefont {Gui}\ \emph {et~al.}(2021)\citenamefont {Gui}, \citenamefont {Sun}, \citenamefont {Wen}, \citenamefont {Tao},\ and\ \citenamefont {Ye}}]{gui2021reviewgenerative}%
  \BibitemOpen
  \bibfield  {author} {\bibinfo {author} {\bibfnamefont {J.}~\bibnamefont {Gui}}, \bibinfo {author} {\bibfnamefont {Z.}~\bibnamefont {Sun}}, \bibinfo {author} {\bibfnamefont {Y.}~\bibnamefont {Wen}}, \bibinfo {author} {\bibfnamefont {D.}~\bibnamefont {Tao}},\ and\ \bibinfo {author} {\bibfnamefont {J.}~\bibnamefont {Ye}},\ }\bibfield  {title} {\bibinfo {title} {A review on generative adversarial networks: {{Algorithms}}, theory, and applications},\ }\href@noop {} {\bibfield  {journal} {\bibinfo  {journal} {IEEE transactions on knowledge and data engineering}\ }\textbf {\bibinfo {volume} {35}},\ \bibinfo {pages} {3313} (\bibinfo {year} {2021})}\BibitemShut {NoStop}%
\bibitem [{\citenamefont {Shorten}\ and\ \citenamefont {Khoshgoftaar}(2019)}]{shorten2019surveyImage}%
  \BibitemOpen
  \bibfield  {author} {\bibinfo {author} {\bibfnamefont {C.}~\bibnamefont {Shorten}}\ and\ \bibinfo {author} {\bibfnamefont {T.~M.}\ \bibnamefont {Khoshgoftaar}},\ }\bibfield  {title} {\bibinfo {title} {A survey on {{Image Data Augmentation}} for {{Deep Learning}}},\ }\href {https://doi.org/10.1186/s40537-019-0197-0} {\bibfield  {journal} {\bibinfo  {journal} {Journal of Big Data}\ }\textbf {\bibinfo {volume} {6}},\ \bibinfo {pages} {60} (\bibinfo {year} {2019})}\BibitemShut {NoStop}%
\bibitem [{\citenamefont {{dos Santos Tanaka}}\ and\ \citenamefont {Aranha}(2019)}]{dossantostanaka2019DataAugmentation}%
  \BibitemOpen
  \bibfield  {author} {\bibinfo {author} {\bibfnamefont {F.~H.~K.}\ \bibnamefont {{dos Santos Tanaka}}}\ and\ \bibinfo {author} {\bibfnamefont {C.}~\bibnamefont {Aranha}},\ }\bibfield  {title} {\bibinfo {title} {Data {{Augmentation Using GANs}}},\ }\href@noop {} {\bibfield  {journal} {\bibinfo  {journal} {Proceedings of Machine Learning Research XXX}\ }\textbf {\bibinfo {volume} {1}},\ \bibinfo {pages} {16} (\bibinfo {year} {2019})}\BibitemShut {NoStop}%
\bibitem [{\citenamefont {Dixon}\ \emph {et~al.}(2020)\citenamefont {Dixon}, \citenamefont {Halperin},\ and\ \citenamefont {Bilokon}}]{dixon2020MachineLearning}%
  \BibitemOpen
  \bibfield  {author} {\bibinfo {author} {\bibfnamefont {M.~F.}\ \bibnamefont {Dixon}}, \bibinfo {author} {\bibfnamefont {I.}~\bibnamefont {Halperin}},\ and\ \bibinfo {author} {\bibfnamefont {P.}~\bibnamefont {Bilokon}},\ }\href {https://doi.org/10.1007/978-3-030-41068-1} {\emph {\bibinfo {title} {Machine {{Learning}} in {{Finance}}: {{From Theory}} to {{Practice}}}}}\ (\bibinfo  {publisher} {Springer International Publishing},\ \bibinfo {address} {Cham},\ \bibinfo {year} {2020})\BibitemShut {NoStop}%
\bibitem [{\citenamefont {Potluru}\ \emph {et~al.}(2024)\citenamefont {Potluru}, \citenamefont {Borrajo}, \citenamefont {Coletta}, \citenamefont {Dalmasso}, \citenamefont {{El-Laham}}, \citenamefont {Fons}, \citenamefont {Ghassemi}, \citenamefont {Gopalakrishnan}, \citenamefont {Gosai}, \citenamefont {Krea{\v c}i{\'c}}, \citenamefont {Mani}, \citenamefont {Obitayo}, \citenamefont {Paramanand}, \citenamefont {Raman}, \citenamefont {Solonin}, \citenamefont {Sood}, \citenamefont {Vyetrenko}, \citenamefont {Zhu}, \citenamefont {Veloso},\ and\ \citenamefont {Balch}}]{potluru2024SyntheticData}%
  \BibitemOpen
  \bibfield  {author} {\bibinfo {author} {\bibfnamefont {V.~K.}\ \bibnamefont {Potluru}}, \bibinfo {author} {\bibfnamefont {D.}~\bibnamefont {Borrajo}}, \bibinfo {author} {\bibfnamefont {A.}~\bibnamefont {Coletta}}, \bibinfo {author} {\bibfnamefont {N.}~\bibnamefont {Dalmasso}}, \bibinfo {author} {\bibfnamefont {Y.}~\bibnamefont {{El-Laham}}}, \bibinfo {author} {\bibfnamefont {E.}~\bibnamefont {Fons}}, \bibinfo {author} {\bibfnamefont {M.}~\bibnamefont {Ghassemi}}, \bibinfo {author} {\bibfnamefont {S.}~\bibnamefont {Gopalakrishnan}}, \bibinfo {author} {\bibfnamefont {V.}~\bibnamefont {Gosai}}, \bibinfo {author} {\bibfnamefont {E.}~\bibnamefont {Krea{\v c}i{\'c}}}, \bibinfo {author} {\bibfnamefont {G.}~\bibnamefont {Mani}}, \bibinfo {author} {\bibfnamefont {S.}~\bibnamefont {Obitayo}}, \bibinfo {author} {\bibfnamefont {D.}~\bibnamefont {Paramanand}}, \bibinfo {author} {\bibfnamefont {N.}~\bibnamefont {Raman}}, \bibinfo {author} {\bibfnamefont {M.}~\bibnamefont {Solonin}}, \bibinfo {author} {\bibfnamefont
  {S.}~\bibnamefont {Sood}}, \bibinfo {author} {\bibfnamefont {S.}~\bibnamefont {Vyetrenko}}, \bibinfo {author} {\bibfnamefont {H.}~\bibnamefont {Zhu}}, \bibinfo {author} {\bibfnamefont {M.}~\bibnamefont {Veloso}},\ and\ \bibinfo {author} {\bibfnamefont {T.}~\bibnamefont {Balch}},\ }\href {https://doi.org/10.48550/arXiv.2401.00081} {\bibinfo {title} {Synthetic {{Data Applications}} in {{Finance}}}} (\bibinfo {year} {2024}),\ \Eprint {https://arxiv.org/abs/2401.00081} {arXiv:2401.00081 [cs]} \BibitemShut {NoStop}%
\bibitem [{\citenamefont {Assefa}(2020)}]{assefa2020GeneratingSynthetic}%
  \BibitemOpen
  \bibfield  {author} {\bibinfo {author} {\bibfnamefont {S.}~\bibnamefont {Assefa}},\ }\bibfield  {title} {\bibinfo {title} {Generating {{Synthetic Data}} in {{Finance}}: {{Opportunities}}, {{Challenges}} and {{Pitfalls}}},\ }\bibfield  {journal} {\bibinfo  {journal} {SSRN Electronic Journal}\ }\href {https://doi.org/10.2139/ssrn.3634235} {10.2139/ssrn.3634235} (\bibinfo {year} {2020})\BibitemShut {NoStop}%
\bibitem [{\citenamefont {Preskill}(2018)}]{preskill2018QuantumComputing}%
  \BibitemOpen
  \bibfield  {author} {\bibinfo {author} {\bibfnamefont {J.}~\bibnamefont {Preskill}},\ }\bibfield  {title} {\bibinfo {title} {Quantum {{Computing}} in the {{NISQ}} era and beyond},\ }\href {https://doi.org/10.22331/q-2018-08-06-79} {\bibfield  {journal} {\bibinfo  {journal} {Quantum}\ }\textbf {\bibinfo {volume} {2}},\ \bibinfo {pages} {79} (\bibinfo {year} {2018})}\BibitemShut {NoStop}%
\bibitem [{\citenamefont {Cerezo}\ \emph {et~al.}(2021)\citenamefont {Cerezo}, \citenamefont {Arrasmith}, \citenamefont {Babbush}, \citenamefont {Benjamin}, \citenamefont {Endo}, \citenamefont {Fujii}, \citenamefont {McClean}, \citenamefont {Mitarai}, \citenamefont {Yuan},\ and\ \citenamefont {Cincio}}]{cerezo2021Variationalquantum}%
  \BibitemOpen
  \bibfield  {author} {\bibinfo {author} {\bibfnamefont {M.}~\bibnamefont {Cerezo}}, \bibinfo {author} {\bibfnamefont {A.}~\bibnamefont {Arrasmith}}, \bibinfo {author} {\bibfnamefont {R.}~\bibnamefont {Babbush}}, \bibinfo {author} {\bibfnamefont {S.~C.}\ \bibnamefont {Benjamin}}, \bibinfo {author} {\bibfnamefont {S.}~\bibnamefont {Endo}}, \bibinfo {author} {\bibfnamefont {K.}~\bibnamefont {Fujii}}, \bibinfo {author} {\bibfnamefont {J.~R.}\ \bibnamefont {McClean}}, \bibinfo {author} {\bibfnamefont {K.}~\bibnamefont {Mitarai}}, \bibinfo {author} {\bibfnamefont {X.}~\bibnamefont {Yuan}},\ and\ \bibinfo {author} {\bibfnamefont {L.}~\bibnamefont {Cincio}},\ }\bibfield  {title} {\bibinfo {title} {Variational quantum algorithms},\ }\href@noop {} {\bibfield  {journal} {\bibinfo  {journal} {Nature Reviews Physics}\ }\textbf {\bibinfo {volume} {3}},\ \bibinfo {pages} {625} (\bibinfo {year} {2021})}\BibitemShut {NoStop}%
\bibitem [{\citenamefont {Lloyd}\ and\ \citenamefont {Weedbrook}(2018)}]{lloyd2018QuantumGenerative}%
  \BibitemOpen
  \bibfield  {author} {\bibinfo {author} {\bibfnamefont {S.}~\bibnamefont {Lloyd}}\ and\ \bibinfo {author} {\bibfnamefont {C.}~\bibnamefont {Weedbrook}},\ }\bibfield  {title} {\bibinfo {title} {Quantum {{Generative Adversarial Learning}}},\ }\href {https://doi.org/10.1103/PhysRevLett.121.040502} {\bibfield  {journal} {\bibinfo  {journal} {Physical Review Letters}\ }\textbf {\bibinfo {volume} {121}},\ \bibinfo {pages} {040502} (\bibinfo {year} {2018})}\BibitemShut {NoStop}%
\bibitem [{\citenamefont {{Dallaire-Demers}}\ and\ \citenamefont {Killoran}(2018)}]{dallaire-demers2018Quantumgenerative}%
  \BibitemOpen
  \bibfield  {author} {\bibinfo {author} {\bibfnamefont {P.-L.}\ \bibnamefont {{Dallaire-Demers}}}\ and\ \bibinfo {author} {\bibfnamefont {N.}~\bibnamefont {Killoran}},\ }\bibfield  {title} {\bibinfo {title} {Quantum generative adversarial networks},\ }\href {https://doi.org/10.1103/PhysRevA.98.012324} {\bibfield  {journal} {\bibinfo  {journal} {Physical Review A}\ }\textbf {\bibinfo {volume} {98}},\ \bibinfo {pages} {012324} (\bibinfo {year} {2018})}\BibitemShut {NoStop}%
\bibitem [{\citenamefont {Aaronson}\ and\ \citenamefont {Arkhipov}(2013)}]{aaronson2013ComputationalComplexity}%
  \BibitemOpen
  \bibfield  {author} {\bibinfo {author} {\bibfnamefont {S.}~\bibnamefont {Aaronson}}\ and\ \bibinfo {author} {\bibfnamefont {A.}~\bibnamefont {Arkhipov}},\ }\bibfield  {title} {\bibinfo {title} {The {{Computational Complexity}} of {{Linear Optics}}},\ }\href {https://doi.org/10.4086/toc.2013.v009a004} {\bibfield  {journal} {\bibinfo  {journal} {Theory of Computing}\ }\textbf {\bibinfo {volume} {9}},\ \bibinfo {pages} {143} (\bibinfo {year} {2013})}\BibitemShut {NoStop}%
\bibitem [{\citenamefont {Wilms}\ \emph {et~al.}(2025)\citenamefont {Wilms}, \citenamefont {Ohff}, \citenamefont {Skolik}, \citenamefont {Eisert}, \citenamefont {Khatri},\ and\ \citenamefont {Reiss}}]{wilms2025Quantumreinforcement}%
  \BibitemOpen
  \bibfield  {author} {\bibinfo {author} {\bibfnamefont {A.}~\bibnamefont {Wilms}}, \bibinfo {author} {\bibfnamefont {L.}~\bibnamefont {Ohff}}, \bibinfo {author} {\bibfnamefont {A.}~\bibnamefont {Skolik}}, \bibinfo {author} {\bibfnamefont {J.}~\bibnamefont {Eisert}}, \bibinfo {author} {\bibfnamefont {S.}~\bibnamefont {Khatri}},\ and\ \bibinfo {author} {\bibfnamefont {D.~A.}\ \bibnamefont {Reiss}},\ }\href {https://doi.org/10.48550/arXiv.2504.16258} {\bibinfo {title} {Quantum reinforcement learning of classical rare dynamics: {{Enhancement}} by intrinsic {{Fourier}} features}} (\bibinfo {year} {2025}),\ \Eprint {https://arxiv.org/abs/2504.16258} {arXiv:2504.16258 [quant-ph]} \BibitemShut {NoStop}%
\bibitem [{\citenamefont {Abbas}\ \emph {et~al.}(2021)\citenamefont {Abbas}, \citenamefont {Sutter}, \citenamefont {Zoufal}, \citenamefont {Lucchi}, \citenamefont {Figalli},\ and\ \citenamefont {Woerner}}]{abbas2021powerquantum}%
  \BibitemOpen
  \bibfield  {author} {\bibinfo {author} {\bibfnamefont {A.}~\bibnamefont {Abbas}}, \bibinfo {author} {\bibfnamefont {D.}~\bibnamefont {Sutter}}, \bibinfo {author} {\bibfnamefont {C.}~\bibnamefont {Zoufal}}, \bibinfo {author} {\bibfnamefont {A.}~\bibnamefont {Lucchi}}, \bibinfo {author} {\bibfnamefont {A.}~\bibnamefont {Figalli}},\ and\ \bibinfo {author} {\bibfnamefont {S.}~\bibnamefont {Woerner}},\ }\bibfield  {title} {\bibinfo {title} {The power of quantum neural networks},\ }\href {https://doi.org/10.1038/s43588-021-00084-1} {\bibfield  {journal} {\bibinfo  {journal} {Nature Computational Science}\ }\textbf {\bibinfo {volume} {1}},\ \bibinfo {pages} {403} (\bibinfo {year} {2021})}\BibitemShut {NoStop}%
\bibitem [{\citenamefont {Molteni}\ \emph {et~al.}(2025)\citenamefont {Molteni}, \citenamefont {Marshall},\ and\ \citenamefont {Dunjko}}]{molteni2025Quantummachine}%
  \BibitemOpen
  \bibfield  {author} {\bibinfo {author} {\bibfnamefont {R.}~\bibnamefont {Molteni}}, \bibinfo {author} {\bibfnamefont {S.~C.}\ \bibnamefont {Marshall}},\ and\ \bibinfo {author} {\bibfnamefont {V.}~\bibnamefont {Dunjko}},\ }\href {https://doi.org/10.48550/arXiv.2504.15964} {\bibinfo {title} {Quantum machine learning advantages beyond hardness of evaluation}} (\bibinfo {year} {2025}),\ \Eprint {https://arxiv.org/abs/2504.15964} {arXiv:2504.15964 [quant-ph]} \BibitemShut {NoStop}%
\bibitem [{\citenamefont {Coyle}\ \emph {et~al.}(2021)\citenamefont {Coyle}, \citenamefont {Henderson}, \citenamefont {Chan Jin~Le}, \citenamefont {Kumar}, \citenamefont {Paini},\ and\ \citenamefont {Kashefi}}]{coyle2021Quantumclassical}%
  \BibitemOpen
  \bibfield  {author} {\bibinfo {author} {\bibfnamefont {B.}~\bibnamefont {Coyle}}, \bibinfo {author} {\bibfnamefont {M.}~\bibnamefont {Henderson}}, \bibinfo {author} {\bibfnamefont {J.}~\bibnamefont {Chan Jin~Le}}, \bibinfo {author} {\bibfnamefont {N.}~\bibnamefont {Kumar}}, \bibinfo {author} {\bibfnamefont {M.}~\bibnamefont {Paini}},\ and\ \bibinfo {author} {\bibfnamefont {E.}~\bibnamefont {Kashefi}},\ }\bibfield  {title} {\bibinfo {title} {Quantum versus classical generative modelling in finance},\ }\href {https://doi.org/10.1088/2058-9565/abd3db} {\bibfield  {journal} {\bibinfo  {journal} {Quantum Science and Technology}\ }\textbf {\bibinfo {volume} {6}},\ \bibinfo {pages} {024013} (\bibinfo {year} {2021})}\BibitemShut {NoStop}%
\bibitem [{\citenamefont {Dogariu}\ \emph {et~al.}(2022)\citenamefont {Dogariu}, \citenamefont {{\c S}tefan}, \citenamefont {Boteanu}, \citenamefont {Lamba}, \citenamefont {Kim},\ and\ \citenamefont {Ionescu}}]{dogariu2022GenerationRealistic}%
  \BibitemOpen
  \bibfield  {author} {\bibinfo {author} {\bibfnamefont {M.}~\bibnamefont {Dogariu}}, \bibinfo {author} {\bibfnamefont {L.-D.}\ \bibnamefont {{\c S}tefan}}, \bibinfo {author} {\bibfnamefont {B.~A.}\ \bibnamefont {Boteanu}}, \bibinfo {author} {\bibfnamefont {C.}~\bibnamefont {Lamba}}, \bibinfo {author} {\bibfnamefont {B.}~\bibnamefont {Kim}},\ and\ \bibinfo {author} {\bibfnamefont {B.}~\bibnamefont {Ionescu}},\ }\bibfield  {title} {\bibinfo {title} {Generation of {{Realistic Synthetic Financial Time-series}}},\ }\href {https://doi.org/10.1145/3501305} {\bibfield  {journal} {\bibinfo  {journal} {ACM Transactions on Multimedia Computing, Communications, and Applications}\ }\textbf {\bibinfo {volume} {18}},\ \bibinfo {pages} {1} (\bibinfo {year} {2022})}\BibitemShut {NoStop}%
\bibitem [{\citenamefont {{\"O}stlund}\ and\ \citenamefont {Rommer}(1995)}]{ostlund1995ThermodynamicLimit}%
  \BibitemOpen
  \bibfield  {author} {\bibinfo {author} {\bibfnamefont {S.}~\bibnamefont {{\"O}stlund}}\ and\ \bibinfo {author} {\bibfnamefont {S.}~\bibnamefont {Rommer}},\ }\bibfield  {title} {\bibinfo {title} {Thermodynamic {{Limit}} of {{Density Matrix Renormalization}}},\ }\href {https://doi.org/10.1103/PhysRevLett.75.3537} {\bibfield  {journal} {\bibinfo  {journal} {Physical Review Letters}\ }\textbf {\bibinfo {volume} {75}},\ \bibinfo {pages} {3537} (\bibinfo {year} {1995})}\BibitemShut {NoStop}%
\bibitem [{\citenamefont {Verstraete}\ \emph {et~al.}(2008)\citenamefont {Verstraete}, \citenamefont {~},\ and\ \citenamefont {{and Cirac}}}]{verstraete2008Matrixproduct}%
  \BibitemOpen
  \bibfield  {author} {\bibinfo {author} {\bibfnamefont {F.}~\bibnamefont {Verstraete}}, \bibinfo {author} {\bibfnamefont {M.}~\bibnamefont {~}, \bibfnamefont {V.}},\ and\ \bibinfo {author} {\bibfnamefont {J.}~\bibnamefont {{and Cirac}}},\ }\bibfield  {title} {\bibinfo {title} {Matrix product states, projected entangled pair states, and variational renormalization group methods for quantum spin systems},\ }\href {https://doi.org/10.1080/14789940801912366} {\bibfield  {journal} {\bibinfo  {journal} {Advances in Physics}\ }\textbf {\bibinfo {volume} {57}},\ \bibinfo {pages} {143} (\bibinfo {year} {2008})}\BibitemShut {NoStop}%
\bibitem [{\citenamefont {Indices}(2025)}]{2025SP500}%
  \BibitemOpen
  \bibfield  {author} {\bibinfo {author} {\bibfnamefont {S.~D.~J.}\ \bibnamefont {Indices}},\ }\href@noop {} {\bibinfo {title} {S\&{{P}} 500{\textregistered}}},\ \bibinfo {howpublished} {https://www.spglobal.com/spdji/en/indices/equity/sp-500/\#overview} (\bibinfo {year} {2025}),\ \bibinfo {note} {last accessed 2025-06-19}\BibitemShut {NoStop}%
\bibitem [{\citenamefont {Black}\ and\ \citenamefont {Scholes}(1973)}]{black1973PricingOptions}%
  \BibitemOpen
  \bibfield  {author} {\bibinfo {author} {\bibfnamefont {F.}~\bibnamefont {Black}}\ and\ \bibinfo {author} {\bibfnamefont {M.}~\bibnamefont {Scholes}},\ }\bibfield  {title} {\bibinfo {title} {The {{Pricing}} of {{Options}} and {{Corporate Liabilities}}},\ }\href {https://doi.org/10.1086/260062} {\bibfield  {journal} {\bibinfo  {journal} {Journal of Political Economy}\ }\textbf {\bibinfo {volume} {81}},\ \bibinfo {pages} {637} (\bibinfo {year} {1973})}\BibitemShut {NoStop}%
\bibitem [{\citenamefont {Cont}(2001)}]{cont2001Empiricalproperties}%
  \BibitemOpen
  \bibfield  {author} {\bibinfo {author} {\bibfnamefont {R.}~\bibnamefont {Cont}},\ }\bibfield  {title} {\bibinfo {title} {Empirical properties of asset returns: Stylized facts and statistical issues},\ }\href {https://doi.org/10.1080/713665670} {\bibfield  {journal} {\bibinfo  {journal} {Quantitative Finance}\ }\textbf {\bibinfo {volume} {1}},\ \bibinfo {pages} {223} (\bibinfo {year} {2001})}\BibitemShut {NoStop}%
\bibitem [{\citenamefont {Eckerli}\ and\ \citenamefont {Osterrieder}(2021)}]{eckerli2021GenerativeAdversarial}%
  \BibitemOpen
  \bibfield  {author} {\bibinfo {author} {\bibfnamefont {F.}~\bibnamefont {Eckerli}}\ and\ \bibinfo {author} {\bibfnamefont {J.}~\bibnamefont {Osterrieder}},\ }\href {https://doi.org/10.48550/arXiv.2106.06364} {\bibinfo {title} {Generative {{Adversarial Networks}} in finance: An overview}} (\bibinfo {year} {2021}),\ \Eprint {https://arxiv.org/abs/2106.06364} {arXiv:2106.06364 [q-fin]} \BibitemShut {NoStop}%
\bibitem [{\citenamefont {Saxena}\ and\ \citenamefont {Cao}(2022)}]{saxena2022GenerativeAdversarial}%
  \BibitemOpen
  \bibfield  {author} {\bibinfo {author} {\bibfnamefont {D.}~\bibnamefont {Saxena}}\ and\ \bibinfo {author} {\bibfnamefont {J.}~\bibnamefont {Cao}},\ }\bibfield  {title} {\bibinfo {title} {Generative {{Adversarial Networks}} ({{GANs}}): {{Challenges}}, {{Solutions}}, and {{Future Directions}}},\ }\href {https://doi.org/10.1145/3446374} {\bibfield  {journal} {\bibinfo  {journal} {ACM Computing Surveys}\ }\textbf {\bibinfo {volume} {54}},\ \bibinfo {pages} {1} (\bibinfo {year} {2022})}\BibitemShut {NoStop}%
\bibitem [{\citenamefont {Arjovsky}\ \emph {et~al.}(2017)\citenamefont {Arjovsky}, \citenamefont {Chintala},\ and\ \citenamefont {Bottou}}]{arjovsky2017WassersteinGenerative}%
  \BibitemOpen
  \bibfield  {author} {\bibinfo {author} {\bibfnamefont {M.}~\bibnamefont {Arjovsky}}, \bibinfo {author} {\bibfnamefont {S.}~\bibnamefont {Chintala}},\ and\ \bibinfo {author} {\bibfnamefont {L.}~\bibnamefont {Bottou}},\ }\bibfield  {title} {\bibinfo {title} {Wasserstein {{Generative Adversarial Networks}}},\ }in\ \href@noop {} {\emph {\bibinfo {booktitle} {Proceedings of the 34th {{International Conference}} on {{Machine Learning}}}}}\ (\bibinfo  {publisher} {PMLR},\ \bibinfo {year} {2017})\ pp.\ \bibinfo {pages} {214--223}\BibitemShut {NoStop}%
\bibitem [{\citenamefont {Villani}(2009)}]{villani2009Wassersteindistances}%
  \BibitemOpen
  \bibfield  {author} {\bibinfo {author} {\bibfnamefont {C.}~\bibnamefont {Villani}},\ }\bibfield  {title} {\bibinfo {title} {The {{Wasserstein}} distances},\ }in\ \href {https://doi.org/10.1007/978-3-540-71050-9_6} {\emph {\bibinfo {booktitle} {Optimal {{Transport}}: {{Old}} and {{New}}}}},\ \bibinfo {editor} {edited by\ \bibinfo {editor} {\bibfnamefont {C.}~\bibnamefont {Villani}}}\ (\bibinfo  {publisher} {Springer},\ \bibinfo {address} {Berlin, Heidelberg},\ \bibinfo {year} {2009})\ pp.\ \bibinfo {pages} {93--111}\BibitemShut {NoStop}%
\bibitem [{\citenamefont {Gulrajani}\ \emph {et~al.}(2017)\citenamefont {Gulrajani}, \citenamefont {Ahmed}, \citenamefont {Arjovsky}, \citenamefont {Dumoulin},\ and\ \citenamefont {Courville}}]{gulrajani2017Improvedtraining}%
  \BibitemOpen
  \bibfield  {author} {\bibinfo {author} {\bibfnamefont {I.}~\bibnamefont {Gulrajani}}, \bibinfo {author} {\bibfnamefont {F.}~\bibnamefont {Ahmed}}, \bibinfo {author} {\bibfnamefont {M.}~\bibnamefont {Arjovsky}}, \bibinfo {author} {\bibfnamefont {V.}~\bibnamefont {Dumoulin}},\ and\ \bibinfo {author} {\bibfnamefont {A.~C.}\ \bibnamefont {Courville}},\ }\bibfield  {title} {\bibinfo {title} {Improved training of wasserstein gans},\ }\href@noop {} {\bibfield  {journal} {\bibinfo  {journal} {Advances in neural information processing systems}\ }\textbf {\bibinfo {volume} {30}} (\bibinfo {year} {2017})}\BibitemShut {NoStop}%
\bibitem [{\citenamefont {Sweke}\ \emph {et~al.}(2021)\citenamefont {Sweke}, \citenamefont {Seifert}, \citenamefont {Hangleiter},\ and\ \citenamefont {Eisert}}]{sweke2021QuantumClassical}%
  \BibitemOpen
  \bibfield  {author} {\bibinfo {author} {\bibfnamefont {R.}~\bibnamefont {Sweke}}, \bibinfo {author} {\bibfnamefont {J.-P.}\ \bibnamefont {Seifert}}, \bibinfo {author} {\bibfnamefont {D.}~\bibnamefont {Hangleiter}},\ and\ \bibinfo {author} {\bibfnamefont {J.}~\bibnamefont {Eisert}},\ }\bibfield  {title} {\bibinfo {title} {On the {{Quantum}} versus {{Classical Learnability}} of {{Discrete Distributions}}},\ }\href {https://doi.org/10.22331/q-2021-03-23-417} {\bibfield  {journal} {\bibinfo  {journal} {Quantum}\ }\textbf {\bibinfo {volume} {5}},\ \bibinfo {pages} {417} (\bibinfo {year} {2021})}\BibitemShut {NoStop}%
\bibitem [{\citenamefont {Benedetti}\ \emph {et~al.}(2018)\citenamefont {Benedetti}, \citenamefont {{Garcia-Pintos}}, \citenamefont {Perdomo}, \citenamefont {{Leyton-Ortega}}, \citenamefont {Nam},\ and\ \citenamefont {{Perdomo-Ortiz}}}]{benedetti2018generativemodeling}%
  \BibitemOpen
  \bibfield  {author} {\bibinfo {author} {\bibfnamefont {M.}~\bibnamefont {Benedetti}}, \bibinfo {author} {\bibfnamefont {D.}~\bibnamefont {{Garcia-Pintos}}}, \bibinfo {author} {\bibfnamefont {O.}~\bibnamefont {Perdomo}}, \bibinfo {author} {\bibfnamefont {V.}~\bibnamefont {{Leyton-Ortega}}}, \bibinfo {author} {\bibfnamefont {Y.}~\bibnamefont {Nam}},\ and\ \bibinfo {author} {\bibfnamefont {A.}~\bibnamefont {{Perdomo-Ortiz}}},\ }\href {https://doi.org/10.1038/s41534-019-0157-8} {\bibinfo {title} {A generative modeling approach for benchmarking and training shallow quantum circuits}},\ \bibinfo {howpublished} {https://arxiv.org/abs/1801.07686v4} (\bibinfo {year} {2018})\BibitemShut {NoStop}%
\bibitem [{\citenamefont {Liu}\ and\ \citenamefont {Wang}(2018)}]{liu2018DifferentiableLearning}%
  \BibitemOpen
  \bibfield  {author} {\bibinfo {author} {\bibfnamefont {J.-G.}\ \bibnamefont {Liu}}\ and\ \bibinfo {author} {\bibfnamefont {L.}~\bibnamefont {Wang}},\ }\href {https://doi.org/10.1103/PhysRevA.98.062324} {\bibinfo {title} {Differentiable {{Learning}} of {{Quantum Circuit Born Machine}}}},\ \bibinfo {howpublished} {https://arxiv.org/abs/1804.04168v1} (\bibinfo {year} {2018})\BibitemShut {NoStop}%
\bibitem [{\citenamefont {Romero}\ and\ \citenamefont {{Aspuru-Guzik}}(2021)}]{romero2021VariationalQuantum}%
  \BibitemOpen
  \bibfield  {author} {\bibinfo {author} {\bibfnamefont {J.}~\bibnamefont {Romero}}\ and\ \bibinfo {author} {\bibfnamefont {A.}~\bibnamefont {{Aspuru-Guzik}}},\ }\bibfield  {title} {\bibinfo {title} {Variational {{Quantum Generators}}: {{Generative Adversarial Quantum Machine Learning}} for {{Continuous Distributions}}},\ }\href {https://doi.org/10.1002/qute.202000003} {\bibfield  {journal} {\bibinfo  {journal} {Advanced Quantum Technologies}\ }\textbf {\bibinfo {volume} {4}},\ \bibinfo {pages} {2000003} (\bibinfo {year} {2021})}\BibitemShut {NoStop}%
\bibitem [{\citenamefont {Anand}\ \emph {et~al.}(2021)\citenamefont {Anand}, \citenamefont {Romero}, \citenamefont {Degroote},\ and\ \citenamefont {{Aspuru-Guzik}}}]{anand2021NoiseRobustness}%
  \BibitemOpen
  \bibfield  {author} {\bibinfo {author} {\bibfnamefont {A.}~\bibnamefont {Anand}}, \bibinfo {author} {\bibfnamefont {J.}~\bibnamefont {Romero}}, \bibinfo {author} {\bibfnamefont {M.}~\bibnamefont {Degroote}},\ and\ \bibinfo {author} {\bibfnamefont {A.}~\bibnamefont {{Aspuru-Guzik}}},\ }\bibfield  {title} {\bibinfo {title} {Noise {{Robustness}} and {{Experimental Demonstration}} of a {{Quantum Generative Adversarial Network}} for {{Continuous Distributions}}},\ }\href {https://doi.org/10.1002/qute.202000069} {\bibfield  {journal} {\bibinfo  {journal} {Advanced Quantum Technologies}\ }\textbf {\bibinfo {volume} {4}},\ \bibinfo {pages} {2000069} (\bibinfo {year} {2021})}\BibitemShut {NoStop}%
\bibitem [{\citenamefont {Barthe}\ \emph {et~al.}(2024)\citenamefont {Barthe}, \citenamefont {Grossi}, \citenamefont {Vallecorsa}, \citenamefont {Tura},\ and\ \citenamefont {Dunjko}}]{barthe2024Parameterizedquantum}%
  \BibitemOpen
  \bibfield  {author} {\bibinfo {author} {\bibfnamefont {A.}~\bibnamefont {Barthe}}, \bibinfo {author} {\bibfnamefont {M.}~\bibnamefont {Grossi}}, \bibinfo {author} {\bibfnamefont {S.}~\bibnamefont {Vallecorsa}}, \bibinfo {author} {\bibfnamefont {J.}~\bibnamefont {Tura}},\ and\ \bibinfo {author} {\bibfnamefont {V.}~\bibnamefont {Dunjko}},\ }\href {https://doi.org/10.48550/arXiv.2402.09848} {\bibinfo {title} {Parameterized quantum circuits as universal generative models for continuous multivariate distributions}} (\bibinfo {year} {2024}),\ \Eprint {https://arxiv.org/abs/2402.09848} {arXiv:2402.09848 [quant-ph]} \BibitemShut {NoStop}%
\bibitem [{\citenamefont {Shen}\ \emph {et~al.}(2024)\citenamefont {Shen}, \citenamefont {Kurkin}, \citenamefont {Salinas}, \citenamefont {Shishenina}, \citenamefont {Dunjko},\ and\ \citenamefont {Wang}}]{shen2024ShadowFrugalExpectationValueSampling}%
  \BibitemOpen
  \bibfield  {author} {\bibinfo {author} {\bibfnamefont {K.}~\bibnamefont {Shen}}, \bibinfo {author} {\bibfnamefont {A.}~\bibnamefont {Kurkin}}, \bibinfo {author} {\bibfnamefont {A.~P.}\ \bibnamefont {Salinas}}, \bibinfo {author} {\bibfnamefont {E.}~\bibnamefont {Shishenina}}, \bibinfo {author} {\bibfnamefont {V.}~\bibnamefont {Dunjko}},\ and\ \bibinfo {author} {\bibfnamefont {H.}~\bibnamefont {Wang}},\ }\href@noop {} {\bibinfo {title} {Shadow-{{Frugal Expectation-Value-Sampling Variational Quantum Generative Model}}}},\ \bibinfo {howpublished} {https://arxiv.org/abs/2412.17039v1} (\bibinfo {year} {2024})\BibitemShut {NoStop}%
\bibitem [{\citenamefont {Kandala}\ \emph {et~al.}(2017)\citenamefont {Kandala}, \citenamefont {Mezzacapo}, \citenamefont {Temme}, \citenamefont {Takita}, \citenamefont {Brink}, \citenamefont {Chow},\ and\ \citenamefont {Gambetta}}]{kandala2017Hardwareefficientvariational}%
  \BibitemOpen
  \bibfield  {author} {\bibinfo {author} {\bibfnamefont {A.}~\bibnamefont {Kandala}}, \bibinfo {author} {\bibfnamefont {A.}~\bibnamefont {Mezzacapo}}, \bibinfo {author} {\bibfnamefont {K.}~\bibnamefont {Temme}}, \bibinfo {author} {\bibfnamefont {M.}~\bibnamefont {Takita}}, \bibinfo {author} {\bibfnamefont {M.}~\bibnamefont {Brink}}, \bibinfo {author} {\bibfnamefont {J.~M.}\ \bibnamefont {Chow}},\ and\ \bibinfo {author} {\bibfnamefont {J.~M.}\ \bibnamefont {Gambetta}},\ }\bibfield  {title} {\bibinfo {title} {Hardware-efficient variational quantum eigensolver for small molecules and quantum magnets},\ }\href@noop {} {\bibfield  {journal} {\bibinfo  {journal} {nature}\ }\textbf {\bibinfo {volume} {549}},\ \bibinfo {pages} {242} (\bibinfo {year} {2017})}\BibitemShut {NoStop}%
\bibitem [{\citenamefont {Takahashi}\ \emph {et~al.}(2019)\citenamefont {Takahashi}, \citenamefont {Chen},\ and\ \citenamefont {{Tanaka-Ishii}}}]{takahashi2019Modelingfinancial}%
  \BibitemOpen
  \bibfield  {author} {\bibinfo {author} {\bibfnamefont {S.}~\bibnamefont {Takahashi}}, \bibinfo {author} {\bibfnamefont {Y.}~\bibnamefont {Chen}},\ and\ \bibinfo {author} {\bibfnamefont {K.}~\bibnamefont {{Tanaka-Ishii}}},\ }\bibfield  {title} {\bibinfo {title} {Modeling financial time-series with generative adversarial networks},\ }\href {https://doi.org/10.1016/j.physa.2019.121261} {\bibfield  {journal} {\bibinfo  {journal} {Physica A: Statistical Mechanics and its Applications}\ }\textbf {\bibinfo {volume} {527}},\ \bibinfo {pages} {121261} (\bibinfo {year} {2019})}\BibitemShut {NoStop}%
\bibitem [{\citenamefont {Wiese}\ \emph {et~al.}(2020)\citenamefont {Wiese}, \citenamefont {~}, \citenamefont {~},\ and\ \citenamefont {{and Kretschmer}}}]{wiese2020QuantGANs}%
  \BibitemOpen
  \bibfield  {author} {\bibinfo {author} {\bibfnamefont {M.}~\bibnamefont {Wiese}}, \bibinfo {author} {\bibfnamefont {K.}~\bibnamefont {~}, \bibfnamefont {Robert}}, \bibinfo {author} {\bibfnamefont {K.}~\bibnamefont {~}, \bibfnamefont {Ralf}},\ and\ \bibinfo {author} {\bibfnamefont {P.}~\bibnamefont {{and Kretschmer}}},\ }\bibfield  {title} {\bibinfo {title} {Quant {{GANs}}: Deep generation of financial time series},\ }\href {https://doi.org/10.1080/14697688.2020.1730426} {\bibfield  {journal} {\bibinfo  {journal} {Quantitative Finance}\ }\textbf {\bibinfo {volume} {20}},\ \bibinfo {pages} {1419} (\bibinfo {year} {2020})}\BibitemShut {NoStop}%
\bibitem [{\citenamefont {Zhang}\ \emph {et~al.}(2019)\citenamefont {Zhang}, \citenamefont {Zhong}, \citenamefont {Dong}, \citenamefont {Wang},\ and\ \citenamefont {Wang}}]{zhang2019StockMarket}%
  \BibitemOpen
  \bibfield  {author} {\bibinfo {author} {\bibfnamefont {K.}~\bibnamefont {Zhang}}, \bibinfo {author} {\bibfnamefont {G.}~\bibnamefont {Zhong}}, \bibinfo {author} {\bibfnamefont {J.}~\bibnamefont {Dong}}, \bibinfo {author} {\bibfnamefont {S.}~\bibnamefont {Wang}},\ and\ \bibinfo {author} {\bibfnamefont {Y.}~\bibnamefont {Wang}},\ }\bibfield  {title} {\bibinfo {title} {Stock {{Market Prediction Based}} on {{Generative Adversarial Network}}},\ }\href {https://doi.org/10.1016/j.procs.2019.01.256} {\bibfield  {journal} {\bibinfo  {journal} {Procedia Computer Science}\ }\bibinfo {series} {2018 {{International Conference}} on {{Identification}}, {{Information}} and {{Knowledge}} in the {{Internet}} of {{Things}}},\ \textbf {\bibinfo {volume} {147}},\ \bibinfo {pages} {400} (\bibinfo {year} {2019})}\BibitemShut {NoStop}%
\bibitem [{\citenamefont {Takahashi}\ and\ \citenamefont {Mizuno}(2024)}]{takahashi2024Generationsynthetic}%
  \BibitemOpen
  \bibfield  {author} {\bibinfo {author} {\bibfnamefont {T.}~\bibnamefont {Takahashi}}\ and\ \bibinfo {author} {\bibfnamefont {T.}~\bibnamefont {Mizuno}},\ }\href {https://doi.org/10.48550/arXiv.2410.18897} {\bibinfo {title} {Generation of synthetic financial time series by diffusion models}} (\bibinfo {year} {2024}),\ \Eprint {https://arxiv.org/abs/2410.18897} {arXiv:2410.18897 [q-fin]} \BibitemShut {NoStop}%
\bibitem [{\citenamefont {Tian}\ \emph {et~al.}(2022)\citenamefont {Tian}, \citenamefont {Sun}, \citenamefont {Du}, \citenamefont {Zhao}, \citenamefont {Liu}, \citenamefont {Zhang}, \citenamefont {Yi}, \citenamefont {Huang}, \citenamefont {Wang}, \citenamefont {Wu}, \citenamefont {Hsieh}, \citenamefont {Liu}, \citenamefont {Yang},\ and\ \citenamefont {Tao}}]{tian2022RecentAdvances}%
  \BibitemOpen
  \bibfield  {author} {\bibinfo {author} {\bibfnamefont {J.}~\bibnamefont {Tian}}, \bibinfo {author} {\bibfnamefont {X.}~\bibnamefont {Sun}}, \bibinfo {author} {\bibfnamefont {Y.}~\bibnamefont {Du}}, \bibinfo {author} {\bibfnamefont {S.}~\bibnamefont {Zhao}}, \bibinfo {author} {\bibfnamefont {Q.}~\bibnamefont {Liu}}, \bibinfo {author} {\bibfnamefont {K.}~\bibnamefont {Zhang}}, \bibinfo {author} {\bibfnamefont {W.}~\bibnamefont {Yi}}, \bibinfo {author} {\bibfnamefont {W.}~\bibnamefont {Huang}}, \bibinfo {author} {\bibfnamefont {C.}~\bibnamefont {Wang}}, \bibinfo {author} {\bibfnamefont {X.}~\bibnamefont {Wu}}, \bibinfo {author} {\bibfnamefont {M.-H.}\ \bibnamefont {Hsieh}}, \bibinfo {author} {\bibfnamefont {T.}~\bibnamefont {Liu}}, \bibinfo {author} {\bibfnamefont {W.}~\bibnamefont {Yang}},\ and\ \bibinfo {author} {\bibfnamefont {D.}~\bibnamefont {Tao}},\ }\href {https://doi.org/10.48550/arXiv.2206.03066} {\bibinfo {title} {Recent {{Advances}} for {{Quantum Neural Networks}} in {{Generative Learning}}}}
  (\bibinfo {year} {2022}),\ \Eprint {https://arxiv.org/abs/2206.03066} {arXiv:2206.03066 [quant-ph]} \BibitemShut {NoStop}%
\bibitem [{\citenamefont {Ngo}\ \emph {et~al.}(2023)\citenamefont {Ngo}, \citenamefont {Nguyen},\ and\ \citenamefont {Thang}}]{ngo2023SurveyRecent}%
  \BibitemOpen
  \bibfield  {author} {\bibinfo {author} {\bibfnamefont {T.~A.}\ \bibnamefont {Ngo}}, \bibinfo {author} {\bibfnamefont {T.}~\bibnamefont {Nguyen}},\ and\ \bibinfo {author} {\bibfnamefont {T.~C.}\ \bibnamefont {Thang}},\ }\bibfield  {title} {\bibinfo {title} {A {{Survey}} of {{Recent Advances}} in {{Quantum Generative Adversarial Networks}}},\ }\href {https://doi.org/10.3390/electronics12040856} {\bibfield  {journal} {\bibinfo  {journal} {Electronics}\ }\textbf {\bibinfo {volume} {12}},\ \bibinfo {pages} {856} (\bibinfo {year} {2023})}\BibitemShut {NoStop}%
\bibitem [{\citenamefont {Zoufal}\ \emph {et~al.}(2019)\citenamefont {Zoufal}, \citenamefont {Lucchi},\ and\ \citenamefont {Woerner}}]{zoufal2019QuantumGenerative}%
  \BibitemOpen
  \bibfield  {author} {\bibinfo {author} {\bibfnamefont {C.}~\bibnamefont {Zoufal}}, \bibinfo {author} {\bibfnamefont {A.}~\bibnamefont {Lucchi}},\ and\ \bibinfo {author} {\bibfnamefont {S.}~\bibnamefont {Woerner}},\ }\bibfield  {title} {\bibinfo {title} {Quantum {{Generative Adversarial Networks}} for learning and loading random distributions},\ }\href {https://doi.org/10.1038/s41534-019-0223-2} {\bibfield  {journal} {\bibinfo  {journal} {npj Quantum Information}\ }\textbf {\bibinfo {volume} {5}},\ \bibinfo {pages} {1} (\bibinfo {year} {2019})}\BibitemShut {NoStop}%
\bibitem [{\citenamefont {Mourya}\ \emph {et~al.}(2025)\citenamefont {Mourya}, \citenamefont {Leipold},\ and\ \citenamefont {Adhikari}}]{mourya2025ContextualQuantum}%
  \BibitemOpen
  \bibfield  {author} {\bibinfo {author} {\bibfnamefont {S.}~\bibnamefont {Mourya}}, \bibinfo {author} {\bibfnamefont {H.}~\bibnamefont {Leipold}},\ and\ \bibinfo {author} {\bibfnamefont {B.}~\bibnamefont {Adhikari}},\ }\href {https://doi.org/10.48550/arXiv.2503.01884} {\bibinfo {title} {Contextual {{Quantum Neural Networks}} for {{Stock Price Prediction}}}} (\bibinfo {year} {2025}),\ \Eprint {https://arxiv.org/abs/2503.01884} {arXiv:2503.01884 [cs]} \BibitemShut {NoStop}%
\bibitem [{\citenamefont {Huang}\ \emph {et~al.}(2021)\citenamefont {Huang}, \citenamefont {Du}, \citenamefont {Gong}, \citenamefont {Zhao}, \citenamefont {Wu}, \citenamefont {Wang}, \citenamefont {Li}, \citenamefont {Liang}, \citenamefont {Lin}, \citenamefont {Xu}, \citenamefont {Yang}, \citenamefont {Liu}, \citenamefont {Hsieh}, \citenamefont {Deng}, \citenamefont {Rong}, \citenamefont {Peng}, \citenamefont {Lu}, \citenamefont {Chen}, \citenamefont {Tao}, \citenamefont {Zhu},\ and\ \citenamefont {Pan}}]{huang2021ExperimentalQuantum}%
  \BibitemOpen
  \bibfield  {author} {\bibinfo {author} {\bibfnamefont {H.-L.}\ \bibnamefont {Huang}}, \bibinfo {author} {\bibfnamefont {Y.}~\bibnamefont {Du}}, \bibinfo {author} {\bibfnamefont {M.}~\bibnamefont {Gong}}, \bibinfo {author} {\bibfnamefont {Y.}~\bibnamefont {Zhao}}, \bibinfo {author} {\bibfnamefont {Y.}~\bibnamefont {Wu}}, \bibinfo {author} {\bibfnamefont {C.}~\bibnamefont {Wang}}, \bibinfo {author} {\bibfnamefont {S.}~\bibnamefont {Li}}, \bibinfo {author} {\bibfnamefont {F.}~\bibnamefont {Liang}}, \bibinfo {author} {\bibfnamefont {J.}~\bibnamefont {Lin}}, \bibinfo {author} {\bibfnamefont {Y.}~\bibnamefont {Xu}}, \bibinfo {author} {\bibfnamefont {R.}~\bibnamefont {Yang}}, \bibinfo {author} {\bibfnamefont {T.}~\bibnamefont {Liu}}, \bibinfo {author} {\bibfnamefont {M.-H.}\ \bibnamefont {Hsieh}}, \bibinfo {author} {\bibfnamefont {H.}~\bibnamefont {Deng}}, \bibinfo {author} {\bibfnamefont {H.}~\bibnamefont {Rong}}, \bibinfo {author} {\bibfnamefont {C.-Z.}\ \bibnamefont {Peng}}, \bibinfo {author} {\bibfnamefont
  {C.-Y.}\ \bibnamefont {Lu}}, \bibinfo {author} {\bibfnamefont {Y.-A.}\ \bibnamefont {Chen}}, \bibinfo {author} {\bibfnamefont {D.}~\bibnamefont {Tao}}, \bibinfo {author} {\bibfnamefont {X.}~\bibnamefont {Zhu}},\ and\ \bibinfo {author} {\bibfnamefont {J.-W.}\ \bibnamefont {Pan}},\ }\bibfield  {title} {\bibinfo {title} {Experimental {{Quantum Generative Adversarial Networks}} for {{Image Generation}}},\ }\href {https://doi.org/10.1103/PhysRevApplied.16.024051} {\bibfield  {journal} {\bibinfo  {journal} {Physical Review Applied}\ }\textbf {\bibinfo {volume} {16}},\ \bibinfo {pages} {024051} (\bibinfo {year} {2021})}\BibitemShut {NoStop}%
\bibitem [{\citenamefont {Zhou}\ \emph {et~al.}(2023)\citenamefont {Zhou}, \citenamefont {Zhang}, \citenamefont {Xie},\ and\ \citenamefont {Wu}}]{zhou2023Hybridquantum}%
  \BibitemOpen
  \bibfield  {author} {\bibinfo {author} {\bibfnamefont {N.-R.}\ \bibnamefont {Zhou}}, \bibinfo {author} {\bibfnamefont {T.-F.}\ \bibnamefont {Zhang}}, \bibinfo {author} {\bibfnamefont {X.-W.}\ \bibnamefont {Xie}},\ and\ \bibinfo {author} {\bibfnamefont {J.-Y.}\ \bibnamefont {Wu}},\ }\bibfield  {title} {\bibinfo {title} {Hybrid quantum--classical generative adversarial networks for image generation via learning discrete distribution},\ }\href {https://doi.org/10.1016/j.image.2022.116891} {\bibfield  {journal} {\bibinfo  {journal} {Signal Processing: Image Communication}\ }\textbf {\bibinfo {volume} {110}},\ \bibinfo {pages} {116891} (\bibinfo {year} {2023})}\BibitemShut {NoStop}%
\bibitem [{\citenamefont {Silver}\ \emph {et~al.}(2023)\citenamefont {Silver}, \citenamefont {Patel}, \citenamefont {Cutler}, \citenamefont {Ranjan}, \citenamefont {Gandhi},\ and\ \citenamefont {Tiwari}}]{silver2023MosaiQQuantum}%
  \BibitemOpen
  \bibfield  {author} {\bibinfo {author} {\bibfnamefont {D.}~\bibnamefont {Silver}}, \bibinfo {author} {\bibfnamefont {T.}~\bibnamefont {Patel}}, \bibinfo {author} {\bibfnamefont {W.}~\bibnamefont {Cutler}}, \bibinfo {author} {\bibfnamefont {A.}~\bibnamefont {Ranjan}}, \bibinfo {author} {\bibfnamefont {H.}~\bibnamefont {Gandhi}},\ and\ \bibinfo {author} {\bibfnamefont {D.}~\bibnamefont {Tiwari}},\ }\bibfield  {title} {\bibinfo {title} {{{MosaiQ}}: {{Quantum Generative Adversarial Networks}} for {{Image Generation}} on {{NISQ Computers}}},\ }in\ \href@noop {} {\emph {\bibinfo {booktitle} {Proceedings of the {{IEEE}}/{{CVF International Conference}} on {{Computer Vision}}}}}\ (\bibinfo {year} {2023})\ pp.\ \bibinfo {pages} {7030--7039}\BibitemShut {NoStop}%
\bibitem [{\citenamefont {Tsang}\ \emph {et~al.}(2023)\citenamefont {Tsang}, \citenamefont {West}, \citenamefont {Erfani},\ and\ \citenamefont {Usman}}]{tsang2023HybridQuantum}%
  \BibitemOpen
  \bibfield  {author} {\bibinfo {author} {\bibfnamefont {S.~L.}\ \bibnamefont {Tsang}}, \bibinfo {author} {\bibfnamefont {M.~T.}\ \bibnamefont {West}}, \bibinfo {author} {\bibfnamefont {S.~M.}\ \bibnamefont {Erfani}},\ and\ \bibinfo {author} {\bibfnamefont {M.}~\bibnamefont {Usman}},\ }\bibfield  {title} {\bibinfo {title} {Hybrid {{Quantum}}--{{Classical Generative Adversarial Network}} for {{High-Resolution Image Generation}}},\ }\href {https://doi.org/10.1109/TQE.2023.3319319} {\bibfield  {journal} {\bibinfo  {journal} {IEEE Transactions on Quantum Engineering}\ }\textbf {\bibinfo {volume} {4}},\ \bibinfo {pages} {1} (\bibinfo {year} {2023})}\BibitemShut {NoStop}%
\bibitem [{\citenamefont {Situ}\ \emph {et~al.}(2020)\citenamefont {Situ}, \citenamefont {He}, \citenamefont {Wang}, \citenamefont {Li},\ and\ \citenamefont {Zheng}}]{situ2020Quantumgenerative}%
  \BibitemOpen
  \bibfield  {author} {\bibinfo {author} {\bibfnamefont {H.}~\bibnamefont {Situ}}, \bibinfo {author} {\bibfnamefont {Z.}~\bibnamefont {He}}, \bibinfo {author} {\bibfnamefont {Y.}~\bibnamefont {Wang}}, \bibinfo {author} {\bibfnamefont {L.}~\bibnamefont {Li}},\ and\ \bibinfo {author} {\bibfnamefont {S.}~\bibnamefont {Zheng}},\ }\bibfield  {title} {\bibinfo {title} {Quantum generative adversarial network for generating discrete distribution},\ }\href {https://doi.org/10.1016/j.ins.2020.05.127} {\bibfield  {journal} {\bibinfo  {journal} {Information Sciences}\ }\textbf {\bibinfo {volume} {538}},\ \bibinfo {pages} {193} (\bibinfo {year} {2020})}\BibitemShut {NoStop}%
\bibitem [{\citenamefont {Kao}\ \emph {et~al.}(2023)\citenamefont {Kao}, \citenamefont {Yang}, \citenamefont {Chiang}, \citenamefont {Hsiao}, \citenamefont {Cao}, \citenamefont {Aliper}, \citenamefont {Ren}, \citenamefont {{Aspuru-Guzik}}, \citenamefont {Zhavoronkov}, \citenamefont {Hsieh},\ and\ \citenamefont {Lin}}]{kao2023ExploringAdvantages}%
  \BibitemOpen
  \bibfield  {author} {\bibinfo {author} {\bibfnamefont {P.-Y.}\ \bibnamefont {Kao}}, \bibinfo {author} {\bibfnamefont {Y.-C.}\ \bibnamefont {Yang}}, \bibinfo {author} {\bibfnamefont {W.-Y.}\ \bibnamefont {Chiang}}, \bibinfo {author} {\bibfnamefont {J.-Y.}\ \bibnamefont {Hsiao}}, \bibinfo {author} {\bibfnamefont {Y.}~\bibnamefont {Cao}}, \bibinfo {author} {\bibfnamefont {A.}~\bibnamefont {Aliper}}, \bibinfo {author} {\bibfnamefont {F.}~\bibnamefont {Ren}}, \bibinfo {author} {\bibfnamefont {A.}~\bibnamefont {{Aspuru-Guzik}}}, \bibinfo {author} {\bibfnamefont {A.}~\bibnamefont {Zhavoronkov}}, \bibinfo {author} {\bibfnamefont {M.-H.}\ \bibnamefont {Hsieh}},\ and\ \bibinfo {author} {\bibfnamefont {Y.-C.}\ \bibnamefont {Lin}},\ }\bibfield  {title} {\bibinfo {title} {Exploring the {{Advantages}} of {{Quantum Generative Adversarial Networks}} in {{Generative Chemistry}}},\ }\href {https://doi.org/10.1021/acs.jcim.3c00562} {\bibfield  {journal} {\bibinfo  {journal} {Journal of Chemical Information and Modeling}\
  }\textbf {\bibinfo {volume} {63}},\ \bibinfo {pages} {3307} (\bibinfo {year} {2023})}\BibitemShut {NoStop}%
\bibitem [{\citenamefont {Herr}\ \emph {et~al.}(2021)\citenamefont {Herr}, \citenamefont {Obert},\ and\ \citenamefont {Rosenkranz}}]{herr2021Anomalydetection}%
  \BibitemOpen
  \bibfield  {author} {\bibinfo {author} {\bibfnamefont {D.}~\bibnamefont {Herr}}, \bibinfo {author} {\bibfnamefont {B.}~\bibnamefont {Obert}},\ and\ \bibinfo {author} {\bibfnamefont {M.}~\bibnamefont {Rosenkranz}},\ }\bibfield  {title} {\bibinfo {title} {Anomaly detection with variational quantum generative adversarial networks},\ }\href {https://doi.org/10.1088/2058-9565/ac0d4d} {\bibfield  {journal} {\bibinfo  {journal} {Quantum Science and Technology}\ }\textbf {\bibinfo {volume} {6}},\ \bibinfo {pages} {045004} (\bibinfo {year} {2021})}\BibitemShut {NoStop}%
\bibitem [{\citenamefont {Fuchs}\ and\ \citenamefont {Horvath}(2023)}]{fuchs2023HybridQuantum}%
  \BibitemOpen
  \bibfield  {author} {\bibinfo {author} {\bibfnamefont {F.}~\bibnamefont {Fuchs}}\ and\ \bibinfo {author} {\bibfnamefont {B.}~\bibnamefont {Horvath}},\ }\href {https://doi.org/10.2139/ssrn.4514510} {\bibinfo {title} {A {{Hybrid Quantum Wasserstein GAN}} with {{Applications}} to {{Option Pricing}}}} (\bibinfo {year} {2023}),\ \Eprint {https://arxiv.org/abs/4514510} {Social Science Research Network:4514510} \BibitemShut {NoStop}%
\bibitem [{\citenamefont {Baglio}(2024)}]{baglio2024Dataaugmentation}%
  \BibitemOpen
  \bibfield  {author} {\bibinfo {author} {\bibfnamefont {J.}~\bibnamefont {Baglio}},\ }\href {https://doi.org/10.48550/arXiv.2405.04401} {\bibinfo {title} {Data augmentation experiments with style-based quantum generative adversarial networks on trapped-ion and superconducting-qubit technologies}} (\bibinfo {year} {2024}),\ \Eprint {https://arxiv.org/abs/2405.04401} {arXiv:2405.04401 [quant-ph]} \BibitemShut {NoStop}%
\bibitem [{\citenamefont {Di~Meglio}\ \emph {et~al.}(2024)\citenamefont {Di~Meglio}, \citenamefont {Jansen}, \citenamefont {Tavernelli}, \citenamefont {Alexandrou}, \citenamefont {Arunachalam}, \citenamefont {Bauer}, \citenamefont {Borras}, \citenamefont {Carrazza}, \citenamefont {Crippa}, \citenamefont {Croft}, \citenamefont {{de Putter}}, \citenamefont {Delgado}, \citenamefont {Dunjko}, \citenamefont {Egger}, \citenamefont {{Fern{\'a}ndez-Combarro}}, \citenamefont {Fuchs}, \citenamefont {Funcke}, \citenamefont {{Gonz{\'a}lez-Cuadra}}, \citenamefont {Grossi}, \citenamefont {Halimeh}, \citenamefont {Holmes}, \citenamefont {K{\"u}hn}, \citenamefont {Lacroix}, \citenamefont {Lewis}, \citenamefont {Lucchesi}, \citenamefont {Martinez}, \citenamefont {Meloni}, \citenamefont {Mezzacapo}, \citenamefont {Montangero}, \citenamefont {Nagano}, \citenamefont {Pascuzzi}, \citenamefont {Radescu}, \citenamefont {Ortega}, \citenamefont {Roggero}, \citenamefont {Schuhmacher}, \citenamefont {Seixas}, \citenamefont {Silvi},
  \citenamefont {Spentzouris}, \citenamefont {Tacchino}, \citenamefont {Temme}, \citenamefont {Terashi}, \citenamefont {Tura}, \citenamefont {T{\"u}ys{\"u}z}, \citenamefont {Vallecorsa}, \citenamefont {Wiese}, \citenamefont {Yoo},\ and\ \citenamefont {Zhang}}]{dimeglio2024QuantumComputing}%
  \BibitemOpen
  \bibfield  {author} {\bibinfo {author} {\bibfnamefont {A.}~\bibnamefont {Di~Meglio}}, \bibinfo {author} {\bibfnamefont {K.}~\bibnamefont {Jansen}}, \bibinfo {author} {\bibfnamefont {I.}~\bibnamefont {Tavernelli}}, \bibinfo {author} {\bibfnamefont {C.}~\bibnamefont {Alexandrou}}, \bibinfo {author} {\bibfnamefont {S.}~\bibnamefont {Arunachalam}}, \bibinfo {author} {\bibfnamefont {C.~W.}\ \bibnamefont {Bauer}}, \bibinfo {author} {\bibfnamefont {K.}~\bibnamefont {Borras}}, \bibinfo {author} {\bibfnamefont {S.}~\bibnamefont {Carrazza}}, \bibinfo {author} {\bibfnamefont {A.}~\bibnamefont {Crippa}}, \bibinfo {author} {\bibfnamefont {V.}~\bibnamefont {Croft}}, \bibinfo {author} {\bibfnamefont {R.}~\bibnamefont {{de Putter}}}, \bibinfo {author} {\bibfnamefont {A.}~\bibnamefont {Delgado}}, \bibinfo {author} {\bibfnamefont {V.}~\bibnamefont {Dunjko}}, \bibinfo {author} {\bibfnamefont {D.~J.}\ \bibnamefont {Egger}}, \bibinfo {author} {\bibfnamefont {E.}~\bibnamefont {{Fern{\'a}ndez-Combarro}}}, \bibinfo {author}
  {\bibfnamefont {E.}~\bibnamefont {Fuchs}}, \bibinfo {author} {\bibfnamefont {L.}~\bibnamefont {Funcke}}, \bibinfo {author} {\bibfnamefont {D.}~\bibnamefont {{Gonz{\'a}lez-Cuadra}}}, \bibinfo {author} {\bibfnamefont {M.}~\bibnamefont {Grossi}}, \bibinfo {author} {\bibfnamefont {J.~C.}\ \bibnamefont {Halimeh}}, \bibinfo {author} {\bibfnamefont {Z.}~\bibnamefont {Holmes}}, \bibinfo {author} {\bibfnamefont {S.}~\bibnamefont {K{\"u}hn}}, \bibinfo {author} {\bibfnamefont {D.}~\bibnamefont {Lacroix}}, \bibinfo {author} {\bibfnamefont {R.}~\bibnamefont {Lewis}}, \bibinfo {author} {\bibfnamefont {D.}~\bibnamefont {Lucchesi}}, \bibinfo {author} {\bibfnamefont {M.~L.}\ \bibnamefont {Martinez}}, \bibinfo {author} {\bibfnamefont {F.}~\bibnamefont {Meloni}}, \bibinfo {author} {\bibfnamefont {A.}~\bibnamefont {Mezzacapo}}, \bibinfo {author} {\bibfnamefont {S.}~\bibnamefont {Montangero}}, \bibinfo {author} {\bibfnamefont {L.}~\bibnamefont {Nagano}}, \bibinfo {author} {\bibfnamefont {V.~R.}\ \bibnamefont {Pascuzzi}},
  \bibinfo {author} {\bibfnamefont {V.}~\bibnamefont {Radescu}}, \bibinfo {author} {\bibfnamefont {E.~R.}\ \bibnamefont {Ortega}}, \bibinfo {author} {\bibfnamefont {A.}~\bibnamefont {Roggero}}, \bibinfo {author} {\bibfnamefont {J.}~\bibnamefont {Schuhmacher}}, \bibinfo {author} {\bibfnamefont {J.}~\bibnamefont {Seixas}}, \bibinfo {author} {\bibfnamefont {P.}~\bibnamefont {Silvi}}, \bibinfo {author} {\bibfnamefont {P.}~\bibnamefont {Spentzouris}}, \bibinfo {author} {\bibfnamefont {F.}~\bibnamefont {Tacchino}}, \bibinfo {author} {\bibfnamefont {K.}~\bibnamefont {Temme}}, \bibinfo {author} {\bibfnamefont {K.}~\bibnamefont {Terashi}}, \bibinfo {author} {\bibfnamefont {J.}~\bibnamefont {Tura}}, \bibinfo {author} {\bibfnamefont {C.}~\bibnamefont {T{\"u}ys{\"u}z}}, \bibinfo {author} {\bibfnamefont {S.}~\bibnamefont {Vallecorsa}}, \bibinfo {author} {\bibfnamefont {U.-J.}\ \bibnamefont {Wiese}}, \bibinfo {author} {\bibfnamefont {S.}~\bibnamefont {Yoo}},\ and\ \bibinfo {author} {\bibfnamefont {J.}~\bibnamefont
  {Zhang}},\ }\bibfield  {title} {\bibinfo {title} {Quantum {{Computing}} for {{High-Energy Physics}}: {{State}} of the {{Art}} and {{Challenges}}},\ }\href {https://doi.org/10.1103/PRXQuantum.5.037001} {\bibfield  {journal} {\bibinfo  {journal} {PRX Quantum}\ }\textbf {\bibinfo {volume} {5}},\ \bibinfo {pages} {037001} (\bibinfo {year} {2024})}\BibitemShut {NoStop}%
\bibitem [{\citenamefont {Paquet}\ and\ \citenamefont {Soleymani}(2022)}]{paquet2022QuantumLeapHybrid}%
  \BibitemOpen
  \bibfield  {author} {\bibinfo {author} {\bibfnamefont {E.}~\bibnamefont {Paquet}}\ and\ \bibinfo {author} {\bibfnamefont {F.}~\bibnamefont {Soleymani}},\ }\bibfield  {title} {\bibinfo {title} {{{QuantumLeap}}: {{Hybrid}} quantum neural network for financial predictions},\ }\href {https://doi.org/10.1016/j.eswa.2022.116583} {\bibfield  {journal} {\bibinfo  {journal} {Expert Systems with Applications}\ }\textbf {\bibinfo {volume} {195}},\ \bibinfo {pages} {116583} (\bibinfo {year} {2022})}\BibitemShut {NoStop}%
\bibitem [{\citenamefont {Kiani}\ \emph {et~al.}(2022)\citenamefont {Kiani}, \citenamefont {De~Palma}, \citenamefont {Marvian}, \citenamefont {Liu},\ and\ \citenamefont {Lloyd}}]{kiani2022Learningquantum}%
  \BibitemOpen
  \bibfield  {author} {\bibinfo {author} {\bibfnamefont {B.~T.}\ \bibnamefont {Kiani}}, \bibinfo {author} {\bibfnamefont {G.}~\bibnamefont {De~Palma}}, \bibinfo {author} {\bibfnamefont {M.}~\bibnamefont {Marvian}}, \bibinfo {author} {\bibfnamefont {Z.-W.}\ \bibnamefont {Liu}},\ and\ \bibinfo {author} {\bibfnamefont {S.}~\bibnamefont {Lloyd}},\ }\bibfield  {title} {\bibinfo {title} {Learning quantum data with the quantum earth mover's distance},\ }\href {https://doi.org/10.1088/2058-9565/ac79c9} {\bibfield  {journal} {\bibinfo  {journal} {Quantum Science and Technology}\ }\textbf {\bibinfo {volume} {7}},\ \bibinfo {pages} {045002} (\bibinfo {year} {2022})}\BibitemShut {NoStop}%
\bibitem [{\citenamefont {Chakrabarti}\ \emph {et~al.}(2019)\citenamefont {Chakrabarti}, \citenamefont {Yiming}, \citenamefont {Li}, \citenamefont {Feizi},\ and\ \citenamefont {Wu}}]{chakrabarti2019QuantumWasserstein}%
  \BibitemOpen
  \bibfield  {author} {\bibinfo {author} {\bibfnamefont {S.}~\bibnamefont {Chakrabarti}}, \bibinfo {author} {\bibfnamefont {H.}~\bibnamefont {Yiming}}, \bibinfo {author} {\bibfnamefont {T.}~\bibnamefont {Li}}, \bibinfo {author} {\bibfnamefont {S.}~\bibnamefont {Feizi}},\ and\ \bibinfo {author} {\bibfnamefont {X.}~\bibnamefont {Wu}},\ }\bibfield  {title} {\bibinfo {title} {Quantum {{Wasserstein Generative Adversarial Networks}}},\ }in\ \href@noop {} {\emph {\bibinfo {booktitle} {Advances in {{Neural Information Processing Systems}}}}},\ Vol.~\bibinfo {volume} {32}\ (\bibinfo  {publisher} {Curran Associates, Inc.},\ \bibinfo {year} {2019})\BibitemShut {NoStop}%
\bibitem [{\citenamefont {Schwander}(2022)}]{schwander2022Quantumgenerative}%
  \BibitemOpen
  \bibfield  {author} {\bibinfo {author} {\bibfnamefont {E.}~\bibnamefont {Schwander}},\ }\emph {\bibinfo {title} {Quantum Generative Modelling for Financial Time Series}},\ \href {https://studenttheses.universiteitleiden.nl/handle/1887/3278321} {Master's thesis},\ \bibinfo  {school} {Leiden University} (\bibinfo {year} {2022})\BibitemShut {NoStop}%
\bibitem [{\citenamefont {Abadi}\ \emph {et~al.}(2016)\citenamefont {Abadi}, \citenamefont {Barham}, \citenamefont {Chen}, \citenamefont {Chen}, \citenamefont {Davis}, \citenamefont {Dean}, \citenamefont {Devin}, \citenamefont {Ghemawat}, \citenamefont {Irving}, \citenamefont {Isard}, \citenamefont {Kudlur}, \citenamefont {Levenberg}, \citenamefont {Monga}, \citenamefont {Moore}, \citenamefont {Murray}, \citenamefont {Steiner}, \citenamefont {Tucker}, \citenamefont {Vasudevan}, \citenamefont {Warden}, \citenamefont {Wicke}, \citenamefont {Yu},\ and\ \citenamefont {Zheng}}]{abadi2016TensorFlowsystem}%
  \BibitemOpen
  \bibfield  {author} {\bibinfo {author} {\bibfnamefont {M.}~\bibnamefont {Abadi}}, \bibinfo {author} {\bibfnamefont {P.}~\bibnamefont {Barham}}, \bibinfo {author} {\bibfnamefont {J.}~\bibnamefont {Chen}}, \bibinfo {author} {\bibfnamefont {Z.}~\bibnamefont {Chen}}, \bibinfo {author} {\bibfnamefont {A.}~\bibnamefont {Davis}}, \bibinfo {author} {\bibfnamefont {J.}~\bibnamefont {Dean}}, \bibinfo {author} {\bibfnamefont {M.}~\bibnamefont {Devin}}, \bibinfo {author} {\bibfnamefont {S.}~\bibnamefont {Ghemawat}}, \bibinfo {author} {\bibfnamefont {G.}~\bibnamefont {Irving}}, \bibinfo {author} {\bibfnamefont {M.}~\bibnamefont {Isard}}, \bibinfo {author} {\bibfnamefont {M.}~\bibnamefont {Kudlur}}, \bibinfo {author} {\bibfnamefont {J.}~\bibnamefont {Levenberg}}, \bibinfo {author} {\bibfnamefont {R.}~\bibnamefont {Monga}}, \bibinfo {author} {\bibfnamefont {S.}~\bibnamefont {Moore}}, \bibinfo {author} {\bibfnamefont {D.~G.}\ \bibnamefont {Murray}}, \bibinfo {author} {\bibfnamefont {B.}~\bibnamefont {Steiner}}, \bibinfo
  {author} {\bibfnamefont {P.}~\bibnamefont {Tucker}}, \bibinfo {author} {\bibfnamefont {V.}~\bibnamefont {Vasudevan}}, \bibinfo {author} {\bibfnamefont {P.}~\bibnamefont {Warden}}, \bibinfo {author} {\bibfnamefont {M.}~\bibnamefont {Wicke}}, \bibinfo {author} {\bibfnamefont {Y.}~\bibnamefont {Yu}},\ and\ \bibinfo {author} {\bibfnamefont {X.}~\bibnamefont {Zheng}},\ }\bibfield  {title} {\bibinfo {title} {{{TensorFlow}}: A system for large-scale machine learning},\ }in\ \href@noop {} {\emph {\bibinfo {booktitle} {Proceedings of the 12th {{USENIX}} Conference on {{Operating Systems Design}} and {{Implementation}}}}},\ \bibinfo {series and number} {{{OSDI}}'16}\ (\bibinfo  {publisher} {USENIX Association},\ \bibinfo {address} {USA},\ \bibinfo {year} {2016})\ pp.\ \bibinfo {pages} {265--283}\BibitemShut {NoStop}%
\bibitem [{\citenamefont {Bradbury}\ \emph {et~al.}(2018)\citenamefont {Bradbury}, \citenamefont {Frostig}, \citenamefont {Hawkins}, \citenamefont {Johnson}, \citenamefont {Leary}, \citenamefont {Maclaurin}, \citenamefont {Necula}, \citenamefont {Paszke}, \citenamefont {Vander{P}las}, \citenamefont {Wanderman-{M}ilne},\ and\ \citenamefont {Zhang}}]{jax2018github}%
  \BibitemOpen
  \bibfield  {author} {\bibinfo {author} {\bibfnamefont {J.}~\bibnamefont {Bradbury}}, \bibinfo {author} {\bibfnamefont {R.}~\bibnamefont {Frostig}}, \bibinfo {author} {\bibfnamefont {P.}~\bibnamefont {Hawkins}}, \bibinfo {author} {\bibfnamefont {M.~J.}\ \bibnamefont {Johnson}}, \bibinfo {author} {\bibfnamefont {C.}~\bibnamefont {Leary}}, \bibinfo {author} {\bibfnamefont {D.}~\bibnamefont {Maclaurin}}, \bibinfo {author} {\bibfnamefont {G.}~\bibnamefont {Necula}}, \bibinfo {author} {\bibfnamefont {A.}~\bibnamefont {Paszke}}, \bibinfo {author} {\bibfnamefont {J.}~\bibnamefont {Vander{P}las}}, \bibinfo {author} {\bibfnamefont {S.}~\bibnamefont {Wanderman-{M}ilne}},\ and\ \bibinfo {author} {\bibfnamefont {Q.}~\bibnamefont {Zhang}},\ }\href {http://github.com/jax-ml/jax} {\bibinfo {title} {{JAX}: composable transformations of {P}ython+{N}um{P}y programs}} (\bibinfo {year} {2018})\BibitemShut {NoStop}%
\bibitem [{\citenamefont {Gray}(2018)}]{gray2018quimbpython}%
  \BibitemOpen
  \bibfield  {author} {\bibinfo {author} {\bibfnamefont {J.}~\bibnamefont {Gray}},\ }\bibfield  {title} {\bibinfo {title} {Quimb: {{A}} python package for quantum information and many-body calculations},\ }\href {https://doi.org/10.21105/joss.00819} {\bibfield  {journal} {\bibinfo  {journal} {Journal of Open Source Software}\ }\textbf {\bibinfo {volume} {3}},\ \bibinfo {pages} {819} (\bibinfo {year} {2018})}\BibitemShut {NoStop}%
\bibitem [{\citenamefont {Hogenboom}(2025)}]{hogenboom2025CasperHogenboomWGAN_financial_timeseries}%
  \BibitemOpen
  \bibfield  {author} {\bibinfo {author} {\bibfnamefont {C.}~\bibnamefont {Hogenboom}},\ }\emph {\bibinfo {title} {WGAN Financial Time-Series}},\ \href {https://github.com/CasperHogenboom/WGAN_financial_time-series} {Master's thesis},\ \bibinfo  {school} {University Maastricht} (\bibinfo {year} {2025})\BibitemShut {NoStop}%
\bibitem [{\citenamefont {Goerg}(2015)}]{goerg2015LambertWay}%
  \BibitemOpen
  \bibfield  {author} {\bibinfo {author} {\bibfnamefont {G.~M.}\ \bibnamefont {Goerg}},\ }\bibfield  {title} {\bibinfo {title} {The {{Lambert Way}} to {{Gaussianize Heavy-Tailed Data}} with the {{Inverse}} of {{Tukey}}'s h {{Transformation}} as a {{Special Case}}},\ }\href {https://doi.org/10.1155/2015/909231} {\bibfield  {journal} {\bibinfo  {journal} {The Scientific World Journal}\ }\textbf {\bibinfo {volume} {2015}},\ \bibinfo {pages} {909231} (\bibinfo {year} {2015})}\BibitemShut {NoStop}%
\bibitem [{\citenamefont {Larocca}\ \emph {et~al.}(2025)\citenamefont {Larocca}, \citenamefont {Thanasilp}, \citenamefont {Wang}, \citenamefont {Sharma}, \citenamefont {Biamonte}, \citenamefont {Coles}, \citenamefont {Cincio}, \citenamefont {McClean}, \citenamefont {Holmes},\ and\ \citenamefont {Cerezo}}]{larocca2025Barrenplateaus}%
  \BibitemOpen
  \bibfield  {author} {\bibinfo {author} {\bibfnamefont {M.}~\bibnamefont {Larocca}}, \bibinfo {author} {\bibfnamefont {S.}~\bibnamefont {Thanasilp}}, \bibinfo {author} {\bibfnamefont {S.}~\bibnamefont {Wang}}, \bibinfo {author} {\bibfnamefont {K.}~\bibnamefont {Sharma}}, \bibinfo {author} {\bibfnamefont {J.}~\bibnamefont {Biamonte}}, \bibinfo {author} {\bibfnamefont {P.~J.}\ \bibnamefont {Coles}}, \bibinfo {author} {\bibfnamefont {L.}~\bibnamefont {Cincio}}, \bibinfo {author} {\bibfnamefont {J.~R.}\ \bibnamefont {McClean}}, \bibinfo {author} {\bibfnamefont {Z.}~\bibnamefont {Holmes}},\ and\ \bibinfo {author} {\bibfnamefont {M.}~\bibnamefont {Cerezo}},\ }\bibfield  {title} {\bibinfo {title} {Barren plateaus in variational quantum computing},\ }\href@noop {} {\bibfield  {journal} {\bibinfo  {journal} {Nature Reviews Physics}\ ,\ \bibinfo {pages} {1}} (\bibinfo {year} {2025})}\BibitemShut {NoStop}%
\bibitem [{\citenamefont {Shi}(2024)}]{shi2024EffectsObservable}%
  \BibitemOpen
  \bibfield  {author} {\bibinfo {author} {\bibfnamefont {C.~C.}\ \bibnamefont {Shi}},\ }\emph {\bibinfo {title} {Effects of {{Observable Choices}} in the {{Performance}} of {{Variational Quantum Generative Models}} - {{CONFIDENTIAL}}}},\ \href@noop {} {Ph.D. thesis},\ \bibinfo  {school} {LIACS, Leiden University} (\bibinfo {year} {2024})\BibitemShut {NoStop}%
\bibitem [{\citenamefont {Cirac}\ \emph {et~al.}(2021)\citenamefont {Cirac}, \citenamefont {{P{\'e}rez-Garc{\'i}a}}, \citenamefont {Schuch},\ and\ \citenamefont {Verstraete}}]{cirac2021Matrixproduct}%
  \BibitemOpen
  \bibfield  {author} {\bibinfo {author} {\bibfnamefont {J.~I.}\ \bibnamefont {Cirac}}, \bibinfo {author} {\bibfnamefont {D.}~\bibnamefont {{P{\'e}rez-Garc{\'i}a}}}, \bibinfo {author} {\bibfnamefont {N.}~\bibnamefont {Schuch}},\ and\ \bibinfo {author} {\bibfnamefont {F.}~\bibnamefont {Verstraete}},\ }\bibfield  {title} {\bibinfo {title} {Matrix product states and projected entangled pair states: {{Concepts}}, symmetries, theorems},\ }\href {https://doi.org/10.1103/RevModPhys.93.045003} {\bibfield  {journal} {\bibinfo  {journal} {Reviews of Modern Physics}\ }\textbf {\bibinfo {volume} {93}},\ \bibinfo {pages} {045003} (\bibinfo {year} {2021})}\BibitemShut {NoStop}%
\bibitem [{\citenamefont {Oseledets}(2011)}]{oseledets2011TensorTrainDecomposition}%
  \BibitemOpen
  \bibfield  {author} {\bibinfo {author} {\bibfnamefont {I.~V.}\ \bibnamefont {Oseledets}},\ }\bibfield  {title} {\bibinfo {title} {Tensor-{{Train Decomposition}}},\ }\href {https://doi.org/10.1137/090752286} {\bibfield  {journal} {\bibinfo  {journal} {SIAM Journal on Scientific Computing}\ }\textbf {\bibinfo {volume} {33}},\ \bibinfo {pages} {2295} (\bibinfo {year} {2011})}\BibitemShut {NoStop}%
\bibitem [{\citenamefont {Berezutskii}\ \emph {et~al.}(2025)\citenamefont {Berezutskii}, \citenamefont {Liu}, \citenamefont {Acharya}, \citenamefont {Ellerbrock}, \citenamefont {Gray}, \citenamefont {Haghshenas}, \citenamefont {He}, \citenamefont {Khan}, \citenamefont {Kuzmin}, \citenamefont {Lyakh}, \citenamefont {Lykov}, \citenamefont {Mandr{\`a}}, \citenamefont {Mansell}, \citenamefont {Melnikov}, \citenamefont {Melnikov}, \citenamefont {Mironov}, \citenamefont {Morozov}, \citenamefont {Neukart}, \citenamefont {Nocera}, \citenamefont {Perlin}, \citenamefont {Perelshtein}, \citenamefont {Steinberg}, \citenamefont {Shaydulin}, \citenamefont {Villalonga}, \citenamefont {Pflitsch}, \citenamefont {Pistoia}, \citenamefont {Vinokur},\ and\ \citenamefont {Alexeev}}]{berezutskii2025Tensornetworks}%
  \BibitemOpen
  \bibfield  {author} {\bibinfo {author} {\bibfnamefont {A.}~\bibnamefont {Berezutskii}}, \bibinfo {author} {\bibfnamefont {M.}~\bibnamefont {Liu}}, \bibinfo {author} {\bibfnamefont {A.}~\bibnamefont {Acharya}}, \bibinfo {author} {\bibfnamefont {R.}~\bibnamefont {Ellerbrock}}, \bibinfo {author} {\bibfnamefont {J.}~\bibnamefont {Gray}}, \bibinfo {author} {\bibfnamefont {R.}~\bibnamefont {Haghshenas}}, \bibinfo {author} {\bibfnamefont {Z.}~\bibnamefont {He}}, \bibinfo {author} {\bibfnamefont {A.}~\bibnamefont {Khan}}, \bibinfo {author} {\bibfnamefont {V.}~\bibnamefont {Kuzmin}}, \bibinfo {author} {\bibfnamefont {D.}~\bibnamefont {Lyakh}}, \bibinfo {author} {\bibfnamefont {D.}~\bibnamefont {Lykov}}, \bibinfo {author} {\bibfnamefont {S.}~\bibnamefont {Mandr{\`a}}}, \bibinfo {author} {\bibfnamefont {C.}~\bibnamefont {Mansell}}, \bibinfo {author} {\bibfnamefont {A.}~\bibnamefont {Melnikov}}, \bibinfo {author} {\bibfnamefont {A.}~\bibnamefont {Melnikov}}, \bibinfo {author} {\bibfnamefont {V.}~\bibnamefont
  {Mironov}}, \bibinfo {author} {\bibfnamefont {D.}~\bibnamefont {Morozov}}, \bibinfo {author} {\bibfnamefont {F.}~\bibnamefont {Neukart}}, \bibinfo {author} {\bibfnamefont {A.}~\bibnamefont {Nocera}}, \bibinfo {author} {\bibfnamefont {M.~A.}\ \bibnamefont {Perlin}}, \bibinfo {author} {\bibfnamefont {M.}~\bibnamefont {Perelshtein}}, \bibinfo {author} {\bibfnamefont {M.}~\bibnamefont {Steinberg}}, \bibinfo {author} {\bibfnamefont {R.}~\bibnamefont {Shaydulin}}, \bibinfo {author} {\bibfnamefont {B.}~\bibnamefont {Villalonga}}, \bibinfo {author} {\bibfnamefont {M.}~\bibnamefont {Pflitsch}}, \bibinfo {author} {\bibfnamefont {M.}~\bibnamefont {Pistoia}}, \bibinfo {author} {\bibfnamefont {V.}~\bibnamefont {Vinokur}},\ and\ \bibinfo {author} {\bibfnamefont {Y.}~\bibnamefont {Alexeev}},\ }\href {https://doi.org/10.48550/arXiv.2503.08626} {\bibinfo {title} {Tensor networks for quantum computing}} (\bibinfo {year} {2025}),\ \Eprint {https://arxiv.org/abs/2503.08626} {arXiv:2503.08626 [quant-ph]} \BibitemShut {NoStop}%
\bibitem [{\citenamefont {Verstraete}\ \emph {et~al.}(2004)\citenamefont {Verstraete}, \citenamefont {Porras},\ and\ \citenamefont {Cirac}}]{verstraete2004DensityMatrix}%
  \BibitemOpen
  \bibfield  {author} {\bibinfo {author} {\bibfnamefont {F.}~\bibnamefont {Verstraete}}, \bibinfo {author} {\bibfnamefont {D.}~\bibnamefont {Porras}},\ and\ \bibinfo {author} {\bibfnamefont {J.~I.}\ \bibnamefont {Cirac}},\ }\bibfield  {title} {\bibinfo {title} {Density {{Matrix Renormalization Group}} and {{Periodic Boundary Conditions}}: {{A Quantum Information Perspective}}},\ }\href {https://doi.org/10.1103/PhysRevLett.93.227205} {\bibfield  {journal} {\bibinfo  {journal} {Physical Review Letters}\ }\textbf {\bibinfo {volume} {93}},\ \bibinfo {pages} {227205} (\bibinfo {year} {2004})}\BibitemShut {NoStop}%
\bibitem [{\citenamefont {Chatfield}(1975)}]{chatfield1975time}%
  \BibitemOpen
  \bibfield  {author} {\bibinfo {author} {\bibfnamefont {C.}~\bibnamefont {Chatfield}},\ }\href@noop {} {\bibinfo {title} {Time series}} (\bibinfo {year} {1975})\BibitemShut {NoStop}%
\bibitem [{\citenamefont {Orlandi}\ \emph {et~al.}(2024)\citenamefont {Orlandi}, \citenamefont {Barbierato},\ and\ \citenamefont {Gatti}}]{orlandi2024EnhancingFinancial}%
  \BibitemOpen
  \bibfield  {author} {\bibinfo {author} {\bibfnamefont {F.}~\bibnamefont {Orlandi}}, \bibinfo {author} {\bibfnamefont {E.}~\bibnamefont {Barbierato}},\ and\ \bibinfo {author} {\bibfnamefont {A.}~\bibnamefont {Gatti}},\ }\bibfield  {title} {\bibinfo {title} {Enhancing {{Financial Time Series Prediction}} with {{Quantum-Enhanced Synthetic Data Generation}}: {{A Case Study}} on the {{S}}\&{{P}} 500 {{Using}} a {{Quantum Wasserstein Generative Adversarial Network Approach}} with a {{Gradient Penalty}}},\ }\href {https://doi.org/10.3390/electronics13112158} {\bibfield  {journal} {\bibinfo  {journal} {Electronics}\ }\textbf {\bibinfo {volume} {13}},\ \bibinfo {pages} {2158} (\bibinfo {year} {2024})}\BibitemShut {NoStop}%
\bibitem [{\citenamefont {Komninos}(2023)}]{komninos2023Quantumcomputing}%
  \BibitemOpen
  \bibfield  {author} {\bibinfo {author} {\bibfnamefont {D.}~\bibnamefont {Komninos}},\ }\emph {\bibinfo {title} {Quantum Computing for Generative Modeling and Applications}},\ \href {https://doi.org/10.26233/heallink.tuc.98643} {Master's thesis},\ \bibinfo  {school} {Technical University of Crete} (\bibinfo {year} {2023})\BibitemShut {NoStop}%
\bibitem [{\citenamefont {Rebentrost}\ \emph {et~al.}(2018)\citenamefont {Rebentrost}, \citenamefont {Gupt},\ and\ \citenamefont {Bromley}}]{rebentrost2018Quantumcomputational}%
  \BibitemOpen
  \bibfield  {author} {\bibinfo {author} {\bibfnamefont {P.}~\bibnamefont {Rebentrost}}, \bibinfo {author} {\bibfnamefont {B.}~\bibnamefont {Gupt}},\ and\ \bibinfo {author} {\bibfnamefont {T.~R.}\ \bibnamefont {Bromley}},\ }\bibfield  {title} {\bibinfo {title} {Quantum computational finance: {{Monte Carlo}} pricing of financial derivatives},\ }\href {https://doi.org/10.1103/PhysRevA.98.022321} {\bibfield  {journal} {\bibinfo  {journal} {Physical Review A}\ }\textbf {\bibinfo {volume} {98}},\ \bibinfo {pages} {022321} (\bibinfo {year} {2018})}\BibitemShut {NoStop}%
\bibitem [{\citenamefont {Woerner}\ and\ \citenamefont {Egger}(2019)}]{woerner2019Quantumrisk}%
  \BibitemOpen
  \bibfield  {author} {\bibinfo {author} {\bibfnamefont {S.}~\bibnamefont {Woerner}}\ and\ \bibinfo {author} {\bibfnamefont {D.~J.}\ \bibnamefont {Egger}},\ }\bibfield  {title} {\bibinfo {title} {Quantum risk analysis},\ }\href {https://doi.org/10.1038/s41534-019-0130-6} {\bibfield  {journal} {\bibinfo  {journal} {npj Quantum Information}\ }\textbf {\bibinfo {volume} {5}},\ \bibinfo {pages} {15} (\bibinfo {year} {2019})}\BibitemShut {NoStop}%
\bibitem [{\citenamefont {MacMahon}\ and\ \citenamefont {Garlaschelli}(2015)}]{macmahon2015CommunityDetection}%
  \BibitemOpen
  \bibfield  {author} {\bibinfo {author} {\bibfnamefont {M.}~\bibnamefont {MacMahon}}\ and\ \bibinfo {author} {\bibfnamefont {D.}~\bibnamefont {Garlaschelli}},\ }\bibfield  {title} {\bibinfo {title} {Community {{Detection}} for {{Correlation Matrices}}},\ }\href {https://doi.org/10.1103/PhysRevX.5.021006} {\bibfield  {journal} {\bibinfo  {journal} {Physical Review X}\ }\textbf {\bibinfo {volume} {5}},\ \bibinfo {pages} {021006} (\bibinfo {year} {2015})}\BibitemShut {NoStop}%
\bibitem [{\citenamefont {Wang}\ \emph {et~al.}(2020)\citenamefont {Wang}, \citenamefont {Hu}, \citenamefont {Sanders},\ and\ \citenamefont {Kais}}]{wang2020QuditsHighDimensional}%
  \BibitemOpen
  \bibfield  {author} {\bibinfo {author} {\bibfnamefont {Y.}~\bibnamefont {Wang}}, \bibinfo {author} {\bibfnamefont {Z.}~\bibnamefont {Hu}}, \bibinfo {author} {\bibfnamefont {B.~C.}\ \bibnamefont {Sanders}},\ and\ \bibinfo {author} {\bibfnamefont {S.}~\bibnamefont {Kais}},\ }\bibfield  {title} {\bibinfo {title} {Qudits and {{High-Dimensional Quantum Computing}}},\ }\bibfield  {journal} {\bibinfo  {journal} {Frontiers in Physics}\ }\textbf {\bibinfo {volume} {8}},\ \href {https://doi.org/10.3389/fphy.2020.589504} {10.3389/fphy.2020.589504} (\bibinfo {year} {2020})\BibitemShut {NoStop}%
\bibitem [{\citenamefont {Ringbauer}\ \emph {et~al.}(2022)\citenamefont {Ringbauer}, \citenamefont {Meth}, \citenamefont {Postler}, \citenamefont {Stricker}, \citenamefont {Blatt}, \citenamefont {Schindler},\ and\ \citenamefont {Monz}}]{ringbauer2022universalqudit}%
  \BibitemOpen
  \bibfield  {author} {\bibinfo {author} {\bibfnamefont {M.}~\bibnamefont {Ringbauer}}, \bibinfo {author} {\bibfnamefont {M.}~\bibnamefont {Meth}}, \bibinfo {author} {\bibfnamefont {L.}~\bibnamefont {Postler}}, \bibinfo {author} {\bibfnamefont {R.}~\bibnamefont {Stricker}}, \bibinfo {author} {\bibfnamefont {R.}~\bibnamefont {Blatt}}, \bibinfo {author} {\bibfnamefont {P.}~\bibnamefont {Schindler}},\ and\ \bibinfo {author} {\bibfnamefont {T.}~\bibnamefont {Monz}},\ }\bibfield  {title} {\bibinfo {title} {A universal qudit quantum processor with trapped ions},\ }\href {https://doi.org/10.1038/s41567-022-01658-0} {\bibfield  {journal} {\bibinfo  {journal} {Nature Physics}\ }\textbf {\bibinfo {volume} {18}},\ \bibinfo {pages} {1053} (\bibinfo {year} {2022})}\BibitemShut {NoStop}%
\bibitem [{\citenamefont {Chen}\ \emph {et~al.}(2022)\citenamefont {Chen}, \citenamefont {Yoo},\ and\ \citenamefont {Fang}}]{chen2022QuantumLong}%
  \BibitemOpen
  \bibfield  {author} {\bibinfo {author} {\bibfnamefont {S.~Y.-C.}\ \bibnamefont {Chen}}, \bibinfo {author} {\bibfnamefont {S.}~\bibnamefont {Yoo}},\ and\ \bibinfo {author} {\bibfnamefont {Y.-L.~L.}\ \bibnamefont {Fang}},\ }\bibfield  {title} {\bibinfo {title} {Quantum {{Long Short-Term Memory}}},\ }in\ \href {https://doi.org/10.1109/ICASSP43922.2022.9747369} {\emph {\bibinfo {booktitle} {{{ICASSP}} 2022 - 2022 {{IEEE International Conference}} on {{Acoustics}}, {{Speech}} and {{Signal Processing}} ({{ICASSP}})}}}\ (\bibinfo  {publisher} {IEEE},\ \bibinfo {address} {Singapore, Singapore},\ \bibinfo {year} {2022})\ pp.\ \bibinfo {pages} {8622--8626}\BibitemShut {NoStop}%
\bibitem [{\citenamefont {Beatty}\ and\ \citenamefont {Stilck~Fran{\c c}a}(2025)}]{beatty2025OrderQuantum}%
  \BibitemOpen
  \bibfield  {author} {\bibinfo {author} {\bibfnamefont {E.}~\bibnamefont {Beatty}}\ and\ \bibinfo {author} {\bibfnamefont {D.}~\bibnamefont {Stilck~Fran{\c c}a}},\ }\bibfield  {title} {\bibinfo {title} {Order p {{Quantum Wasserstein Distances}} from {{Couplings}}},\ }\bibfield  {journal} {\bibinfo  {journal} {Annales Henri Poincar{\'e}}\ }\href {https://doi.org/10.1007/s00023-025-01557-z} {10.1007/s00023-025-01557-z} (\bibinfo {year} {2025})\BibitemShut {NoStop}%
\end{thebibliography}%
\appendix

\section{Architecture of the discriminator}
\label{appendix:cnn_achitecture}
We trained the classical discriminator in our QGAN simulations with a convolutional neural network.
Table~\ref{tab:CNN-specifics} summarizes its properties and hyperparameters.
This choice is motivated by~\cite{takahashi2019Modelingfinancial}, where it successfully was applied as a discriminator of a GAN that generates financial time series. 
\begin{table}[htbp!]
    \vspace{5mm}
    \centering
    \includegraphics[width=0.9\linewidth]{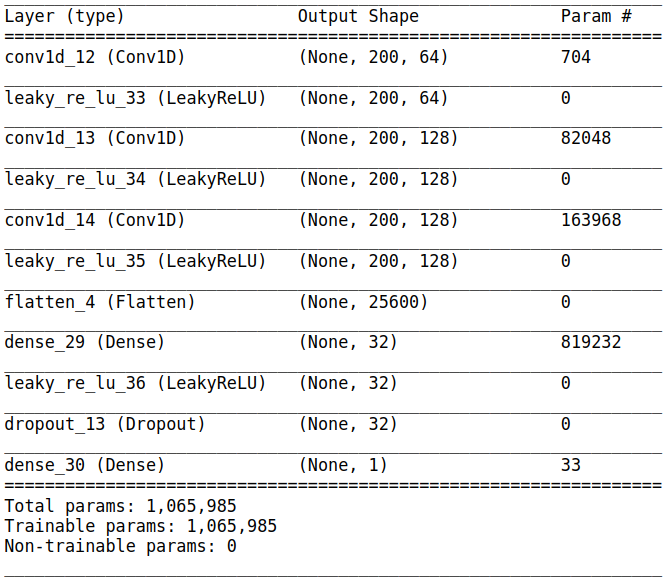}
    \caption{Hyperparameters and properties of the convolutional neural network used as the discriminator in the QGANs.}
    \label{tab:CNN-specifics}
\end{table}

\section{Comparison of foreign exchange currency pairs generated by quantum circuit Born machine}
\label{appendix:fx_comparison}
\begin{figure}[htbp!]
    \centering
    \includegraphics[width=1\linewidth]{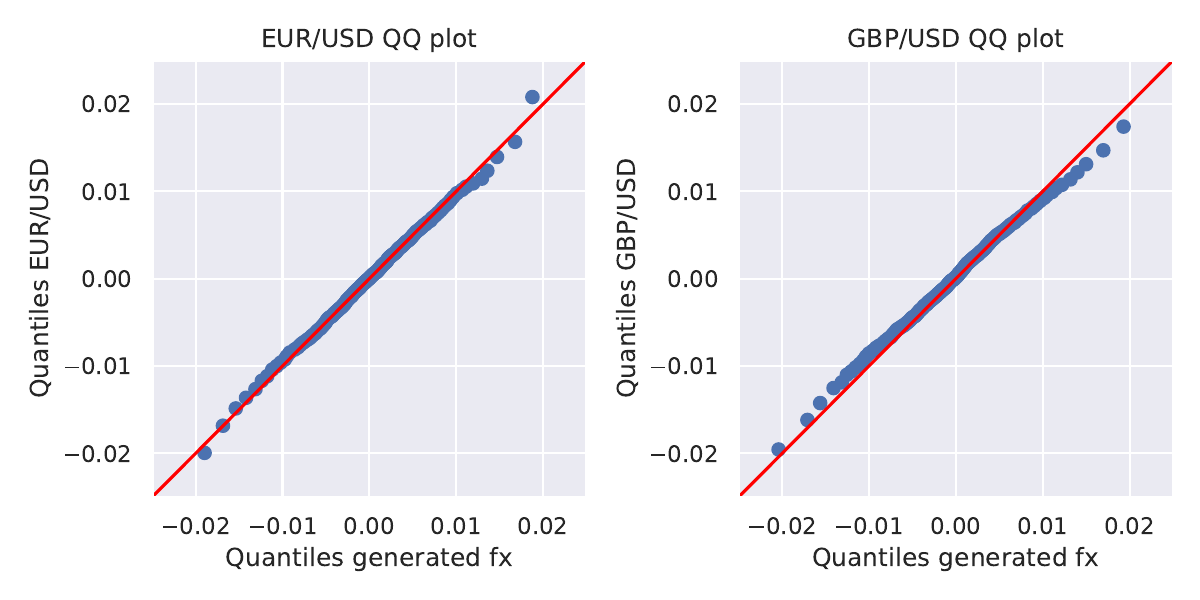}
    \caption{Quantile-quantile plot comparing samples from the trained QGAN model with a PQC of $4$ qubits and $4$ layers to the target distribution of EUR/USD and GBP/USD log returns.}
    \label{fig:fx_plot}
\end{figure}

In~\cite{coyle2021Quantumclassical}, a QGAN is constructed where the quantum generator was used as a quantum circuit Born machine. 
It was trained to generate distributions of foreign exchange pairs, producing samples that better matched the true distributions than those from a classical restricted Boltzmann machine with a comparable model size. \\
We also trained our QGAN, where the quantum circuit consisting of $4$ qubits and $4$ layers is simulated with the full-state approach in reproducing the same pairs of foreign exchanges as in~\cite{coyle2021Quantumclassical}.
We trained the single-qubit Pauli-$X$ and Pauli-$Z$ observables on the distributions of the EUR/USD and the GBP/USD foreign exchange log returns, respectively.
Figure~\ref{fig:fx_plot} shows the quantile-quantile plot comparing samples from our trained model with the target distribution.
Our trained QGAN samples match the target distribution more closely than the results for the quantum circuit Born machine and the classical restricted Boltzmann machine shown in Figure~10 of~\cite{coyle2021Quantumclassical}, while using fewer qubits than used for the quantum circuit Born machine.
This difference to the results from the quantum circuit Born machine comes from to the discrete nature of that model, which has naturally a higher imprecision of generated samples compared to the expectation value sampler used in our model.


\section{Full-state simulation: alternative circuit architecture}
\label{appendix:full-state-simulation-connected}
\begin{figure*}[htbp!]
    \centering
    \begin{tikzpicture}
        \node[inner sep=0pt] (img) at (0,0)
            {\includegraphics[width=0.8\linewidth]{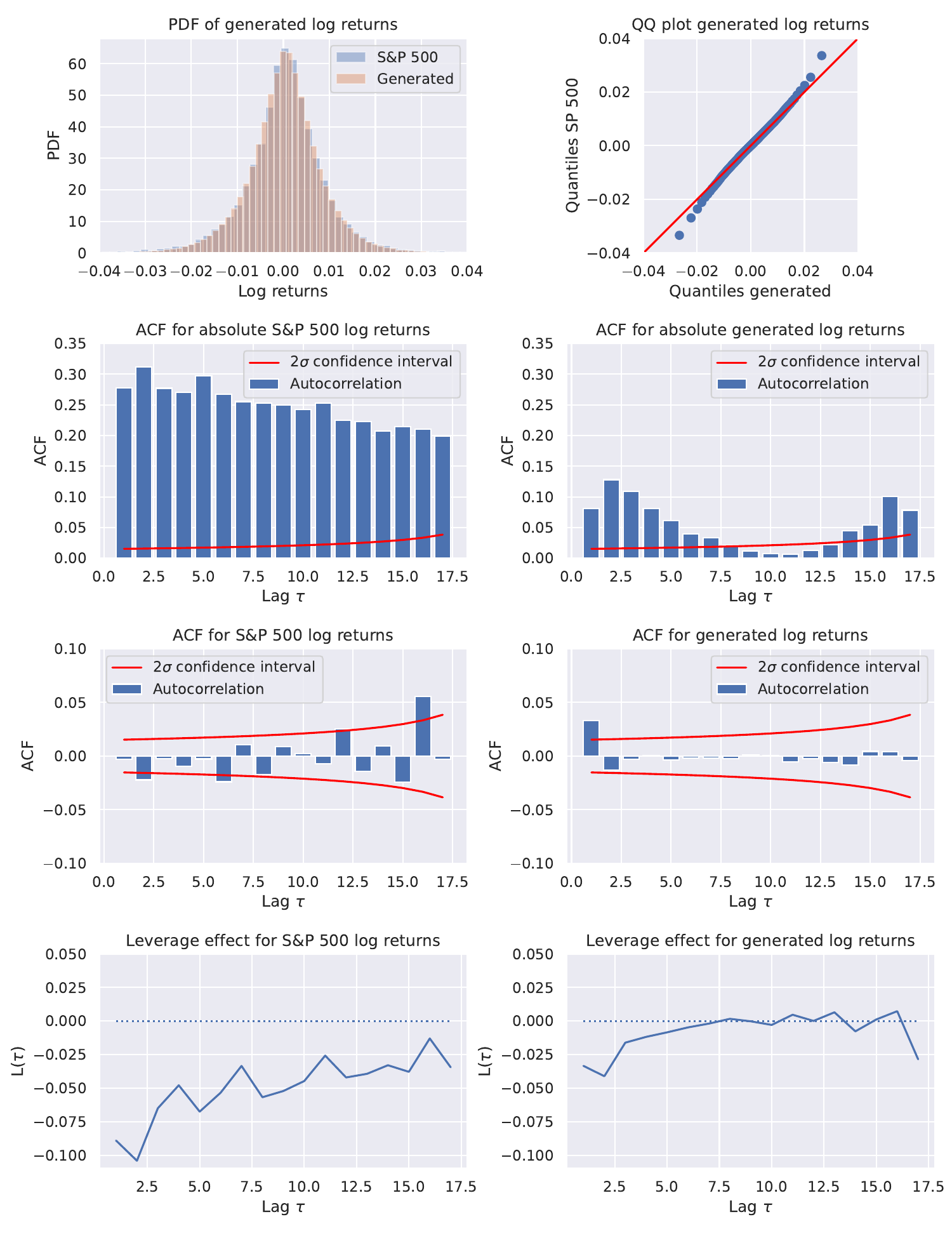}};

        \node at (-5.8,8.9) {\textbf{(a)}};
        \node at (1.2,8.9) {\textbf{(b)}};

        \node at (-5.8,4.5) {\textbf{(c)}};
        \node at (1.2,4.5) {\textbf{(d)}};

        \node at (-5.8,-0.1) {\textbf{(e)}};
        \node at (1.2,-0.1) {\textbf{(f)}};

        \node at (-5.8,-4.7) {\textbf{(g)}};
        \node at (1.2,-4.7) {\textbf{(h)}};

    \end{tikzpicture}
    \caption{Metrics of the stylized facts for a synthetic time series of window size $20$ generated by a QGAN, compared to the metrics of the S\&P~500 index. 
    The generator of the QGAN is a PQC consisting of $10$ qubits and $4$ layers, simulated with the full-state approach. 
    Contrary to the PQC used in Figure~\ref{fig:full-state-simulation}, we added an additional CNOT gate between the first and the 10th qubit in each layer.
    In \textbf{(a)}, we plot the probability density functions and in \textbf{(b)} the quantile-quantile plot of both the S\&P~500 index and the generated time series.
    In \textbf{(c)}-\textbf{(h)}, we plot the metrics absolute autocorrelation, linear autocorrelation and the leverage effect, as an indication of the stylized facts as described in Section~\ref{subsec:financialtimeseries}. 
    The Subfigures \textbf{(c)},  \textbf{(e)} and \textbf{(g)} show the metrics of the S\&P~500 index and the Subfigures \textbf{(d)}, \textbf{(f)} and \textbf{(h)} the metrics of the generated time series, respectively. Confidence intervals are calculated as in~\cite{chatfield1975time}.}
    \label{fig:full-state-simulation-open-boundary}
\end{figure*}
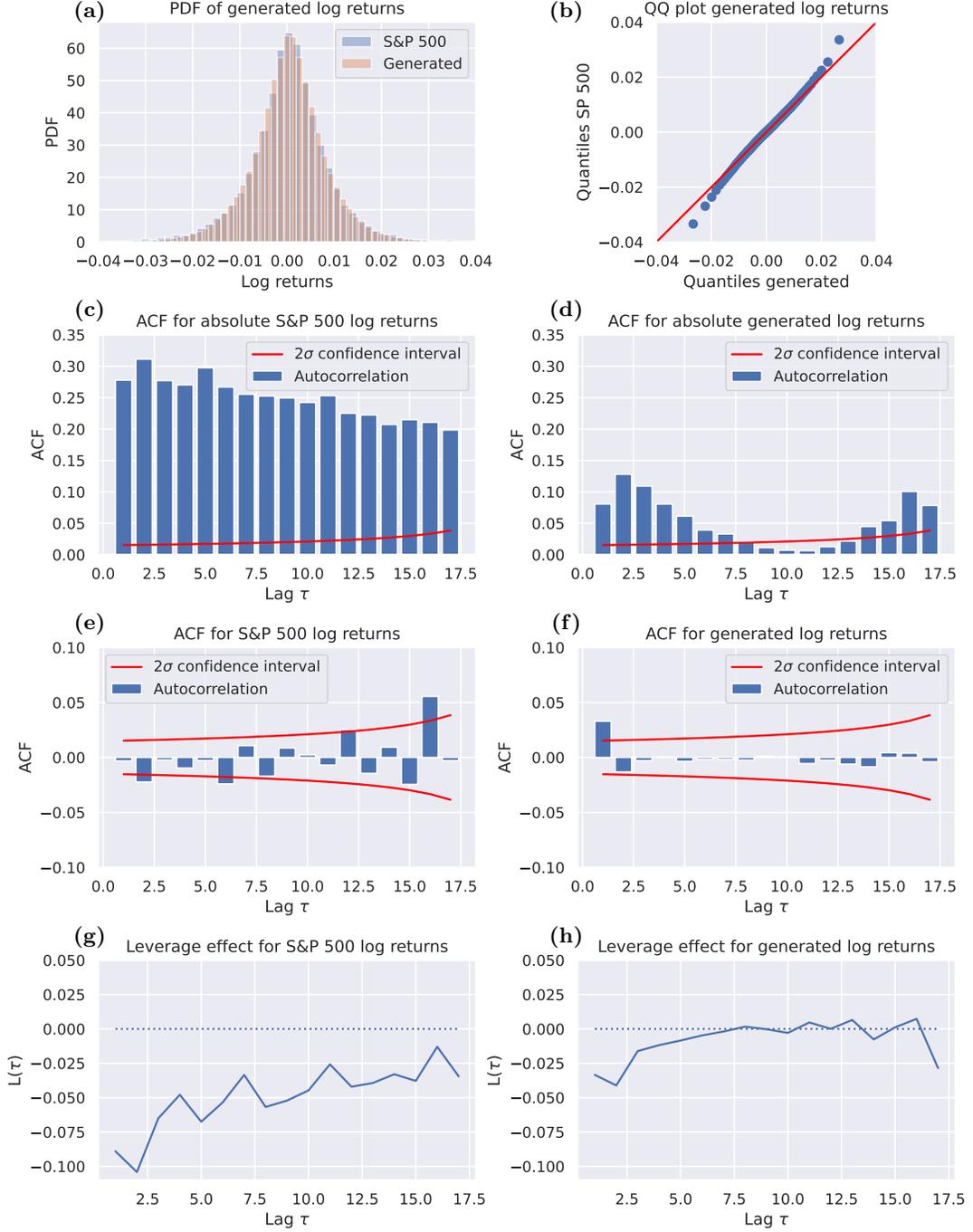

In addition to the PQC shown in Figure~\ref{fig:pqc}, we trained a QGAN using a modified PQC architecture simulated with the full-state approach. 
In order to increase long-range qubit correlations, we added a CNOT gate between the first and 10th qubit in each layer of the PQC (results in Figure~\ref{fig:full-state-simulation-open-boundary}).
Subfigure \textbf{(d)} shows that this architectural change increases the absolute autocorrelation at larger time lags.
The metrics of the generated time series are shown in column \textbf{(b)} of Table~\ref{tab: table of metrics of differnt sims}.

\section{MPS simulations for different numbers of layers and bond dimensions}
\label{appendix: table of mps sims}
\begin{sidewaysfigure}
 \vspace{7cm}
    \includegraphics[scale=0.43]{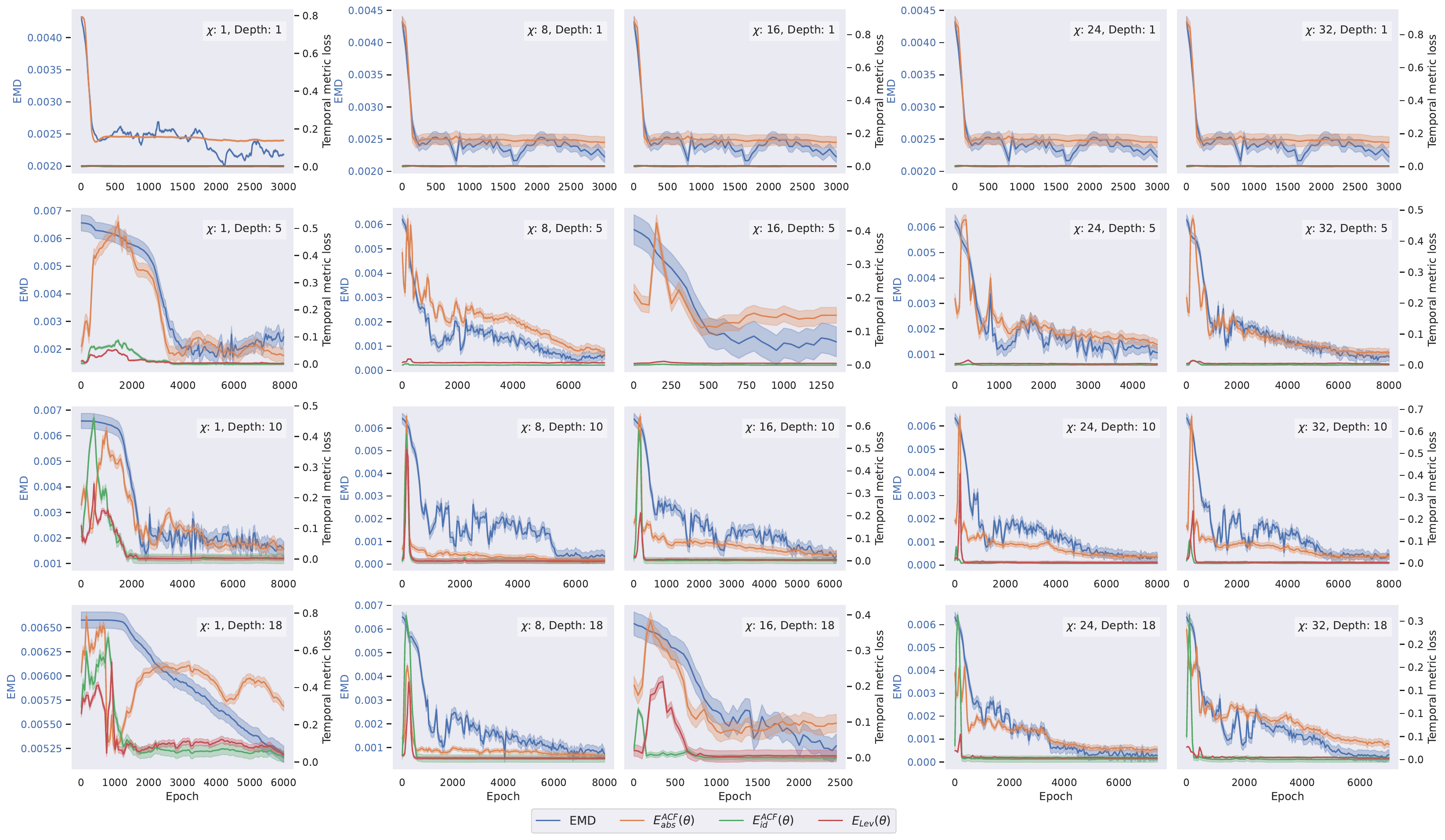}
    \caption{Wasserstein loss as defined in Equation \eqref{eq:loss-discriminator} (here called the EMD) and metrics corresponding to the temporal correlations as described in Section~\ref{subsec:financialtimeseries} in the training of the QGAN in the MPS simulation with $10$ qubits, $1,5,10$ and $18$ layers and bond dimensions $\chi$ of $1,8,16,24$ and $32$, depending on the number of epochs. We show mean and standard deviation of $5$ training runs.}
    \label{fig:mps_sims_table}
\end{sidewaysfigure}

In Figure~\ref{fig:mps_sims_table}, we show the quantitative metrics of training a QGAN where the PQC consisting of $10$ qubits are simulated with the MPS approach for $1,5,10$ and $18$ layers and bond dimensions of $1,8,16,24$ and $32$.
See Section~\ref{subsec:mps simulation}.
Note that a bond dimension of $32$ is giving an exact MPS approximation of the $10$-qubit state.

\end{document}